\documentclass[vecphys,envcountchap]{svmult}

\usepackage{makeidx}         
\usepackage{graphicx}        
\usepackage{multicol}        
\usepackage[bottom]{footmisc}

\usepackage{amsmath}
\usepackage[english]{babel}
\usepackage{epsfig}
\usepackage{color}

\newcommand{\beann}{\begin{eqnarray*}} \newcommand{\eeann}{\end{eqnarray*}}
\newcommand{\bea}{\begin{eqnarray}} \newcommand{\eea}{\end{eqnarray}}
\newcommand{\labs}{\left\vert} \newcommand{\rabs}{\right\vert}
\newcommand{\lab}{\left\langle} \newcommand{\rab}{\right\rangle}

\newcommand{\lrb}{\left(} \newcommand{\rrb}{\right)}
\newcommand{\lsb}{\left[} \newcommand{\rsb}{\right]}
\newcommand{\DSJ}{DSJ}
\newcommand{\SSJ}{SSJ}

\def\ii{\textrm{i}\,}
\def\ee{\textrm{e}\,}

\def\ie{\textit{i.e.}}
\def\Hc{\textrm{H.c.}}

\def\oc{\omega_{\textrm{c}}}
\def\KB{k_{\textrm{B}}}

\begin{document}

\title*{Green function techniques \\ in the treatment of quantum transport \\ at the molecular scale}
\titlerunning{Green function techniques in the treatment of quantum transport}

\author{D. A. Ryndyk, R. Guti\'{e}rrez, B. Song, and G. Cuniberti}

\institute{Dmitry\,A. Ryndyk \at Institute for Theoretical Physics, University of Regensburg, \\
D-93040 Regensburg, Germany, \\
\email{dmitry.ryndyk@physik.uni-regensburg.de}
\and Rafael Guti\'{e}rrez \at Institute for Material Science and Max Bergmann Center of Biomaterials, \\
Dresden University of Technology, D-01062 Dresden, Germany, \\
\email{rafael.gutierrez@tu-dresden.de} \and Bo
Song \at Institute for Material Science and Max Bergmann Center of Biomaterials, \\
Dresden University of Technology, D-01062 Dresden, Germany, \\
\email{bo.song@tu-dresden.de}
\and Gianaurelio Cuniberti \at Institute for Material Science and Max Bergmann Center of Biomaterials, \\
Dresden University of Technology, D-01062 Dresden, Germany, \\
\email{g.cuniberti@tu-dresden.de} }

\maketitle

\abstract{ The theoretical investigation of charge (and spin) transport at nanometer
length scales requires the use of advanced and powerful techniques able to deal with
the dynamical properties of the relevant physical systems, to explicitly include
out-of-equilibrium situations typical for electrical/heat transport as well as to
take into account interaction effects in a systematic way.  Equilibrium Green
function techniques and their extension to  non-equilibrium situations via the
Keldysh formalism build one of the pillars of current state-of-the-art approaches to
quantum transport which have been implemented in both model Hamiltonian formulations
and first-principle methodologies. In this chapter we offer a tutorial overview of
the applications of Green functions to deal with some fundamental aspects of charge
transport at the nanoscale, mainly focusing on applications to model Hamiltonian
formulations. }

\newpage
\section{Introduction}

The natural limitations that are expected to arise by the further miniaturization
attempts of semiconductor-based electronic devices have led in the past two decades
to the emergence of the new field of molecular electronics, where electronic
functions are going to be performed at the single-molecule level, see recent
overview in Refs.
\cite{Reed00sciam,Joachim00nature,Ratner02matertoday,Nitzan03science,
Cuniberti05book,Joachim05pnas}. The original conception which lies at the bottom of
this fascinating field can be traced back to the paper by Ari Aviram and Mark Ratner
in 1974 \cite{Aviram74cpl}, where a single-molecule rectifying diode was proposed.
Obviously, one of the core issues at stake in molecular electronics is to clarify
the question whether  single molecules (or more complex molecular aggregates) can
support an electric current. To achieve this goal, extremely refined experimental
techniques are required in order to probe the response of such a nano-object to
external fields. The meanwhile paradigmatic situation is that of a single molecule
contacted by two metallic electrodes between which a bias voltage is applied.
\nocite{Reed97science,Chen99science,Park00nature,Park02nature,Liang02nature,
Smit02nature,Kushmerick04nanolett,Yu04nanolett,Yu04prl,Xiao04nanolett,Elbing05pnas,
Venkataraman06nanolett,Poot06nanolett,Osorio07advmat,Loertscher07prl,
Porath97prb,Stipe97prl,Hahn01prl,Kim02prl,Ho02jcp,Nilius02science,Nilius03prl,Hla04prl,
Liu04jcp,Qiu04prl,Wu04prl,Repp04science,Repp05prl,Repp05prl2,Nazin05prl,Pradhan05prl,Choi06prl,
Martin06prl,Wu06science,Olsson07prl,Bannani07science,Liljeroth07science}

\subsubsection*{Recent experiments}

Enormous progress has been achieved  in the experimental realization of such
nano-devices, we only mention the development of controllable single-molecule
junctions \cite{Reed97science}-\cite{Loertscher07prl} and scanning tunneling
microscopy based techniques \cite{Porath97prb}-\cite{Liljeroth07science}. With their
help, a plethora of interesting phenomena like rectification~\cite{Elbing05pnas},
negative differential conductance~\cite{Chen99science,Repp05prl}, Coulomb
blockade~\cite{Porath97prb,Park00nature,Park02nature,Yu04nanolett,Yu04prl,Osorio07advmat},
Kondo effect~\cite{Park02nature,Liang02nature}, vibrational
effects~\cite{Park00nature,Hahn01prl,Smit02nature,Kushmerick04nanolett,Yu04prl,Liu04jcp,Qiu04prl,
Wu04prl,Repp05prl,Repp05prl2,Osorio07advmat}, and nanoscale memory
effects~\cite{Repp04science,Choi06prl,Martin06prl,Olsson07prl,Liljeroth07science},
among others, have been demonstrated.

The traditional semiconductor nanoelectronics also remains in front of modern
research, in particular due to recent experiments with small quantum dots, where
cotunneling effects were observed \cite{Schleser05prl,Sigrist06prl,Sigrist07prl}, as
well as new rectification effects in double quantum dots, interpreted as spin
blockade~\cite{Ono02science,Johnson05prb,Muralidharan07prb,Inarrea07prb}. Note, that
semiconductor experiments are very well controlled at present time, so they play an
important role as a benchmark for the theory.

Apart from single molecules, carbon nanotubes have also found extensive applications
and have been the target of experimental and theoretical studies over the last
years, see Ref.~\cite{Charlier07rmp} for a very recent review. The expectations to
realize electronics at the molecular scale also reached into the domain of
bio-molecular systems, thus opening new perspectives for the field due to the
specific self-recognition and self-assembling properties of biomolecules. For
instance, DNA oligomers have been already used as templates in molecular electronic
circuits \cite{Richter01apl,Mertig02nanolett,Keren03science}. Much less clear is,
however, if bio-molecules, and more specifically short DNA oligomers could also act
as wiring systems. Their electrical response properties are much harder to disclose
and there is still much controversy about the factors that determine charge
migration through such systems
\cite{Meggers98jacs,Kelley99science,Porath00nature,Treadway02chemphys,Xu04nanolett,
Gutierrez06inbook1,Gutierrez06inbook2,Porath04topics,Shapir08nature}.

\subsubsection*{Theoretical methods}

The theoretical treatment of transport at the nanoscale (see introduction
in~\cite{Datta95book,Haug96book,Ferry97book,Dittrich98book,Bruus04book,Datta05book})
requires the combined use of different techniques which range from minimal model
Hamiltonians, passing through semi-empirical methods up to full first-principle
methodologies. We mention here some important contributions, while we have no
possibility to cite all relevant papers.

Model Hamiltonians can in a straightforward way select, out of the many variables
that can control charge migration those which are thought to be the most relevant
ones for a specific molecule-electrode set-up. They contain, however, in a sometimes
not well-controlled way, many free parameters; hence, they can point at generic
effects, but they must be complemented with other methodologies able to yield
microscopic specific information. Semi-empirical methods can deal with rather large
systems due to the use of special subsets of electronic states to construct
molecular Hamiltonians as well as to the approximate treatment of interactions, but
often have the drawback of not being transferable. {\it Ab initio} approaches,
finally, can deal in a very precise manner with the electronic and atomic structure
of the different constituents of a molecular junction (metallic electrodes,
molecular wire, the interface) but it is not {\it apriori} evident that they can
also be applied to strong non-equilibrium situations.

From a more formal standpoint, there are roughly two main theoretical frameworks
that can be used to study  quantum transport in nanosystems at finite voltage:
generalized master equations (GME) \cite{Weiss99book,Breuer02book} and
nonequilibrium Green function (NGF) techniques
\cite{Kadanoff62book,Keldysh64,Langreth76inbook,Rammer86RMP,Haug96book}. The former
also lead to more simple rate equations  in the case where (i) the electrode-system
coupling can be considered as a weak perturbation, and (ii) off-diagonal elements of
the reduced density matrix in the eigenstate representation (coherences) can be
neglected due to very short decoherence times. Both approaches, the GME and NGF
techniques, can yield formally exact expressions for many observables. For
non-interacting systems, one can even solve analytically many models. However, once
interactions are introduced - and these are the most interesting cases containing a
very rich physics - different approximation schemes have to be introduced to make
the problems tractable.

In this chapter, we will review mainly the technique of non-equilibrium Green
functions. This approach is able to deal with a very broad variety of physical
problems related to quantum transport at the molecular scale. It can deal with
strong non-equilibrium situations via an extension of the conventional GF formalism
to the Schwinger-Keldysh contour~\cite{Keldysh64} and it can also include
interaction effects (electron-electron, electron-vibron, etc) in a systematic way
(diagrammatic perturbation theory, equation of motion techniques). Proposed first
time for the mesoscopic structures in the early seventies by Caroli et
al.~\cite{Caroli71jpcss1,Caroli71jpcss2,Caroli71jpcss3,Caroli71jpcss4}, this
approach was formulated in an elegant way by Meir, Wingreen and
Jauho~\cite{Meir92prl,Wingreen93prb,Jauho94prb,Haug96book,Jauho06jpcs}, who derived
exact expression for nonequilibrium current through an interacting nanosystem placed
between large noninteracting leads in terms of the nonequilibrium Green functions of
the nanosystem. Still, the problem of calculation of these Green functions is not
trivial. We consider some possible approaches in the case of electron-electron and
electron-vibron interactions. Moreover, as we will show later on, it can reproduce
results obtained within the master equation approach in the weak coupling limit to
the electrodes (Coulomb blockade), but it can also go {\it beyond} this limit and
cover intermediate coupling (Kondo effect) and strong coupling (Fabry-Perot)
domains. It thus offer the possibility of dealing with different physical regimes in
a unified way.

Now we review briefly some results obtained recently in the main directions of
modern research: general nanoscale quantum transport theory, atomistic transport
theory and applications to particular single-molecule systems.

\subsubsection*{General nanoscale quantum transport theory}

On the way to interpretation of modern experiments with single-molecule junctions
and STM spectroscopy of single molecules on surfaces, two main theoretical problems
are to be solved. The first is development of appropriate models based on ab initio
formulation. The second is effective and scalable theory of quantum transport
through multilevel interacting systems. We first consider the last problem, assuming
that the model Hamiltonian is known. Quantum transport through {\em noninteracting}
system can be considered using the famous Landauer-B\"{u}ttiker
method~\cite{Landauer57ibm,Landauer70philmag,Economou81prl,FisherLee81prb,Buttiker85prb,
Buttiker86prl,Landauer88ibm,Buttiker88ibm,Stone88ibm,Baranger89prb}, which
establishes the fundamental relation between the wave functions (scattering
amplitudes) of a system and its conducting properties. The method can be applied to
find the current through a noninteracting system or through an {\em effectively
noninteracting} system, for example if the mean-field description is valid and the
inelastic scattering is not essential. Such type of an electron transport is called
coherent, because there is no phase-breaking and quantum interference is preserved
during the electron motion across the system. In fact, coherence is initially
assumed in many {\em ab initio} based transport methods (DFT+NGF, and others), so
that the Landauer-B\"{u}ttiker method is now routinely applied to any basic
transport calculation through nanosystems and single molecules. Besides, it is
directly applicable in many semiconductor quantum dot systems with weak
electron-electron interactions. Due to simplicity and generality of this method, it
is now widely accepted and is in the base of our understanding of coherent
transport.

However, the peculiarity of single-molecule transport is just essential role of
electron-electron and electron-vibron interactions, so that Landauer-B\"{u}ttiker
method is not enough usually to describe essential physics even qualitatively.
During last years many new methods were developed to describe transport at finite
voltage, with focus on correlation and inelastic effects, in particular in the cases
when Coulomb blockade, Kondo effect and vibronic effects take place.

{\bf Vibrons} (the localized phonons) are very important because molecules are
flexible. The theory of electron-vibron interaction has a long history, but many
questions it implies are not answered up to now. While the isolated electron-vibron
model can be solved exactly by the so-called polaron or Lang-Firsov transformation
\cite{Lang63jetp,Hewson74jjap,Mahan90book}, the coupling to the leads produces a
true many-body problem. The inelastic resonant tunneling of {\em single} electrons
through the localized state coupled to phonons was considered in
Refs.~\cite{Glazman88jetp,Wingreen88prl,Wingreen89prb,Jonson89prb,Cizek04prb,Cizek05czechjp}.
There the exact solution in the single-particle approximation was derived, ignoring
completely the Fermi sea in the leads. At strong electron-vibron interaction and
weak couplings to the leads the satellites of the main resonant peak are formed in
the spectral function.

The essential progress in calculation of transport properties in the strong
electron-vibron interaction limit has been made with the help of the master equation
approach~\cite{Braig03prb,Aji03condmat,Mitra04prb,Koch05prl,Koch06prb,Koch06prb2,Wegewijs05condmat,
Zazunov06prb1,Ryndyk08preprint}. This method, however, is valid only in the limit of
very weak molecule-to-lead coupling and neglects all spectral effects, which are the
most important at finite coupling to the leads.

At strong coupling to the leads and the finite level width the master equation
approach can no longer be used, and we apply alternatively the nonequilibrium Green
function technique which have been recently developed to treat vibronic effects in a
perturbative or self-consistent way in the cases of weak and intermediate
electron-vibron
interaction~\cite{Tikhodeev01surfsci,Mii02surfsci,Mii03prb,Tikhodeev04prb,Galperin04nanolett,
Galperin04jcp,Galperin05jphyschemb,Frederiksen04master,Frederiksen04prl,Hartung04master,
Ryndyk05prb,Ryndyk06prb,Ryndyk07prb,Paulsson05prb,Paulsson06jpconf,Arseyev05jetplett,Arseyev06jetplett,Zazunov06prb}.

The case of intermediate and strong electron-vibron interaction {\em at intermediate
coupling to the leads} is the most interesting, but also the most difficult. The
existing approaches are
mean-field~\cite{Hewson79jphysc,Galperin05nanolett,DAmico08njp}, or start from the
exact solution for the isolated system and then treat tunneling as a
perturbation~\cite{Kral97,Lundin02prb,Zhu03prb,Alexandrov03prb,Flensberg03prb,Galperin06prb,Zazunov07prb}.
The fluctuations beyond mean-field approximations were considered in
Refs.\,~\cite{Mitra05prl,Mozyrsky06prb}

In parallel, the related activity was in the field of single-electron shuttles and
quantum
shuttles~\cite{Gorelik98prl,Boese01epl,Fedorets02epl,Fedorets03prb,Fedorets04prl,McCarthy03prb,
Novotny03prl,Novotny04prl,Blanter04prl,Smirnov04prb,Chtchelkatchev04prb}. Finally,
based on the Bardeen's tunneling Hamiltonian method
\cite{Bardeen61prl,Harrison61pr,Cohen62prl,Prange63pr,Duke69book} and Tersoff-Hamann
approach~\cite{Tersoff83prl,Tersoff85prb}, the theory of inelastic electron
tunneling spectroscopy (IETS) was
developed~\cite{Persson87prl,Gata93prb,Tikhodeev01surfsci,Mii02surfsci,Mii03prb,Tikhodeev04prb,Raza07condmat}.

The recent review of the electron-vibron problem and its relation to the molecular
transport see in Ref.~\cite{Galperin07jpcm}.

\textbf{Coulomb interaction} is the other important ingredient of the models,
describing single molecules. It is in the origin of such fundamental effects as
Coulomb blockade and Kondo efect. The most convenient and simple enough is
Anderson-Hubbard model, combining the formulations of Anderson impurity
model~\cite{Anderson61prev} and Hubbard many-body model
\cite{Hubbard63procroysoc,Hubbard64procroysoc1,Hubbard64procroysoc2}. To analyze
such strongly correlated system several complementary methods can be used: master
equation and perturbation in tunneling, equation-of-motion method, self-consistent
Green functions, renormalization group and different numerical methods.

When the coupling to the leads is weak, electron-electron interaction results in
Coulomb blockade, the sequential tunneling is described by the master equation
method~\cite{Averin86jltp,Averin91inbook,Grabert92book,vanHouten92inbook,
Kouwenhoven97inbook,Schoeller97inbook,Schoen97inbook,vanderWiel02rmp} and small
cotunneling current in the blockaded regime can be calculated by the next-order
perturbation theory~\cite{Averin89pla,Averin90prl,Averin92inbook}. This theory was
used successfully to describe electron tunneling via discrete quantum states in
quantum dots~\cite{Averin91prb,Beenakker91prb,vonDelft01pr,Bonet02prb}. Recently
there were several attempts to apply master equation to multi-level models of
molecules, in particular describing benzene
rings~\cite{Hettler03prl,Muralidharan06prb,Begemann08prb}.

To describe consistently cotunneling, level broadening and higher-order (in
tunneling) processes, more sophisticated methods to calculate the reduced density
matrix were developed, based on the Liouville - von Neumann
equation~\cite{Petrov04chemphys,Petrov05chemphys,Petrov06prb,Elste05prb,Harbola06prb,
Pedersen07prb,Mayrhofer07epjb,Begemann08prb} or real-time diagrammatic
technique~\cite{Schoeller94prb,Koenig96prl,Koenig96prb,Koenig97prl,Koenig98prb,
Thielmann03prb,Thielmann05prl,Aghassi06prb}. Different approaches were reviewed
recently in Ref.\,\cite{Timm08prb}.

The equation-of-motion (EOM) method is one of the basic and powerful ways to find
the Green functions of interacting quantum systems. In spite of its simplicity it
gives the appropriate results for strongly correlated nanosystems, describing
qualitatively and in some cases quantitatively such important transport phenomena as
Coulomb blockade and Kondo effect in quantum dots. The results of the EOM method
could be calibrated with other available calculations, such as the master equation
approach in the case of weak coupling to the leads, and the perturbation theory in
the case of strong coupling to the leads and weak electron-electron interaction.

In the case of a single site junction with two (spin-up and spin-down) states and
Coulomb interaction between these states (Anderson impurity model), the \emph{linear
conductance} properties have been successfully studied by means of the EOM approach
in the cases related to Coulomb blockade\cite{Lacroix81jphysf,Meir91prl} and the
Kondo effect~\cite{Meir93prl}. Later the same method was applied to some two-site
models~\cite{Niu95prb,Pals96jpcm,Lamba00prb,Song07prb}. Multi-level systems were
started to be considered only recently~\cite{Palacios97prb,Yi02prb}. Besides, there
are some difficulties in building the lesser GF in the nonequilibrium case (at
finite bias voltages) by means of the EOM
method~\cite{Niu99jpcm,Swirkowicz03prb,Bulka04prb}.

The diagrammatic method was also used to analyze the Anderson impurity model. First
of all, the perturbation theory can be used to describe weak electron-electron
interaction and even some features of the Kondo effect~\cite{Fujii03prb}. The family
of nonperturbative current-conserving self-consistent approximations for Green
functions has a long history and goes back to the Schwinger functional derivative
technique, Kadanoff-Baym approximations and Hedin equations in the equilibrium
many-body
theory~\cite{Schwinger51pnas,Martin59prev,Baym61prev,Baym62prev,Hedin65prev,
Gunnarsson88rpp,White92prb,Onida02rmp}.
Recently GW approximation was investigated together with other
methods~\cite{Thygesen07jcp,Thygesen08prl,Thygesen08prb,Wang08prb}. It was shown
that dynamical correlation effects and self-consistency can be very important at
finite bias.

Finally, we want to mention briefly three important fields of research, that we do
not consider in the present review: the theory of Kondo
effect~\cite{Glazman88jetplett,Ng88prl,Hershfield92prb,Meir93prl,Wingreen94prb,Aguado00prl,Kim01prb,Plihal01prb},
spin-dependent
transport~\cite{Sanvito06jctn,Rocha06prb,Naber07jpdap,Ke08prl,Ning08prl}, and
time-dependent
transport~\cite{Jauho94prb,Grifoni98pr,Kohler05pr,Sanchez06prb,Maciejko06prb}.

\subsubsection*{Atomistic transport theory}

Atomistic transport theory utilizes semi-empirical
(tight-binding~\cite{Cuniberti02cp,Todorov02jpcm}) or {\em ab initio} based methods.
In all cases the microscopic structure is taken into account with different level of
accuracy.

The most popular is the approach combining density-functional theory (DFT) and NGF
and known as
DFT+NGF~\cite{Taylor01prb1,Taylor01prb2,Damle01prb,Xue01jcp,Xue02cp,Xue03prb1,Xue03prb2,
DiCarlo02pb,Frauenheim02jpcm,Brandbyge02prb,Taylor03prb,Lee04prb,Pecchia04repprogphys,Ke04prb,Ke05prb,
Ke05jcp1,Ke05jcp2,Liu06jcp,Ke07jcp,Pump07ss,DelValle07naturenanotech,Pecchia07prb,Pecchia08njp}.
This method, however is not free from internal problems. First of all, it is
essentially mean-filed method neglecting strong local correlations and inelastic
scattering. Second, density-functional theory is a ground state theory and e.g.~the
transmission calculated using static DFT eigenvalues will display peaks  at the
Kohn-Sham excitation energies, which in general do not coincide with the true
excitation energies. Extensions to include excited states as in time-dependent
density-functional theory, though very
promising~\cite{Sai05prl,Kurth05prb,DiVentra07prl}, are not fully developed up to
date.

To improve DFT-based models several approaches were suggested, including inelastic
electron-vibron
interaction~\cite{Frederiksen04prl,Pecchia04nanolett,Sergueev05prl,Paulsson05prb,Paulsson06nanolett,
Frederiksen07prb,Frederiksen07prb2,Gagliardi08njp,Raza08condmat,Raza08prb} or
Coulomb interaction beyond mean-field level~\cite{Ferretti05prb}, or based on the
LDA+U approache~\cite{Anisimov91prb}. The principally different alternative to DFT
is to use {\em an initio quantum chemistry} based many-body quantum transport
approach~\cite{Delaney04prl,Delaney04intjquantumchem,Fagas06prb,Albrecht06jap}.

Finally, transport in bio-molecules attracted more attention, in particular
electrical conductance of
DNA~\cite{Cuniberti02prb,Gutierrez03eurolett,Gutierrez05nanolett,Gutierrez05prb,Gutierrez06prb}.

\subsubsection*{Outline}

The review is organized as follows. In Sections {\bf 2} we will first introduce the
Green functions for non-interacting systems, and present few examples of transport
through non-interacting regions. Then we review the master equation approach and its
application to describe Coulomb blockade and vibron-mediated Franck-Condon blockade.
In Section {\bf 3} the Keldysh NGF technique is developed in detail. In equilibrium
situations or within the linear response regime,  dynamic response and static
correlation functions are related via the fluctuation-dissipation theorem. Thus,
solving Dyson equation for the retarded GF is enough to obtain the correlation
functions. In strong out-of-equilibrium situations, however, dynamic response and
correlation functions have to be calculated simultaneously and are not related by
fluctuation-dissipation theorems. The Kadanof-Baym-Keldysh approach yield  a
compact, powerful formulation to derive Dyson and kinetic equations for
non-equilibrium systems. In Sec. {\bf 4} we present different applications of the
Green function techniques. We show how Coulomb blockade can be described within the
Anderson-Hubbard model, once an appropriate truncation of the equation of motion
hierarchy is performed (Sec. 4.A). Further, the paradigmatic case of transport
through a single electronic level coupled to a local vibrational mode is discussed
in detail within the context of the self-consistent Born approximation. It is shown
that already this simple model can display non-trivial physics (Sec. 4.B). Finally,
the case of an electronic system interacting with a bosonic bath is discussed in
Sec. 4.C where it is shown that the presence of an environment with a continuous
spectrum can  modified the low-energy analytical structure of the Green function and
lead to dramatic changes in the electrical response of the system. We point at the
relevance of this situation to discuss transport experiments in short DNA oligomers.
We have not addressed the problem of the (equilibrium or non-equilibrium)  Kondo
effect, since this issue alone would require a chapter on its own due to the
non-perturbative character of the processes leading to the formation of the Kondo
resonance.

In view of the broadness of the topic, the authors were forced to do a very subjective
selection of the topics to be included in this review as well as of the most relevant
literature. We thus apologize for the omission of many interesting studies which could
not be dealt with in the restricted space at our disposal. We refer the interested
reader to the other contributions in this book and the cited papers.

\section{From coherent transport to sequential tunneling (basics)}

\subsection{Coherent transport: single-particle Green functions}

Nano-scale and molecular-scale systems are naturally described by the discrete-level
models, for example eigenstates of quantum dots, molecular orbitals, or atomic
orbitals. But the leads are very large (infinite) and have a continuous energy
spectrum. To include the lead effects systematically, it is reasonable to start from
the discrete-level representation for the whole system. It can be made by the
tight-binding (TB) model, which was proposed to describe quantum systems in which
the localized electronic states play an essential role, it is widely used as an
alternative to the plane wave description of electrons in solids, and also as a
method to calculate the electronic structure of molecules in quantum chemistry.

A very effective method to describe scattering and transport is the Green function (GF)
method. In the case of non-interacting systems and coherent transport  single-particle
GFs are used. In this section we consider the matrix Green function method for coherent
transport through discrete-level systems.

\paragraph*{(i) Matrix (tight-binding) Hamiltonian}

The main idea of the method is to represent the wave function of a particle as a linear
combination of some known {\em localized} states $\psi_\alpha({\bf r},\sigma)$, where
$\alpha$ denote the set of quantum numbers, and $\sigma$ is the spin index (for example,
atomic orbitals, in this particular case the method is called LCAO -- linear combination
of atomic orbitals)
\begin{equation}\label{TB-psi}
  \psi(\xi)=\sum_\alpha c_\alpha\psi_\alpha(\xi),
\end{equation}
here and below we use $\xi\equiv(\bf r,\sigma)$ to denote both spatial coordinates and
spin.

Using the Dirac notations $|\alpha\rangle\equiv\psi_\alpha(\xi)$ and assuming that
$\psi_\alpha(\xi)$ are orthonormal functions
$\langle\alpha|\beta\rangle=\delta_{\alpha\beta}$ we can write the {\em single-particle
matrix (tight-binding) Hamiltonian} in the Hilbert space formed by $\psi_\alpha(\xi)$
\begin{equation}\label{TB-Hamiltonian}
  \hat H=\sum_\alpha (\epsilon_\alpha+e\varphi_\alpha)|\alpha\rangle\langle \alpha|+
  \sum_{\alpha\beta}t_{\alpha\beta}|\alpha\rangle\langle \beta|,
\end{equation}
the first term in this Hamiltonian describes the states with energies $\epsilon_\alpha$,
$\varphi_\alpha$ is the electrical potential, the second term should be included if the
states $|\alpha\rangle$ are not eigenstates of the Hamiltonoian. In the TB model
$t_{\alpha\beta}$ is the hopping matrix element between states $|\alpha\rangle$ and
$|\beta\rangle$, which is nonzero, as a rule, for nearest neighbor cites. The
two-particle interaction is described by the Hamiltonian
\begin{equation}\label{TB-Hamiltonian-2}
  \hat H=
  \sum_{\alpha\beta,\delta\gamma}V_{\alpha\beta,\delta\gamma}
  |\alpha\rangle|\beta\rangle\langle \delta|\langle \gamma|,
\end{equation}
in the two-particle Hilbert space, and so on.

The energies and hopping matrix elements in this Hamiltomian can be calculated, if the
single-particle real-space Hamiltomian $\hat h(\xi)$ is known:
\begin{equation}
\epsilon_\alpha\delta_{\alpha\beta}+t_{\alpha\beta}= \int\psi^*_\alpha(\xi)\hat h(\xi)
\psi_\beta(\xi) d\xi.
\end{equation}

This approach was developed originally as an approximate method, if the wave functions of
isolated atoms are taken as a basis wave functions $\psi_\alpha(\xi)$, but also can be
formulated exactly with the help of Wannier functions. Only in the last case the
expansion (\ref{TB-psi}) and the Hamiltonian (\ref{TB-Hamiltonian}) are exact, but some
extension to the arbitrary basis functions is possible. In principle, the TB model is
reasonable only when {\em local} states can be orthogonalized. The method is useful to
calculate the conductance of complex quantum systems in combination with {\em ab initio}
methods. It is particular important to describe small molecules, when the atomic orbitals
form the basis.

In the mathematical sense, the TB model is a discrete (grid) version of the
continuous Schr\"odinger equation, thus it is routinely used in numerical
calculations.

To solve the single-particle problem it is convenient to introduce a new representation,
where the coefficients $c_\alpha$ in the expansion (\ref{TB-psi}) are the components of a
vector wave function (we assume here that all states $\alpha$ are numerated by integers)
\begin{equation}\label{TB-Psi}
  \Psi=\left(\begin{array}{c} c_1 \\ c_2 \\ \vdots \\ c_N \end{array}\right),
\end{equation}
and the eigenstates $\Psi_\lambda$ are to be found from the matrix Schr\"odinger equation
\begin{equation}\label{TB-Schroedinger}
  {\bf H}\Psi_\lambda=E_\lambda\Psi_\lambda,
\end{equation}
with the matrix elements of the single-particle Hamiltonian
\begin{equation}
H_{\alpha\beta}=\left\{\begin{array}{ll}
\epsilon_\alpha+e\varphi_\alpha, & \alpha=\beta, \\
t_{\alpha\beta}, & \alpha\neq\beta.
\end{array}\right.
\end{equation}
\begin{figure}[t]
\begin{center}
\epsfxsize=0.5\hsize
\epsfbox{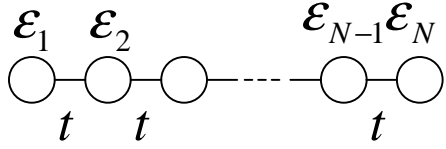}
\caption{A finite linear chain of single-level sites.}
\label{fig-3-1}
\end{center}
\end{figure}

Now let us consider some typical systems, for which the matrix method is appropriate
starting point. The simplest example is a single quantum dot, the basis is formed by the
{\em eigenstates}, the corresponding Hamiltonian is diagonal
\begin{equation}
  {\bf H}=\left(\begin{array}{ccccc}
  \epsilon_1 & 0 & 0 & \cdots & 0 \\
  0 & \epsilon_2 & 0 & \cdots & 0 \\
  \vdots & \ddots & \ddots & \ddots & \vdots \\
  0 & \cdots & 0 & \epsilon_{N-1} & 0 \\
  0 & \cdots & 0 & 0 & \epsilon_N
  \end{array}\right).
\end{equation}

The next typical example is a linear chain of single-state sites with only
nearest-neighbor couplings (Fig.\,\ref{fig-3-1})
\begin{equation}
  {\bf H}=\left(\begin{array}{ccccc}
  \epsilon_1 & t & 0 & \cdots & 0 \\
  t & \epsilon_2 & t & \cdots & 0 \\
  \vdots & \ddots & \ddots & \ddots & \vdots \\
  0 & \cdots & t & \epsilon_{N-1} & t \\
  0 & \cdots & 0 & t & \epsilon_N
  \end{array}\right).
\end{equation}

The method is applied as well to consider the semi-infinite leads. Although the matrices
are formally infinitely-dimensional in this case, we shall show below, that the problem
is reduced to the finite-dimensional problem for the quantum system of interest, and the
semi-infinite leads can be integrated out.

Finally, in the second quantized form the tight-binding Hamiltonian is
\begin{equation}\label{TB-H-SC}
  \hat H=\sum_\alpha\left(\epsilon_\alpha+e\varphi_\alpha\right)c^\dagger_\alpha c_\alpha+
  \sum_{\alpha\neq\beta}t_{\alpha\beta}c^\dagger_\alpha c_\beta.
\end{equation}

\paragraph*{(ii) Matrix Green functions and contact self-energies}

The solution of single-particle quantum problems, formulated with the help of a matrix
Hamiltonian, is possible along the usual line of finding the wave-functions on a lattice,
solving the Schr\"odinger equation (\ref{TB-Schroedinger}). The other method, namely
matrix Green functions, considered in this section, was found to be more convenient for
transport calculations, especially when interactions are included.

The retarded {\em single-particle} matrix Green function ${\bf G}^R(\epsilon)$ is
determined by the equation
\begin{equation}\label{TB-GR-EQ}
  \left[(\epsilon+i\eta){\bf I}-{\bf H}\right]{\bf G}^R={\bf I},
\end{equation}
where $\eta$ is an infinitesimally small positive number $\eta=0^+$.

For an isolated noninteracting system the Green function is simply obtained
after the matrix inversion
\begin{equation}
  {\bf G}^R=\left[(\epsilon+i\eta){\bf I}-{\bf H}\right]^{-1}.
\end{equation}
Let us consider the trivial example of a two-level system with the Hamiltonian
\begin{equation}
{\bf H}=\left(\begin{array}{cc}\epsilon_1&t\\ t&\epsilon_2\end{array}\right).
\end{equation}
The retarded GF is easy found to be ($\tilde\epsilon=\epsilon+i\eta$)
\begin{equation}
{\bf G}^R(\epsilon)=\frac{1}{(\tilde\epsilon-\epsilon_1)(\tilde\epsilon-\epsilon_2)+t^2}
\left(\begin{array}{cc}\tilde\epsilon-\epsilon_2&t\\
t&\tilde\epsilon-\epsilon_1\end{array}\right).
\end{equation}

\begin{figure}[t]
\begin{center}
\epsfxsize=0.6\hsize
\epsfbox{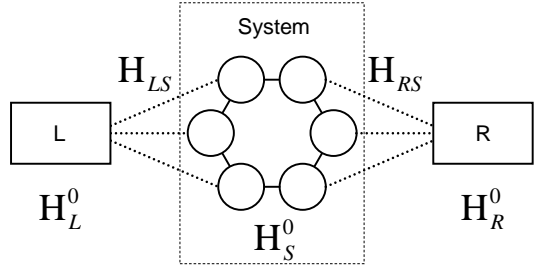}
\caption{A quantum system coupled to the left and right leads.}
\label{fig-3-2}
\end{center}
\end{figure}

Now let us consider the case, when the system of interest is coupled to two contacts
(Fig.\,\ref{fig-3-2}). We assume here that the contacts are also described by the
tight-binding model and by the matrix GFs. Actually, the semi-infinite contacts should be
described by the matrix of infinite dimension. We shall consider the semi-infinite
contacts in the next section.

Let us present the full Hamiltonian of the considered system in a following block
form
\begin{equation}
{\bf H}=\left(\begin{array}{ccc}{\bf H}^0_L & {\bf H}_{LS} & 0 \\
{\bf H}^\dag_{LS} & {\bf H}^0_S & {\bf H}^\dag_{RS} \\
0 & {\bf H}_{RS} & {\bf H}^0_R
\end{array}\right),
\end{equation}
where ${\bf H}^0_L$, ${\bf H}^0_S$, and ${\bf H}^0_R$ are Hamiltonians of the left lead,
the system, and the right lead separately. And the off-diagonal terms describe
system-to-lead coupling. The Hamiltonian should be hermitian, so that
\begin{equation}
  {\bf H}_{SL}={\bf H}^\dag_{LS},\ \ \ {\bf H}_{SR}={\bf H}^\dag_{RS}.
\end{equation}

The Eq.\,(\ref{TB-GR-EQ}) can be written as
\begin{equation}\label{TB-matrix}
\left(\begin{array}{ccc}{\bf E}-{\bf H}^0_L & -{\bf H}_{LS} & 0 \\
-{\bf H}^\dag_{LS} & {\bf E}-{\bf H}^0_S & -{\bf H}^\dag_{RS} \\
0 & -{\bf H}_{RS} & {\bf E}-{\bf H}^0_R
\end{array}\right)
\left(\begin{array}{ccc}{\bf G}_{L} & {\bf G}_{LS} & 0 \\
{\bf G}_{SL} & {\bf G}_{S} & {\bf G}_{SR} \\
0 & {\bf G}_{RS} & {\bf G}_{R}
\end{array}\right)={\bf I},
\end{equation}
where we introduce the matrix ${\bf E}=(\epsilon+i\eta){\bf I}$, and represent the matrix
Green function in a convenient form, the notation of retarded function is omitted in
intermediate formulas. Now our first goal is to find the system Green function ${\bf
G}_{S}$ which defines all quantities of interest. From the matrix equation
(\ref{TB-matrix})
\begin{align}
\left({\bf E}-{\bf H}^0_L\right){\bf G}_{LS}-{\bf H}_{LS}{\bf G}_{S}=0, \\
-{\bf H}^\dag_{LS}{\bf G}_{LS}+\left({\bf E}-{\bf H}^0_S\right){\bf G}_{S}-
{\bf H}^\dag_{RS}{\bf G}_{RS}={\bf I}, \\
-{\bf H}_{RS}{\bf G}_{S}+\left({\bf E}-{\bf H}^0_R\right){\bf G}_{RS}=0.
\end{align}
From the first and the third equations one has
\begin{align}
& {\bf G}_{LS}=\left({\bf E}-{\bf H}^0_L\right)^{-1}{\bf H}_{LS}{\bf G}_{S}, \\
& {\bf G}_{RS}=\left({\bf E}-{\bf H}^0_R\right)^{-1}{\bf H}_{RS}{\bf G}_{S},
\end{align}
and substituting it into the second equation we arrive at the equation
\begin{equation}\label{TB-Dyson}
  \left({\bf E}-{\bf H}^0_S-{\bf\Sigma}\right){\bf G}_{S}={\bf I},
\end{equation}
where we introduce the {\em contact self-energy} (which should be also called retarded,
we omit the index in this chapter)
\begin{equation}\label{TB-Sigma}
  {\bf\Sigma}={\bf H}^\dag_{LS}\left({\bf E}-{\bf H}^0_L\right)^{-1}{\bf H}_{LS}+
  {\bf H}^\dag_{RS}\left({\bf E}-{\bf H}^0_R\right)^{-1}{\bf H}_{RS}.
\end{equation}

Finally, we found, that the retarded GF of a nanosystem coupled to the leads is
determined by the expression
\begin{equation}
  {\bf G}^R_{S}(\epsilon)=\left[(\epsilon+i\eta){\bf I}-{\bf H}^0_S-{\bf\Sigma}\right]^{-1},
\end{equation}
the effects of the leads are included through the self-energy.

Here we should stress the important property of the self-energy (\ref{TB-Sigma}), it is
determined only by the coupling Hamiltonians and the retarded GFs of the {\em isolated}
leads ${\bf G}^{0R}_{i}=\left({\bf E}-{\bf H}^0_R\right)^{-1}$ ($i=L,R$)
\begin{equation}\label{TB-Sigma2}
  {\bf\Sigma}_i={\bf H}^\dag_{iS}\left({\bf E}-{\bf H}^0_i\right)^{-1}{\bf H}_{iS}=
  {\bf H}^\dag_{iS}{\bf G}^{0R}_{i} {\bf H}_{iS},
\end{equation}
it means, that the contact self-energy is independent of the state of the nanosystem
itself and describes completely the influence of the leads. Later we shall see that this
property conserves also for interacting system, if the leads are noninteracting.

Finally, we should note, that the Green functions considered in this
section, are {\em single-particle} GFs, and can be used only for noninteracting
systems.

\paragraph*{(iii) Semi-infinite leads}

\begin{figure}[t]
\begin{center}
\epsfxsize=0.6\hsize
\epsfbox{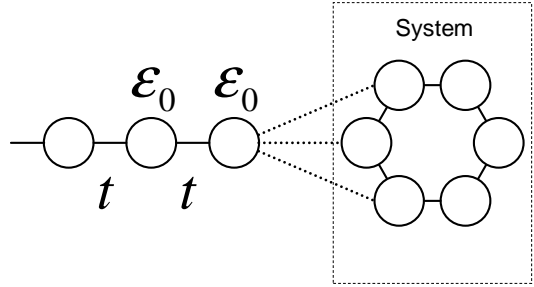}
\caption{A quantum system coupled to a semi-infinite 1D lead.}
\label{fig-3-3}
\end{center}
\end{figure}

Let us consider now a nanosystem coupled to a semi-infinite lead (Fig.\,\ref{fig-3-3}).
The direct matrix inversion can not be performed in this case. The spectrum of a
semi-infinite system is continuous. We should transform the expression (\ref{TB-Sigma2})
into some other form.

To proceed, we use the relation between the Green function and the eigenfunctions
$\Psi_\lambda$ of a system, which are solutions of the Schr\"odinger equation
(\ref{TB-Schroedinger}). Let us define $\Psi_\lambda(\alpha)\equiv c_\lambda$ in the
eigenstate $|\lambda\rangle$ in the sense of definition (\ref{TB-Psi}), then
\begin{equation}\label{TB-G-through-ES}
  G^R_{\alpha\beta}(\epsilon)=\sum_\lambda\frac{\Psi_\lambda(\alpha)\Psi^*_\lambda(\beta)}
  {\epsilon+i\eta-E_\lambda},
\end{equation}
where $\alpha$ is the TB state (site) index, $\lambda$ denotes the eigenstate,
$E_\lambda$ is the energy of the eigenstate. The summation in this formula can be easy
replaced by the integration in the case of a continuous spectrum. It is important to
notice, that the eigenfunctions $\Psi_\lambda(\alpha)$ should be calculated for the
separately taken semi-infinite lead, because the Green function of isolated lead is
substituted into the contact self-energy.

For example, for the semi-infinite 1D chain of single-state sites ($n,m=1,2,...$)
\begin{equation}\label{TB-G-1D}
  G^R_{nm}(\epsilon)=\int_{-\pi}^\pi\frac{dk}{2\pi}\frac{\Psi_k(n)\Psi^*_k(m)}
  {\epsilon+i\eta-E_k},
\end{equation}
with the eigenfunctions $\Psi_k(n)=\sqrt{2}\sin kn$, $E_k=\epsilon_0+2t\cos k$.

Let us consider a simple situation, when the nanosystem is coupled only to the end site
of the 1D lead (Fig.\,\ref{fig-3-3}). From (\ref{TB-Sigma2}) we obtain the matrix
elements of the self-energy
\begin{equation}
  \Sigma_{\alpha\beta}=V^*_{1\alpha}V_{1\beta}G^{0R}_{11},
\end{equation}
where the matrix element $V_{1\alpha}$ describes the coupling between the end site of the
lead ($n=m=1$) and the state $|\alpha\rangle$ of the nanosystem.

\begin{figure}[t]
\begin{center}
\epsfxsize=0.6\hsize
\epsfbox{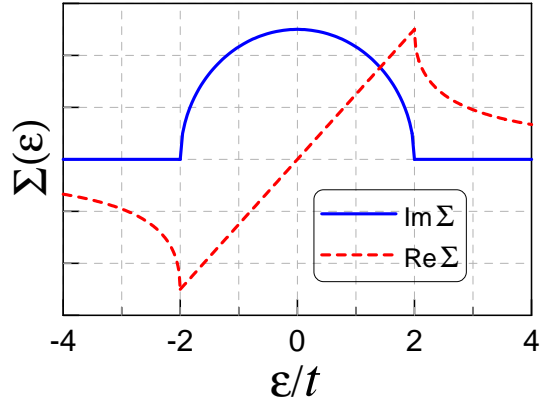}
\caption{(Color) Real and imaginary parts of the contact self-energy as a function of energy
         for a one-band one-dimensional lead.}
\label{fig-3-4}
\end{center}
\end{figure}

To make clear the main physical properties of the lead self-energy, let us analyze
in detail the semi-infinite 1D lead with the Green function (\ref{TB-G-1D}). The
integral can be calculated analytically (\cite{Datta05book}, p.\,213,
\cite{Cuniberti02cp})
\begin{equation}
  G^R_{11}(\epsilon)=\frac{1}{\pi}\int_{-\pi}^\pi\frac{\sin^2k dk}
  {\epsilon+i\eta-\epsilon_0-2t\cos k}=-\frac{\exp(iK(\epsilon))}{t},
\end{equation}
$K(\epsilon)$ is determined from $\epsilon=\epsilon_0+2t\cos K$. Finally, we obtain
the following expressions for the real and imaginary part of the self-energy
\begin{eqnarray}
&\displaystyle {\rm Re}\Sigma_{\alpha\alpha}=\frac{|V_{1\alpha}|^2}{t}
\left(\kappa-\sqrt{\kappa^2-1}\left[\theta(\kappa-1)-\theta(-\kappa-1)\right]\right),
\\
&\displaystyle {\rm Im}\Sigma_{\alpha\alpha}=-\frac{|V_{1\alpha}|^2}{t}
\sqrt{1-\kappa^2}\theta(1-|\kappa|), \\
&\displaystyle \kappa=\frac{\epsilon-\epsilon_0}{2t}.
\end{eqnarray}
The real and imaginary parts of the self-energy, given by these expressions, are
shown in Fig.\,\ref{fig-3-4}. There are several important general conclusion that we
can make looking at the formulas and the curves.

(a) The self-energy is a complex function, the real part describes the energy shift
of the level, and the imaginary part describes broadening. The {\em finite}
imaginary part appears as a result of the continuous spectrum in the leads. The
broadening is described traditionally by the matrix
\begin{equation}\label{TB-Gamma}
  {\bf\Gamma}=i\left({\bf\Sigma}-{\bf\Sigma}^\dag\right),
\end{equation}
called {\em level-width function}.

(b) In the wide-band limit ($t\rightarrow\infty$), at the energies
$\epsilon-\epsilon_0\ll t$, it is possible to neglect the real part of the
self-energy, and the only effect of the leads is level broadening. So that the
self-energy of the left (right) lead is
\begin{equation}\label{TB-Sigma-WBL}
  {\bf\Sigma}_{L(R)}=-i\frac{{\bf\Gamma}_{L(R)}}{2}.
\end{equation}

\paragraph*{(iv) Transmission, conductance, current}

After all, we want again to calculate the current through the nanosystem. We assume, as
before, that the contacts are equilibrium, and there is the voltage $V$ applied between
the left and right contacts. The calculation of the current in a general case is more
convenient to perform using the full power of the nonequilibrium Green function method.
Here we present a simplified approach, valid for noninteracting systems only, following
Paulsson \cite{Paulsson02condmat}.

Let us come back to the Schr\"odinger equation (\ref{TB-Schroedinger}) in the matrix
representation, and write it in the following form
\begin{equation}\label{TB-Schroedinger2}
\left(\begin{array}{ccc}{\bf H}^0_L & {\bf H}_{LS} & 0 \\ {\bf H}^\dag_{LS} & {\bf
H}^0_S & {\bf H}^\dag_{RS} \\ 0 & {\bf H}_{RS} & {\bf H}^0_R
\end{array}\right)
\left(\begin{array}{c} \Psi_L \\ \Psi_S  \\ \Psi_R \end{array}\right)=
E\left(\begin{array}{c} \Psi_L \\ \Psi_S  \\ \Psi_R \end{array}\right),
\end{equation}
where $\Psi_L$, $\Psi_S$, and $\Psi_R$ are vector wave functions of the left lead, the
nanosystem, and the right lead correspondingly.

Now we find the solution in the scattering form (which is difficult to call true
scattering because we do not define explicitly the geometry of the leads). Namely, in the
left lead $\Psi_L=\Psi^0_L+\Psi^1_L$, where $\Psi^0_L$ is the eigenstate of ${\bf
H}^0_L$, and is considered as known initial wave. The "reflected" wave $\Psi^1_L$, as
well as the transmitted wave in the right lead $\Psi_R$, appear only as a result of the
interaction between subsystems. The main trick is, that we find a {\em retarded} solution.

Solving the equation (\ref{TB-Schroedinger2}) with these
conditions, the solution is
\begin{eqnarray}\label{TB-Psi-1}
& \Psi_L=\left(1+{\bf G}^{0R}_L{\bf H}_{LS}{\bf G}^R_S{\bf H}^\dagger_{LS}\right)\Psi^0_L, \\
& \Psi_R={\bf G}^{0R}_R{\bf H}_{RS}{\bf G}^R_S{\bf H}^\dagger_{LS}\Psi^0_L  \\
& \Psi_S={\bf G}^R_S{\bf H}^\dagger_{LS}\Psi^0_L. \label{TB-Psi-3}
\end{eqnarray}
The physical sense of this expressions is quite transparent, they describe the quantum
amplitudes of the scattering processes. Three functions $\Psi_L$, $\Psi_S$, and $\Psi_R$
are equivalent together to the scattering state in the Landauer-B\"uttiker theory. Note,
that ${\bf G}^R_S$ here is the full GF of the nanosystem including the lead
self-energies.

Now the next step. We want to calculate the current. The partial (for some particular
eigenstate $\Psi^0_{L\lambda}$) current from the lead to the system is
\begin{equation}\label{TB-j}
  j_{i=L,R}=\frac{ie}{\hbar}\left(\Psi^\dagger_i{\bf H}_{iS}\Psi_S-\Psi^\dagger_S{\bf
  H}^\dagger_{iS}\Psi_i\right).
\end{equation}

To calculate the total current we should substitute the expressions for the wave
functions (\ref{TB-Psi-1})-(\ref{TB-Psi-3}), and summarize all contributions
\cite{Paulsson02condmat}. As a result the Landauer formula is obtained. We present
the calculation for the transmission function. First, after substitution of the wave
functions we have for the partial current going through the system
\begin{align}\label{TB-j2}
  j_\lambda=j_L=-j_R= & -\frac{ie}{\hbar}\left(\Psi^\dagger_R{\bf H}_{RS}\Psi_S-\Psi^\dagger_S{\bf
  H}^\dagger_{RS}\Psi_R\right)= \nonumber \\
&  -\frac{ie}{\hbar}\left(\Psi^{0\dagger}_L{\bf H}_{LS}{\bf G}^A_S{\bf H}^\dagger_{RS}
   \left({\bf G}^{0\dagger}_R-{\bf G}^{0}_R\right)
   {\bf H}_{RS}{\bf G}^R_S{\bf H}^\dagger_{LS}\Psi^0_L\right)= \nonumber \\
&  \frac{e}{\hbar}\left(\Psi^{0\dagger}_L{\bf H}_{LS}{\bf G}^A_S
   {\bf\Gamma}_R{\bf G}^R_S{\bf H}^\dagger_{LS}\Psi^0_L\right).
\end{align}

The full current of all possible left eigenstates is given by
\begin{equation}
  I=\sum_\lambda j_\lambda=\sum_\lambda \frac{e}{\hbar}\left(\Psi^{0\dagger}_{L\lambda}
  {\bf H}_{LS}{\bf G}^A_S{\bf\Gamma}_R{\bf G}^R_S{\bf
  H}^\dagger_{LS}\Psi^0_{L\lambda}\right)f_L(E_\lambda),
\end{equation}
the distribution function $f_L(E_\lambda)$ describes the population of the left states,
the distribution function of the right lead is absent here, because we consider only the
current from the left to the right.

The same current is given by the Landauer formula through the transmission function $\bar
T(E)$
\begin{equation}
  I=\frac{e}{h}\int_{-\infty}^{\infty}\overline{T}(E)f_L(E)dE.
\end{equation}

If one compares these two expressions for the current, the transmission function at some
energy is obtained as
\begin{align}
   \overline{T}(E)=
   2\pi\sum_\lambda\delta(E-E_\lambda)\left(\Psi^{0\dagger}_{L\lambda}{\bf H}_{LS}{\bf G}^A_S
   {\bf\Gamma}_R{\bf G}^R_S{\bf H}^\dagger_{LS}\Psi^0_{L\lambda}\right) \nonumber \\
   =2\pi\sum_\lambda\sum_\delta\delta(E-E_\lambda)\left(\Psi^{0\dagger}_{L\lambda}{\bf H}_{LS}
   \Psi_{\delta}\right)\left(\Psi^{\dagger}_{\delta}
   {\bf G}^A_S{\bf\Gamma}_R{\bf G}^R_S{\bf H}^\dagger_{LS}\Psi^0_{L\lambda}\right) \nonumber \\
   =\sum_\delta\left(\Psi^{\dagger}_{\delta}
   {\bf G}^A_S{\bf\Gamma}_R{\bf G}^R_S{\bf H}^\dagger_{LS}\left(2\pi\sum_\lambda\delta(E-E_\lambda)\Psi^0_{L\lambda}
   \Psi^{0\dagger}_{L\lambda}\right){\bf H}_{LS}\Psi_{\delta}\right) \nonumber \\
   ={\rm Tr}\left({\bf\Gamma}_L{\bf G}^A_S{\bf\Gamma}_R{\bf G}^R_S\right).
\end{align}
To evaluate the sum in brackets we used the eigenfunction expansion
(\ref{TB-G-through-ES}) for the left contact.

We obtained the new representation for the transmission formula, which is very
convenient for numerical calculations
\begin{equation}
  \overline{T}={\rm Tr}\left(\hat t\hat t^\dagger\right)=
  {\rm Tr}\left({\bf\Gamma}_L{\bf G}^A{\bf\Gamma}_R{\bf
  G}^R\right).
\end{equation}

Finally, one important remark, at finite voltage the diagonal energies in the
Hamiltonians ${\bf H}^0_L$, ${\bf H}^0_S$, and ${\bf H}^0_R$ are shifted
$\epsilon_\alpha\rightarrow\epsilon_\alpha+e\varphi_\alpha$. Consequently, the
energy dependencies of the self-energies defined by (\ref{TB-Sigma2}) are also
changed and the lead self-energies are voltage dependent. However, it is convenient
to define the self-energies using the Hamiltonians at zero voltage, in that case the
voltage dependence should be explicitly shown in the transmission formula
\begin{equation}
  \overline{T}(E)={\rm Tr}\left[{\bf\Gamma}_L(E-e\varphi_L){\bf G}^R(\epsilon)
  {\bf\Gamma}_R(E-e\varphi_R){\bf G}^A(\epsilon)\right],
\end{equation}
where $\varphi_R$ and $\varphi_L$ are electrical potentials of the right and left leads.

With known transmission function, the current $I$ at finite voltage $V$ can be calculated
by the usual Landauer-B\"utiker formulas (without spin degeneration, otherwise it should
be multiplied additionally by 2)
\begin{equation}\label{Landauer}
  I(V)=\frac{e}{h}\int_{-\infty}^{\infty}\overline{T}(E)
  \left[f_L(E)-f_R(E)\right]dE,
\end{equation}
where the equilibrium distribution functions of the contacts should be written with
corresponding chemical potentials $\mu_i$, and electrical potentials $\varphi_i$
\begin{equation}\label{fFD}
  f_L(E)=\frac{1}{\exp\left(\frac{E-\mu_L-e\varphi_L}{T}\right)+1},\ \ \
  f_R(E)=\frac{1}{\exp\left(\frac{E-\mu_R-e\varphi_R}{T}\right)+1}.
\end{equation}

The zero-voltage conductance $G$ is
\begin{equation}
  G=\left.\frac{dI}{dV}\right|_{V=0}=-\frac{e^2}{h}\int_{-\infty}^{\infty}\overline{T}(E)
  \frac{\partial f^0(E)}{\partial E}dE,
\end{equation}
where $f^0(E)$ is the equilibriumfermi-function
\begin{equation}
   f^0(E)=\frac{1}{\exp\left(\frac{E-\mu}{T}\right)+1}.
\end{equation}

\newpage
\subsection{Interacting nanosystems and master equation method}

The single-particle matrix Green function methods, considered in the previous
section, can be applied only in the case of {\em noninteracting} electrons and
without inelastic scattering. In the case of interacting systems, the other
approach, known as the method of tunneling (or transfer) Hamiltonian (TH), plays an
important role, and is widely used to describe tunneling in superconductors, in
ferromagnets, effects in small tunnel junctions such as Coulomb blockade (CB), etc. The
main advantage of this method is that it is easely combined  with powerful methods of
many-body theory. Besides, it is very convenient even for noninteracting electrons,
when the coupling between subsystems is weak, and the tunneling process can be
described by rather simple matrix elements.

\subsubsection{Tunneling and master equation}

\paragraph*{(i) Tunneling (transfer) Hamiltonian}

The main idea is to represent the Hamiltonian of the system (we consider first a
single contact between two subsystems) as a sum of three parts: "left" $\hat H_L$,
"right" $\hat H_R$, and "tunneling" $\hat H_T$
\begin{equation}\label{TH_1}
  \hat H=\hat H_L+\hat H_R+\hat H_T,
\end{equation}
$\hat H_L$ and $\hat H_R$ determine "left" $|Lk\rangle$ and "right" $|Rq\rangle$
states
\begin{eqnarray}
& \hat H_L\psi_{k}(\xi)=E_k\psi_{k}(\xi), & \\ & \hat
H_R\psi_{q}(\xi)=E_q\psi_{q}(\xi), &
\end{eqnarray}
below in this lecture we use the index $k$ for left states and the index $q$ for
right states. $\hat H_T$ determines "transfer" between these states and is {\em
defined} through matrix elements $V_{kq}=\langle Lk|\hat H_T|Rq\rangle$. With these
definitions the single-particle tunneling Hamiltonian is
\begin{equation}\label{TH_2}
  \hat H=\sum_{k\in L}E_k|k\rangle\langle k|+\sum_{q\in R}E_q|q\rangle\langle q|+
  \sum_{kq}\left[V_{qk}|q\rangle\langle k|+V^*_{qk}|k\rangle\langle q|\right].
\end{equation}

The method of the tunneling Hamiltonian was introduced by Bardeen \cite{Bardeen61prl},
developed by Harrison \cite{Harrison61pr}, and formulated in most familiar second
quantized form by Cohen, Falicov, and Phillips \cite{Cohen62prl}. In spite of many
very successful applications of the TH method, it was many times criticized for it's
phenomenological character and incompleteness, beginning from the work of Prange
\cite{Prange63pr}. However, in the same work Prange showed that the tunneling
Hamiltonian is well defined in the sense of the perturbation theory. These
developments and discussions were summarized by Duke \cite{Duke69book}. Note, that
the formulation equivalent to the method of the tunneling Hamiltonian can be derived
exactly from the tight-binding approach.

Indeed, the tight-binding model assumes that the left and right states can be clearly
separated, also when they are orthogonal. The difference with the continuous case
is, that we restrict the Hilbert space introducing the tight-binding model, so that
the solution is not exact in the sense of the continuous Schr\"odinger equation.
But, in fact, we only consider physically relevant states, neglecting high-energy
states not participating in transport.

Compare the tunneling Hamiltonian (\ref{TH_2}) and the tight-binding Hamiltonian
(\ref{TB-Hamiltonian}), divided into left and right parts
\begin{equation}\label{TH-TB-Hamiltonian2}
  \hat H=\sum_{\alpha\beta\in L}\tilde\epsilon_{\alpha\beta}|\alpha\rangle\langle\beta|+
  \sum_{\delta\gamma\in R}\tilde\epsilon_{\delta\gamma}|\delta\rangle\langle\gamma|+
  \sum_{\alpha\in L,\,\delta\in R}\big[V_{\delta\alpha}|\delta\rangle\langle\alpha|+
  V^*_{\delta\alpha}|\alpha\rangle\langle\delta|\big].
\end{equation}
The first two terms are the Hamiltonians of the left and right parts, the third term
describes the left-right (tunneling) coupling. The equivalent matrix representation
of this Hamiltonian is
\begin{equation}
{\bf H}=\left(\begin{array}{cc}{\bf H}^0_L & {\bf H}_{LR} \\ {\bf H}^\dag_{LR} &
{\bf H}^0_R
\end{array}\right).
\end{equation}

The Hamiltonians (\ref{TH_2}) and (\ref{TH-TB-Hamiltonian2}) are essentially the
same, only the first one is written in the eigenstate basis $|k\rangle$,
$|q\rangle$, while the second in the tight-binding basis $|\alpha\rangle$,
$|\beta\rangle$ of the left lead and $|\delta\rangle$, $|\gamma\rangle$ of the right
lead. Now we want to transform the TB Hamiltonian (\ref{TH-TB-Hamiltonian2}) into
the eigenstate representation.

Canonical transformations from the tight-binding (atomic orbitals) representation to
the eigenstate (molecular orbitals) representation play an important role, and we
consider it in detail. Assume, that we find two unitary matrices ${\bf S}_L$ and
${\bf S}_R$, such that the Hamiltonians of the left part ${\bf H}^0_L$ and of the
right part ${\bf H}^0_R$ can be diagonalized by the canonical transformations
\begin{eqnarray}
& \displaystyle \bar{{\bf H}}^0_L={\bf S}^{-1}_L{\bf H}^0_L{\bf S}_L, \\ &
\displaystyle \bar{{\bf H}}^0_R={\bf S}^{-1}_R{\bf H}^0_R{\bf S}_R.
\end{eqnarray}

The left and right eigenstates can be written as
\begin{eqnarray}
& \displaystyle |k\rangle=\sum_\alpha S_{Lk\alpha}|\alpha\rangle, \\ & \displaystyle
|q\rangle=\sum_\delta S_{Rq\delta}|\delta\rangle,
\end{eqnarray}
and the first two free-particle terms of the Hamiltonian (\ref{TH_2}) are
reproduced. The tunneling terms are transformed as
\begin{eqnarray}
& \displaystyle \bar{{\bf H}}_{LR}={\bf S}^{-1}_L{\bf H}_{LR}{\bf S}_R, \\ &
\displaystyle \bar{{\bf H}}^\dag_{LR}={\bf S}^{-1}_R{\bf H}^\dag_{LR}{\bf S}_L,
\end{eqnarray}
or explicitly
\begin{equation}
  \sum_{\alpha\in L,\,\delta\in R}V_{\delta\alpha}|\delta\rangle\langle\alpha|=
  \sum_{kq}V_{qk}|q\rangle\langle k|,
\end{equation}
where
\begin{equation}
  V_{qk}=\sum_{\alpha\in L,\,\delta\in R}V_{\delta\alpha}S_{L\alpha k}S_{R\delta q}.
\end{equation}
The last expression solve the problem of transformation of the tight-binding matrix
elements into tunneling matrix elements.

For applications the tunneling Hamiltonian (\ref{TH_2}) should be formulated in the
second quantized form. We introduce creation and annihilation {\em Schr\"odinger}
operators $c^\dag_{Lk}$, $c_{Lk}$, $c^\dag_{Rq}$, $c_{Rq}$. Using the usual rules we
obtain
\begin{equation}
  \hat H=\hat H_L\left(\left\{c^{\dag}_{k};c_{k}\right\}\right)+
  \hat H_R\left(\left\{c^{\dag}_{q};c_{q}\right\}\right)+
  \hat H_T\left(\left\{c^{\dag}_{k};c_{k};c^{\dag}_{q};c_{q}\right\}\right),
\end{equation}
\begin{equation}\label{TH_3}
  \hat H=\sum_{k}(\epsilon_{k}+e\varphi_L(t))
  c^{\dag}_{k}c_{k}+\sum_{q}(\epsilon_{q}+e\varphi_R(t))
  c^{\dag}_{q}c_{q}+\sum_{kq}\left[V_{qk}c^\dagger_q c_k+V^*_{qk}c^\dagger_k
  c_q\right].
\end{equation}

It is assumed that left $c_k$ and right $c_q$ operators describe independent states
and are anticommutative. For nonorthogonal states of the Hamiltonian $\hat H_l+\hat
H_R$ it is not exactly so. But if we consider $\hat H_L$ and $\hat H_R$ as two
independent Hamiltonians with independent Hilbert spaces we resolve this problem.
Thus we again should consider (\ref{TH_3}) not as a true Hamiltonian, but as the
formal expression describing the current between left and right states. In the weak
coupling case the small corrections to the commutation relations are of the order of
$|V_{qk}|$ and can be neglected. If the tight-binding formulation is possible,
(\ref{TH_3}) is exact within the framework of this formulation. In general the
method of tunneling Hamiltonian can be considered as a {\em phenomenological}
microscopic approach, which was proved to give reasonable results in many cases,
e.g. in description of tunneling between superconductors and Josephson effect.

\paragraph*{(ii) Tunneling current}

The current from the state $k$ into the state $q$ is given by the golden rule
\begin{equation}
  J_{k\rightarrow q}=e\Gamma_{qk}=\frac{2\pi e}{\hbar}|V_{qk}|^2f_L(k)\left(1-f_R(q)\right)
  \delta(E_k-E_q),
\end{equation}
the probability $(1-f_R(E_q))$ that the right state is unoccupied should be
included, it is different from the scattering approach because left and right states
are two independent states!

Then we write the total current as the sum of all partial currents from left states
to right states and vice versa (note that the terms $f_L(k)f_R(q)$ are cancelled)
\begin{equation}\label{TH-J-GR}
  J=\frac{2\pi e}{\hbar}\sum_{kq}
  |V_{qk}|^2\left[f(k)-f(q)\right]
  \delta(E_{q}-E_{k}).
\end{equation}

For tunneling between two equilibrium leads distribution functions are simply
Fermi-Dirac functions (\ref{fFD}) and current can be finally written in the well
known form (To do this one should multiply the integrand on
$1=\int\delta(E-E_q)dE$.)
\begin{equation}
  J =\frac{e}{h}\int_{-\infty}^{\infty}T(E,V)
  \left[f_L(E)-f_R(E)\right]dE,
 \end{equation}
with
\begin{equation}
  T(E,V)=(2\pi)^2\sum_{qk}|V_{kq}|^2\delta(E-E_{k}-e\varphi_L)
  \delta(E-E_{q}-e\varphi_R).
\end{equation}

This expression is equivalent to the Landauer formula (\ref{Landauer}), but the
transmission function is related now to the tunneling matrix element.

Now let us calculate the tunneling current as the time derivative of the number of
particles operator in the left lead $\hat N_L=\sum_k c^{\dag}_kc_k$. Current from
the left to right contact is
\begin{equation}
  J(t)=-e\left\langle\left(\frac{dN_L}{dt}\right)\right\rangle_S=-\frac{ie}{\hbar}
  \left\langle\left[\hat H_T,N_L\right]_-\right\rangle_S,
\end{equation}
where $\langle...\rangle_S$ is the average over time-dependent Schr\"odinger state.
$\hat N_L$ commute with both left and right Hamiltonians, but not with the tunneling
Hamiltonian
\begin{equation}
\left[\hat H_T,N_L\right]_-= \sum_{k'}\sum_{kq}\left[\left(V_{qk}
c^{\dag}_{q}c_{k}+V^*_{qk} c_{q}c_{k}^{\dag}\right)c^{\dag}_{k'}c_{k'}\right]_-,
\end{equation}
using commutation relations
$$c_{k}c^{\dag}_{k'}c_{k'}-c^{\dag}_{k'}c_{k'}c_{k}=
c_{k}c^{\dag}_{k'}c_{k'}+c^{\dag}_{k'}c_{k}c_{k'}=
(c_{k}c^{\dag}_{k'}+\delta_{kk'}-c_{k}c^{\dag}_{k'})c_{k'}= \delta_{kk'}c_{k},$$
we obtain
\begin{equation}
  J(t)=\frac{ie}{\hbar}\sum_{kq}\left[V_{qk}
  \left\langle c^{\dag}_{q}c_k \right\rangle_S-
  V^*_{qk}\left\langle c^{\dag}_kc_{q} \right\rangle_S
  \right].
\end{equation}

Now we switch to the Heisenberg picture, and average over initial time-independent
{\em equilibrium} state
\begin{equation}
{\left\langle\color{black}\hat{O}({t})\right\rangle}=
Sp\left({\hat{\rho}_{eq}}\hat{O}({t})\right),\ \ \ \
\hat{\rho}_{eq}=\frac{e^{-H_{eq}/T}}{Sp\left(e^{-H_{eq}/T}\right)}.
\end{equation}

One obtains
\begin{equation}
  J(t)=\frac{ie}{\hbar}\sum_{kq}\left[V_{qk}
  \left\langle c^{\dag}_{q}(t)c_{k}(t) \right\rangle-
  V^*_{qk}\left\langle c^{\dag}_{k}(t)c_{q}(t) \right\rangle
  \right].
\end{equation}

It can be finally written as
\begin{equation*}
  J(t)=\frac{2e}{\hbar}{\rm Im}\left(\sum_{kq} V_{qk}
  {\rho_{kq}(t)}\right)=
  \frac{2e}{\hbar}{\rm Re}\left(\sum_{kq} V_{qk}
  {G^<_{kq}(t,t)}\right).
\end{equation*}

We define "left-right" density matrix or more generally lesser Green function
\begin{equation*}
G^<_{kq}(t_1,t_2)=i\left\langle c^\dag_{q}(t_2)c_k(t_1)\right\rangle.
\end{equation*}

Later we show that these expressions for the tunneling current give the same answer
as was obtained above by the golden rule in the case of noninteracting leads.

\paragraph*{(iii) Sequential tunneling and the master equation}

Let us come back to our favorite problem -- transport through a quantum system.
There is one case (called {\em sequential tunneling}), when the simple formulas
discussed above can be applied even in the case of resonant tunneling

Assume that a noninteracting nanosystem is coupled weakly to a thermal bath (in
addition to the leads). The effect of the thermal bath is to break phase coherence
of the electron inside the system during some time $\tau_{ph}$, called {\em
decoherence or phase-breaking time}. $\tau_{ph}$ is an important time-scale in the
theory, it should be compared with the so-called "tunneling time" -- the
characteristic time for the electron to go from the nanosystem to the lead, which
can be estimated as an inverse level-width function $\Gamma^{-1}$. So that the
criteria of sequential tunneling is
\begin{equation}\label{TH-seq}
  \Gamma\tau_{ph}\ll 1.
\end{equation}
The finite decoherence time is due to some inelastic scattering mechanism inside the
system, but typically this time is shorter than the energy relaxation time
$\tau_\epsilon$, and the distribution function of electrons inside the system can be
nonequilibrium (if the finite voltage is applied), this transport regime is well
known in semiconductor superlattices and quantum-cascade structures.

In the sequential tunneling regime the tunneling events between the left lead and
the nanosystem and between the left lead and the nanosystem are independent and the
current from the left (right) lead to the nanosystem is given by the golden rule
expression (\ref{TH-J-GR}). Let us modify it to the case of tunneling from the lead
to a {\em single level} $|\alpha\rangle$ of a quantum system
\begin{equation}\label{TH-J-GR2}
  J=\frac{2\pi e}{\hbar}\sum_{k}
  |V_{\alpha k}|^2\left[f(k)-P_\alpha\right]
  \delta(E_\alpha-E_{k}),
\end{equation}
where we introduce the probability $P_\alpha$ to find the electron in the state
$|\alpha\rangle$ with the energy $E_\alpha$.

\paragraph*{(iv) Rate equations for noninteracting systems}

Rate equation method is a simple approach base on the balance of incoming and
outgoing currents. Assuming that the contacts are equilibrium we obtain for the left
and right currents
\begin{equation}\label{TH-J-GR3}
  J_{i=L(R)}=e\Gamma_{i\alpha}\left[f_i^0(E_\alpha)-P_\alpha\right],
\end{equation}
where
\begin{equation}\label{TH-Gi}
\Gamma_{i\alpha} =\frac{2\pi}{\hbar}\sum_{k} |V_{\alpha k}|^2\delta(E_\alpha-E_{k}).
\end{equation}

In the stationary state $J=J_L=-J_R$, and from this condition the level population
$P_\alpha$ is found to be
\begin{equation}\label{Th-P}
  P_\alpha=\frac{\Gamma_{L\alpha}f^0_L(E_\alpha)+\Gamma_{R\alpha}f^0_R(E_\alpha)}
  {\Gamma_{L\alpha}+\Gamma_{R\alpha}},
\end{equation}
with the current
\begin{equation}\label{TH-J-GR4}
  J=e\frac{\Gamma_{L\alpha}\Gamma_{R\alpha}}
  {\Gamma_{L\alpha}+\Gamma_{R\alpha}}\big(f^0_L(E_\alpha)-f^0_R(E_\alpha)\big).
\end{equation}
It is interesting to note that this expression is exactly the same, as one can
obtain for the resonant tunneling through a single level without any scattering. It
should be not forgotten, however, that we did not take into account additional level
broadening due to scattering.

\paragraph*{(v) Master equation for interacting systems}

Now let us formulate briefly a more general approach to transport through
interacting nanosystems weakly coupled to the leads in the sequential tunneling
regime, namely the master equation method. Assume, that the system can be in several
states $|\lambda\rangle$, which are the eigenstates of an isolated system and
introduce the distribution function $P_\lambda$ -- the probability to find the
system in the state $|\lambda\rangle$. Note, that these states are {\em
many-particle} states, for example for a two-level quantum dot the possible states
are $|\lambda\rangle=|00\rangle$, $|10\rangle$, $01|\rangle$, and $|11\rangle$. The
first state is empty dot, the second and the third with one electron, and the last
one is the double occupied state. The other non-electronic degrees of freedom can be
introduce on the same ground in this approach. The only restriction is that some
full set of eigenstates should be used
\begin{equation}
  \sum_\lambda P_\lambda=1.
\end{equation}

The next step is to treat tunneling as a perturbation. Following this idea, the
transition rates $\Gamma^{\lambda\lambda'}$ from the state $\lambda'$ to the state
$\lambda$ are calculated using the Fermi golden rule
\begin{equation}\label{TH-ME-Golden}
  \Gamma^{fi}=\frac{2\pi}{\hbar}
  \left|\left\langle f|\hat H_T|i\right\rangle\right|^2\delta(E_f-E_i).
\end{equation}

Then, the kinetic (master) equation can be written as
\begin{equation}
  \frac{dP_\lambda}{dt}=\sum_{\lambda'}\Gamma^{\lambda\lambda'}P_{\lambda'}-
  \sum_{\lambda'}\Gamma^{\lambda'\lambda}P_\lambda,
\end{equation}
where the first term describes tunneling transition {\em into the state}
$|\lambda\rangle$, and the second term -- tunneling transition {\em out of the
state} $|\lambda\rangle$.

In the stationary case the probabilities are determined from
\begin{equation}
  \sum_{\lambda'}\Gamma^{\lambda\lambda'}P_{\lambda'}=
  \sum_{\lambda'}\Gamma^{\lambda'\lambda}P_\lambda.
\end{equation}

For noninteracting electrons the transition rates are determined by the
single-electron tunneling rates, and are nonzero only for the transitions between
the states with the number of electrons different by one. For example, transition
from the state $|\lambda'\rangle$ with empty electron level $\alpha$ into the state
$|\lambda\rangle$ with filled state $\alpha$ is described by
\begin{equation}
  \Gamma^{n_\alpha=1\ n_\alpha=0}=
  \Gamma_{L\alpha}f^0_L(E_\alpha)+\Gamma_{R\alpha}f^0_R(E_\alpha),
\end{equation}
where $\Gamma_{L\alpha}$ and $\Gamma_{R\alpha}$ are left and right level-width
functions (\ref{TH-Gi}).

For interacting electrons the calculation is a little bit more complicated. One
should establish the relation between {\em many-particle} eigenstates of the system
and {\em single-particle} tunneling. To do this, let us note, that the states
$|f\rangle$ and $|i\rangle$ in the golden rule formula (\ref{TH-ME-Golden}) are
actually the states of the whole system, including the leads. We denote the initial
and final states as
\begin{eqnarray}
& |i\rangle=|\hat k_i,\lambda'\rangle=|\hat k_i\rangle|\lambda'\rangle, \\ &
|f\rangle=|\hat k_f,\lambda\rangle=|\hat k_f\rangle|\lambda\rangle,
\end{eqnarray}
where $\hat k$ is the occupation of the single-particle states in the lead. The
parameterization is possible, because we apply the perturbation theory, and isolated
lead and nanosystem are independent.

The important point is, that the leads are actually in the equilibrium mixed state,
the single electron states are populated with probabilities, given by the
Fermi-Dirac distribution function. Taking into account all possible single-electron
tunneling processes, we obtain the incoming tunneling rate
\begin{align}
  \Gamma^{\lambda\lambda'}_{in} =
  \frac{2\pi}{\hbar} & \sum_{ik\sigma} f_i^0(E_{ik\sigma})
  \left|\left\langle i\bar k,\lambda\left|\bar H_{T}\right|ik,\lambda'\right\rangle\right|^2
  \delta(E_{\lambda'}+E_{ik\sigma}-E_{\lambda}),
\end{align}
where we use the short-hand notations: $|ik,\lambda'\rangle$ is the state with
occupied $k$-state in the $i-$th lead, while $|i\bar k,\lambda\rangle$ is the state
with unoccupied $k$-state in the $i-$th lead, and all other states are assumed to be
unchanged, $E_\lambda$ is the energy of the state $\lambda$ .

To proceed, we introduce the following Hamiltonian, describing single electron
tunneling and charging of the nanosystem state
\begin{equation}\label{TH-cX-H}
  \hat H_T=\sum_{k\lambda\lambda'}\left[V_{\lambda\lambda'k}c_kX^{\lambda\lambda'}
  +V^*_{\lambda\lambda'k}c^\dagger_kX^{\lambda'\lambda}\right],
\end{equation}
the Hubbard operators $X^{\lambda\lambda'}=|\lambda\rangle\langle\lambda'|$ describe
transitions between eigenstates of the nanosystem.

Substituting this Hamiltonian one obtains
\begin{equation}
  \Gamma^{\lambda\lambda'}_{in} =
  \frac{2\pi}{\hbar}\sum_{ik\sigma} f_i^0(E_{ik\sigma})
  \left|V_{ik\sigma}\right|^2
  \left|V_{\lambda\lambda'k}\right|^2\delta(E_{\lambda'}+E_{ik\sigma}-E_{\lambda}).
\end{equation}

In the important limiting case, when the matrix element $V_{\lambda\lambda'k}$ is
$k$-independent, the sum over $k$ can be performed, and finally
\begin{equation}
  \Gamma^{\lambda\lambda'}_{in} =
  \sum_{i=L,R}\Gamma_{i}(E_{\lambda}-E_{\lambda'})\left|V_{\lambda\lambda'}\right|^2
  f_i^0(E_{\lambda}-E_{\lambda'}).
\end{equation}

Similarly, the outgoing rate is
\begin{equation}
  \Gamma^{\lambda\lambda'}_{out} =
  \sum_{i=L,R}\Gamma_{i}(E_{\lambda'}-E_{\lambda})\left|V_{\lambda\lambda'}\right|^2
  \left(1-f_i^0(E_{\lambda'}-E_{\lambda})\right).
\end{equation}

The current (from the left or right lead to the system) is
\begin{equation}
  J_{i=L,R}(t)=e\sum_{\lambda\lambda'}\left(\Gamma^{\lambda\lambda'}_{i\,in}P_{\lambda'}-
  \Gamma^{\lambda\lambda'}_{i\,out}P_{\lambda'}\right).
\end{equation}

This system of equations solves the transport problem in the sequential tunneling
regime.

\subsubsection{Electron-electron interaction and Coulomb blockade}

\paragraph*{(i) Anderson-Hubbard and constant-interaction models}

To take into account both discrete energy levels of a system and the
electron-electron interaction, it is convenient to start from the general
Hamiltonian
\begin{equation}\label{CB-H1}
  \hat H=\sum_{\alpha\beta}\tilde\epsilon_{\alpha\beta}d^\dag_\alpha d_\beta+
  \frac{1}{2}\sum_{\alpha\beta\gamma\delta}V_{\alpha\beta,\gamma\delta}
  d^\dag_\alpha d^\dag_\beta d_\gamma d_\delta.
\end{equation}
The first term of this Hamiltonian is a free-particle discrete-level model
(\ref{TB-H-SC}) with $\tilde\epsilon_{\alpha\beta}$ including electrical potentials.
And the second term describes all possible interactions between electrons and is
equivalent to the real-space Hamiltonian
\begin{equation}\label{CB-H2}
  \hat H_{ee}=\frac{1}{2}\int d\xi\int d\xi'\hat\psi^\dagger(\xi)\hat\psi^\dagger(\xi')
  V(\xi,\xi')\hat\psi(\xi')\hat\psi(\xi),
\end{equation}
where $\hat\psi(\xi)$ are field operators
\begin{equation}
  \hat\psi(\xi)=\sum_\alpha\psi_\alpha(\xi)d_\alpha,
\end{equation}
$\psi_\alpha(\xi)$ are the basis single-particle functions, we remind, that spin
quantum numbers are included in $\alpha$, and spin indices are included in
$\xi\equiv{\bf r},\sigma$ as variables.

The matrix elements are defined as
\begin{equation}
  V_{\alpha\beta,\gamma\delta}=
  \int d\xi\int d\xi'\psi^*_\alpha(\xi)\psi^*_{\beta}(\xi')
  V(\xi,\xi')\psi_{\gamma}(\xi)\psi_\delta(\xi').
\end{equation}

For pair Coulomb interaction $V(|{\bf r}|)$ the matrix elements are
\begin{equation}\label{CB-H3}
  V_{\alpha\beta,\gamma\delta}=\sum_{\sigma\sigma'}
  \int d{\bf r}\int d{\bf r'}\psi^*_\alpha({\bf r},\sigma)\psi^*_{\beta}({\bf r'},\sigma')
  V(|{\bf r-r'}|)\psi_{\gamma}({\bf r},\sigma)\psi_\delta({\bf r'},\sigma').
\end{equation}

Assume now, that the basis states $|\alpha\rangle$ are the states with definite spin
quantum number $\sigma_\alpha$. It means, that only one spin component of the wave
function, namely $\psi_\alpha(\sigma_\alpha)$ is nonzero, and
$\psi_\alpha(\bar\sigma_\alpha)=0$. In this case the only nonzero matrix elements
are those with $\sigma_\alpha=\sigma_\gamma$ and $\sigma_\beta=\sigma_\delta$, they
are
\begin{equation}\label{CB-H4}
  V_{\alpha\beta,\gamma\delta}=
  \int d{\bf r}\int d{\bf r'}\psi^*_\alpha({\bf r})\psi^*_{\beta}({\bf r'})
  V(|{\bf r-r'}|)\psi_{\gamma}({\bf r})\psi_\delta({\bf r'}).
\end{equation}

In the case of delocalized basis states $\psi_\alpha({\bf r})$, the main matrix
elements are those with $\alpha=\gamma$ and $\beta=\delta$, because the wave
functions of two different states with the same spin are orthogonal in real space
and their contribution is small. It is also true for the systems with localized wave
functions $\psi_\alpha({\bf r})$, when the overlap between two different states is
weak. In these cases it is enough to replace the interacting part by the
Anderson-Hubbard Hamiltonian, describing only density-density interaction
\begin{equation}\label{CB-AH}
  \hat H_{AH}=\frac{1}{2}\sum_{\alpha\neq\beta}U_{\alpha\beta}\hat n_\alpha \hat n_\beta.
\end{equation}
with the Hubbard interaction defined as
\begin{equation}\label{CB-H5}
  U_{\alpha\beta}=
  \int d{\bf r}\int d{\bf r'}|\psi_\alpha({\bf r})|^2|\psi_{\beta}({\bf r'})|^2
  V(|{\bf r-r'}|).
\end{equation}

In the limit of a single-level quantum dot (which is, however, a two-level system
because of spin degeneration) we get the Anderson impurity model (AIM)
\begin{equation}\label{CB-AIM}
  \hat H_{AIM}=\sum_{\sigma=\uparrow\downarrow}\epsilon_\sigma d^\dag_\sigma d_\sigma+
  U\hat n_\uparrow \hat n_\downarrow.
\end{equation}

The other important limit is the constant interaction model (CIM), which is valid
when many levels interact with similar energies, so that approximately, assuming
$U_{\alpha\beta}=U$ for any states $\alpha$ and $\beta$
\begin{equation}\label{CB-CIM1}
  \hat H_{AH}=\frac{1}{2}\sum_{\alpha\neq\beta}U_{\alpha\beta}\hat n_\alpha \hat n_\beta
  \approx
  \frac{U}{2}\left(\sum_{\alpha}\hat n_\alpha\right)^2-
  \frac{U}{2}\left(\sum_{\alpha}\hat n_\alpha^2\right)=
  \frac{U\hat N(\hat N-1)}{2}.
\end{equation}
where we used $\hat n^2=\hat n$.

Thus, the CIM reproduces the charging energy considered above, and the Hamiltonian
of an isolated system is
\begin{equation}\label{CB-CI2}
  \hat H_{CIM}=\sum_{\alpha\beta}\tilde\epsilon_{\alpha\beta}d^\dag_\alpha d_\beta+E(N).
\end{equation}

Note, that the equilibrium compensating charge density can be easily introduced into
the AH Hamiltonian
\begin{equation}\label{CB-AH2}
  \hat H_{AH}=\frac{1}{2}\sum_{\alpha\neq\beta}U_{\alpha\beta}
  \left(\hat n_\alpha-\bar n_\alpha\right)\left(\hat n_\beta-\bar n_\beta\right).
\end{equation}

\paragraph*{(ii) Coulomb blockade in quantum dots}

Here we want to consider the Coulomb blockade in intermediate-size quantum dots,
where the typical energy level spacing $\Delta\epsilon$ is not too small to neglect
it completely, but the number of levels is large enough, so that one can use the
constant-interaction model (\ref{CB-CI2}), which we write in the eigenstate basis as
\begin{equation}
  \hat H_{CIM}=\sum_\alpha\tilde\epsilon_\alpha d^\dag_\alpha d_\alpha+E(n),
\end{equation}
where the charging energy $E(n)$ is determined in the same way as previously, for
example by the expression (\ref{CB-CIM1}). Note, that for quantum dots the usage of
classical capacitance is not well established, although for large quantum dots it is
possible. Instead, we shift the energy levels in the dot
$\tilde\epsilon_\alpha=\epsilon_\alpha+e\varphi_\alpha$ by the electrical potential
\begin{equation}\label{CB-QD-phi}
  \varphi_\alpha=V_G+V_R+\eta_\alpha(V_L-V_R),
\end{equation}
where $\eta_\alpha$ are some coefficients, dependent on geometry. This method can be
easily extended to include any self-consistent effects on the mean-field level by
the help of the Poisson equation (instead of classical capacitances). Besides, if
all $\eta_\alpha$ are the same, our approach reproduce again the the classical
expression
\begin{equation}
  \hat E_{CIM}=\sum_\alpha\epsilon_\alpha n_\alpha+E(n)+en\varphi_{ext}.
\end{equation}

The {\em addition energy} now depends not only on the charge of the molecule, but
also on the state $|\alpha\rangle$, in which the electron is added
\begin{equation}\label{CB-Ec5}
  \Delta E_{n\alpha}^+(n,n_\alpha=0\rightarrow n+1,n_\alpha=1)=E(n+1)-E(n)+\epsilon_\alpha,
\end{equation}
we can assume in this case, that the single particle energies are additive to the
charging energy, so that the full quantum eigenstate of the system is $|n,\hat
n\rangle$, where the set $\hat n\equiv\{n_\alpha\}$ shows weather the particular
single-particle state $|\alpha\rangle$ is empty or occupied. Some arbitrary state
$\hat n$ looks like
\begin{equation}\label{CB-QD-n}
\hat
n\equiv\{n_\alpha\}\equiv\big(n_1,n_2,n_3,n_4,n_5,...\big)=\big(1,1,0,1,0,...\big).
\end{equation}
Note, that the distribution $\hat n$ defines also $n=\sum_\alpha n_\alpha$. It is
convenient, however, to keep notation $n$ to remember about the charge state of a
system, below we use both notations $|n,\hat n\rangle$ and short one $|\hat
n\rangle$ as equivalent.

The other important point is that the distribution function $f_n(\alpha)$ in the
charge state $|n\rangle$ is not assumed to be equilibrium, as previously (this
condition is not specific to quantum dots with discrete energy levels, the
distribution function in metallic islands can also be nonequilibrium. However, in
the parameter range, typical for classical Coulomb blockade, the tunneling time is
much smaller than the energy relaxation time, and quasiparticle nonequilibrium
effects are usually neglected).

With these new assumptions, the theory of sequential tunneling is quite the same, as
was considered in the previous section. The master equation is
\cite{Beenakker91prb,Averin91prb,vanHouten92inbook,vonDelft01pr}
\begin{align}\label{CB-SET-M3}
  \frac{dp(n,\hat n,t)}{dt} & =
  \sum_{\hat n'} \left(\Gamma^{n\,n-1}_{\hat n\hat n'}p(n-1,\hat n',t)+
  \Gamma^{n\,n+1}_{\hat n\hat n'}p(n+1,\hat n',t) \right)- \nonumber \\
& \sum_{\hat n'}\left(\Gamma^{n-1\,n}_{\hat n'\hat n}+
  \Gamma^{n+1\,n}_{\hat n'\hat n}\right)p(n,\hat n,t)+
  I\left\{p(n,\hat n,t)\right\},
\end{align}
where $p(n,\hat n,t)$ is now the probability to find the system in the state
$|n,\hat n\rangle$, $\Gamma^{n\,n-1}_{\hat n\hat n'}$ is the transition rate from
the state with $n-1$ electrons and single level occupation $\hat n'$ into the state
with $n$ electrons and single level occupation $\hat n$. The sum is over all states
${\hat n'}$, which are different by one electron from the state ${\hat n}$. The last
term is included to describe possible inelastic processes inside the system and
relaxation to the equilibrium function $p_{eq}(n,\hat n)$. In principle, it is not
necessary to introduce such type of dissipation in calculation, because the current
is in any case finite. But the dissipation may be important in large systems and at
finite temperatures. Besides, it is necessary to describe the limit of classical
single-electron transport, where the distribution function of qausi-particles is
assumed to be equilibrium. Below we shall not take into account this term, assuming
that tunneling is more important.

While all considered processes are, in fact, single-particle tunneling processes, we
arrive at
\begin{align}\label{CB-SET-M4}
  \frac{dp(\hat n,t)}{dt}=
  \sum_{\beta} & \left(\delta_{n_\beta 1}\Gamma^{n\,n-1}_{\beta}p(\hat n,n_\beta=0,t)+
  \delta_{n_\beta 0}\Gamma^{n\,n+1}_{\beta}p(\hat n,n_\beta=1,t) \right)- \nonumber \\
& \sum_{\beta}\left(\delta_{n_\beta 1}\Gamma^{n-1\,n}_{\beta}+
  \delta_{n_\beta 0}\Gamma^{n+1\,n}_{\beta}\right)p(\hat n,t),
\end{align}
where the sum is over single-particle states. The probability $p(\hat
n,n_\beta=0,t)$ is the probability of the state equivalent to ${\hat n}$, but
without the electron in the state $\beta$. Consider, for example, the first term in
the right part. Here the delta-function $\delta_{n_\beta 1}$ shows, that this term
should be taken into account only if the single-particle state $\beta$ in the
many-particle state ${\hat n}$ is occupied, $\Gamma^{n\,n-1}_{\beta}$ is the
probability of tunneling from the lead to this state, $p(\hat n,n_\beta=0,t)$ is the
probability of the state ${\hat n'}$, from which the system can come into the state
${\hat n}$.

The transitions rates are defined by the same golden rule expressions, as before,
but with explicitly shown single-particle state $\alpha$
\begin{align}\label{CB-GR2}
  \Gamma_{L\alpha}^{n+1\,n}=
  \frac{2\pi}{\hbar} & \left|\left\langle n+1,n_\alpha=1|\hat H_{TL}|n,n_\alpha=0\right\rangle\right|^2
  \delta(E_i-E_f)=
  \nonumber \\
&  \frac{2\pi}{\hbar}\sum_{k}
  \left|V_{k\alpha}\right|^2
  f_k\delta(\Delta E_{n\alpha}^+-E_k),
\end{align}
\begin{align}\label{CB-GR2a}
  \Gamma_{L\alpha}^{n-1\,n}=
  \frac{2\pi}{\hbar} & \left|\left\langle n-1,n_\alpha=0|\hat H_{TL}|n,n_\alpha=1\right\rangle\right|^2
  \delta(E_i-E_f)=
  \nonumber \\
&  \frac{2\pi}{\hbar}\sum_{k}
  \left|V_{k\alpha}\right|^2
  \left(1-f_k\right)\delta(\Delta E_{n-1\,\alpha}^+-E_k),
\end{align}
there is no occupation factors $(1-f_\alpha)$, $f_\alpha$ because this state is
assumed to be empty in the sense of the master equation (\ref{CB-SET-M4}). The
energy of the state is now included into the addition energy.

Using again the level-width function
\begin{equation}
\Gamma_{i=L,R\,\alpha}(E) =\frac{2\pi}{\hbar}\sum_{k}
|V_{ik,\alpha}|^2\delta(E-E_{k}).
\end{equation}
we obtain
\begin{eqnarray}
& \Gamma_{\alpha}^{n+1\,n}=
  \Gamma_{L\alpha}f^0_L(\Delta E_{n\alpha}^+)+\Gamma_{R\alpha}f^0_R(\Delta
  E_{n\alpha}^+), \\[0.3cm]
& \Gamma_{\alpha}^{n-1\,n}=
  \Gamma_{L\alpha}\left(1-f^0_L(\Delta E_{n-1\,\alpha}^+)\right)+
  \Gamma_{R\alpha}\left(1-f^0_R(\Delta E_{n-1\,\alpha}^+)\right).
\end{eqnarray}

Finally, the current from the left or right contact to a system is
\begin{equation}\label{CB-J2}
  J_{i=L,R}=e\sum_\alpha\sum_{\hat n}p(\hat n)\Gamma_{i\alpha}\left(
  \delta_{n_\alpha 0}f^0_i(\Delta E_{n\alpha}^+)-
  \delta_{n_\alpha 1}(1-f^0_i(\Delta E_{n\alpha}^+)) \right).
\end{equation}
The sum over $\alpha$ takes into account all possible single particle tunneling
events, the sum over states $\hat n$ summarize probabilities $p(\hat n)$ of these
states.

\paragraph*{(iii) Linear conductance}

\begin{figure}[t]
\begin{center}
\epsfxsize=0.6\hsize
\epsfbox{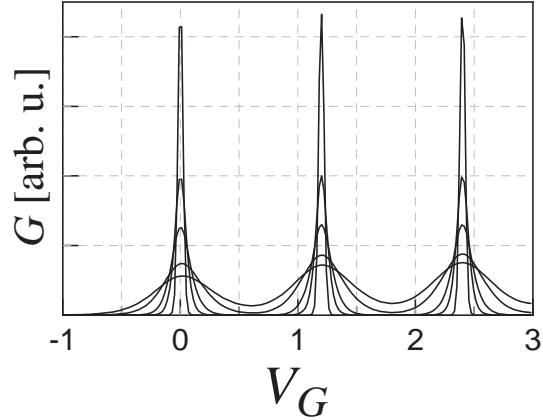}
\caption{Linear conductance of a QD as a function of the gate voltage at different
         temperatures $T=0.01E_C$, $T=0.03E_C$, $T=0.05E_C$, $T=0.1E_C$, $T=0.15E_C$ (lower curve).}
\label{fig-4-10}
\end{center}
\end{figure}

The linear conductance can be calculated analytically
\cite{Beenakker91prb,vanHouten92inbook}. Here we present the final result:
\begin{equation}\label{CB-QD-G}
  G=\frac{e^2}{T}\sum_{\alpha}\sum_{n=1}^\infty\frac{\Gamma_{L\alpha}\Gamma_{R\alpha}}
  {\Gamma_{L\alpha}+\Gamma_{R\alpha}} P_{eq}(n,n_\alpha=1)\left[1-f^0(\Delta E_{n-1\,\alpha}^+)\right],
\end{equation}
where $P_{eq}(n,n_\alpha=1)$ is the joint probability that the quantum dot contains
$n$ electrons and the level $\alpha$ is occupied
\begin{equation}
  P_{eq}(n,n_\alpha=1)=\sum_{\hat n}p_{eq}(\hat n)
  \delta\left(n-\sum_\beta n_\beta\right)\delta_{n_\alpha 1},
\end{equation}
and the equilibrium probability (distribution function) is determined by the Gibbs
distribution in the grand canonical ensemble:
\begin{equation}
  p_{eq}(\hat n)=\frac{1}{Z}\exp\left[-\frac{1}{T}\left(\sum_\alpha\tilde\epsilon_\alpha+
  E(n)\right)\right].
\end{equation}

A typical behaviour of the conductance as a function of the gate voltage at
different temperatures is shown in Fig.~\ref{fig-4-10}. In the resonant tunneling
regime at low temperatures $T\ll\Delta\epsilon$ the peak height is strongly
temperature-dependent. It is changed by classical temperature dependence (constant
height) at $T\gg\Delta\epsilon$.

\paragraph*{(iv) Transport at finite bias voltage}

At finite bias voltage we find new manifestations of the interplay between
single-electron tunneling and resonant free-particle tunneling.

\begin{figure}[t]
\begin{center}
\epsfxsize=0.6\hsize
\epsfbox{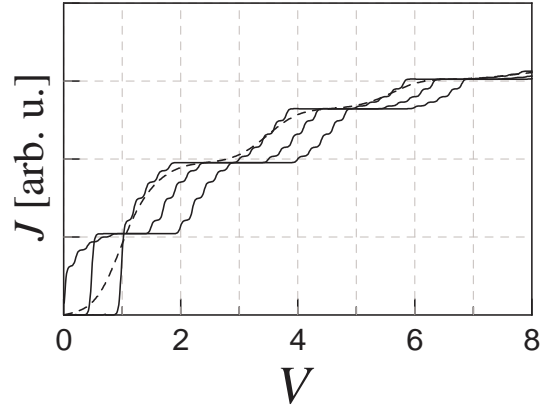}
\caption{Coulomb staircase.} \label{fig-4-11}
\end{center}
\end{figure}

\begin{figure}
\begin{center}
\epsfxsize=0.6\hsize
\epsfbox{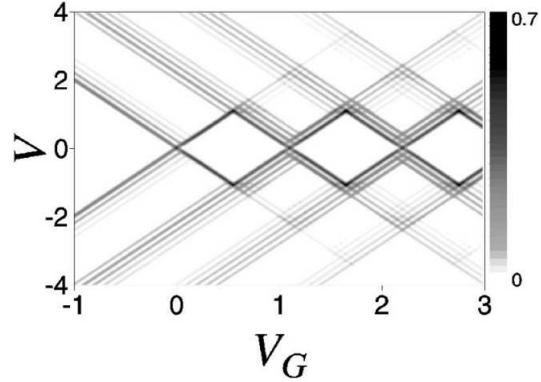}
\caption{Contour plot of the differential conductance.} \label{fig-4-12}
\end{center}
\end{figure}

Now, let us consider the current-voltage curve of the differential conductance
(Fig.~\ref{fig-4-12}).  First of all, Coulomb staircase is reproduced, which is more
pronounced, than for metallic islands, because the density of states is limited by
the available single-particle states and the current is saturated. Besides, small
additional steps due to discrete energy levels appear. This characteristic behaviour
is possible for large enough dots with $\Delta\epsilon\ll E_C$. If the level spacing
is of the oder of the charging energy $\Delta\epsilon\sim E_C$, the Coulomb blockade
steps and discrete-level steps look the same, but their statistics (position and
height distribution) is determined by the details of the single-particle spectrum
and interactions \cite{vonDelft01pr}.

Finally, let us consider the contour plot of the differential conductance
(Fig.~\ref{fig-4-12}). Ii is essentially different from those for the metallic
island. First, it is not symmetric in the gate voltage, because the energy spectrum
is restricted from the bottom, and at negative bias all the levels are above the
Fermi-level (the electron charge is negative, and a negative potential means a
positive energy shift). Nevertheless, existing stability patterns are of the same
origin and form the same structure. The qualitatively new feature is additional
lines correspondent to the additional discrete-level steps in the voltage-current
curves.In general, the current and conductance of quantum dots demonstrate all
typical features of discrete-level systems: current steps, conductance peaks.
Without Coulomb interaction the usual picture of resonant tunneling is reproduced.
In the limit of dense energy spectrum $\Delta\epsilon\rightarrow 0$ the sharp
single-level steps are merged into the smooth Coulomb staircase.

\subsubsection{Vibrons and Franck-Condon blockade}

\paragraph*{(i) Linear vibrons}

\begin{figure}[b]
\begin{center}
\epsfxsize=0.6\hsize
\epsfbox{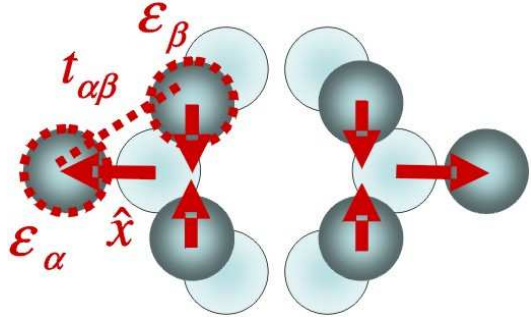}
\caption{(Color) A local molecular vibration. The empty circles show the equilibrium
positions of the atoms. The energies $\epsilon_\alpha$, $\epsilon_\beta$ and the
overlap integral $t_{\alpha\beta}$ are perturbed.} \label{fig-5-1}
\end{center}
\end{figure}

Vibrons are quantum local vibrations of nanosystems (Fig.~\ref{fig-5-1}), especially
important in flexible molecules. In the linear regime the small displacements of the
system can be expressed as linear combinations of the coordinates of the normal
modes $x_q$, which are described by a set of independent linear oscillators with the
Hamiltonian
\begin{equation}\label{V-H1}
  \hat H^{(0)}_V=\sum_q\left(\frac{\hat p^2_q}{2m_q}+\frac{1}{2}m_q\omega^2_q\hat x^2_q\right).
\end{equation}
The parameters $m_q$ are determined by the microscopic theory, and $\hat p_q$
(\mbox{$\hat p_q=-i\hbar\frac{\partial}{\partial x_q}$} in the $x$-representation)
is the momentum conjugated to $\hat x_q$,
%
  $\left[\hat x_q,\hat p_q\right]_-=i\hbar$.

Let us outline briefly a possible way to calculate the normal modes of a molecule,
and the relation between the positions of individual atoms and collective variables.
We assume, that the atomic configuration of a system is determined mainly by the
elastic forces, which are insensitive to the {\em transport} electrons. The dynamics
of this system is determined by the {\em atomic} Hamiltonian
\begin{equation}\label{V-Hi1}
  \hat H_{at}=\sum_n\frac{P_n^2}{2M_n}+W\left(\{{\bf R}_n\}\right),
\end{equation}
where $W\left(\{{\bf R}_n\}\right)$ is the elastic energy, which includes also the
static external forces and can be calculated by some {\em ab initio} method. Now
define new generalized variables $q_i$ with corresponding momentum $p_i$ (as the
generalized coordinates not only atomic positions, but also any other convenient
degrees of freedom can be considered, for example, molecular rotations,
center-of-mass motion, etc.)
\begin{equation}\label{V-Hi2}
  \hat H_{at}=\sum_i\frac{p_i^2}{2m_i}+W\left(\{q_i\}\right),
\end{equation}
"masses" $m_i$ should be considered as some parameters. The equilibrium coordinates
$q^0_i$ are defined from the energy minimum, the set of equations is
\begin{equation}
  \frac{\partial W\left(\{q^0_i\}\right)}{\partial q_i}=0.
\end{equation}

The equations for linear oscillations are obtained from the next order expansion in
the deviations $\Delta q_i=q_i-q^0_i$
\begin{equation}\label{V-Hi3}
  \hat H_{at}=\sum_i\frac{p_i^2}{2m_i}+\sum_{ij}
  \frac{\partial^2 W\left(\{q^0_j\}\right)}{\partial q_i\partial q_j}\Delta q_i\Delta q_j.
\end{equation}

This Hamiltonian describes a set of coupled oscillators. Finally, applying the
canonical transformation from $\Delta q_i$ to new variables $x_q$ ($q$ is now the
index of independent modes)
\begin{equation}
x_q=\sum_iC_{qi}q_i
\end{equation}
we derive the Hamiltonian (\ref{V-H1}) together with the frequencies $\omega_q$ of
vibrational modes.

It is useful to introduce the creation and annihilation operators
\begin{eqnarray}\label{V-a}
&\displaystyle a^\dag_q=\frac{1}{\sqrt{2}}\left(\sqrt{\frac{m_q\omega_q}{\hbar}}\hat
x_q+
  \frac{i}{\sqrt{m_q\omega_q\hbar}}\hat p_q\right), \\
&\displaystyle a_q=\frac{1}{\sqrt{2}}\left(\sqrt{\frac{m_q\omega_q}{\hbar}}\hat x_q-
  \frac{i}{\sqrt{m_q\omega_q\hbar}}\hat p_q\right),
\end{eqnarray}
in this representation the Hamiltonian of free vibrons is ($\hbar=1$)
\begin{equation}\label{V-H3}
 \hat H^{(0)}_{V}=\sum_q\omega_qa_q^\dag a_q.
\end{equation}

\paragraph*{(ii) Electron-vibron Hamiltonian}

A system without vibrons is described as before by a basis set of states
$|\alpha\rangle$ with energies $\epsilon_\alpha$ and inter-state overlap integrals
$t_{\alpha\beta}$, the model Hamiltonian of a noninteracting system is
\begin{equation}\label{H_M}
 \hat H^{(0)}_S=\sum_\alpha\left(\epsilon_\alpha+e\varphi_\alpha(t)\right)
 d^{\dag}_\alpha d_\alpha +\sum_{\alpha\neq\beta}t_{\alpha\beta}
 d^{\dag}_\alpha d_\beta,
\end{equation}
where $d^{\dag}_\alpha$,$d_\alpha$ are creation and annihilation operators in the
states $|\alpha\rangle$, and $\varphi_\alpha(t)$ is the (self-consistent) electrical
potential (\ref{CB-QD-phi}). The index $\alpha$ is used to mark single-electron
states (atomic orbitals) including the spin degree of freedom.

To establish the Hamiltonian describing the interaction of electrons with vibrons in
nanosystems, we can start from the generalized Hamiltonian
\begin{equation}
 \hat H_S=\sum_\alpha\tilde\epsilon_\alpha\left(\left\{x_q\right\}\right)
 d^{\dag}_\alpha d_\alpha +\sum_{\alpha\neq\beta}t_{\alpha\beta}\left(\left\{x_q\right\}\right)
 d^{\dag}_\alpha d_\beta,
\end{equation}
where the parameters are some functions of the vibronic normal coordinates $x_q$.
Note that we consider now only the electronic states, which were excluded previously
from the Hamiltonian (\ref{V-Hi1}), it is important to prevent double counting.

Expanding to the first order near the equilibrium state we obtain
\begin{equation}
 \hat H_{ev}=\sum_\alpha\sum_q
 \frac{\partial\tilde\epsilon_\alpha(0)}{\partial x_q}x_q
 d^{\dag}_\alpha d_\alpha +\sum_{\alpha\neq\beta}\sum_q
 \frac{\partial t_{\alpha\beta}(0)}{\partial x_q}x_q
 d^{\dag}_\alpha d_\beta,
\end{equation}
where $\tilde\epsilon_\alpha(0)$ and $t_{\alpha\beta}(0)$ are unperturbed values of
the energy and the overlap integral. In the quantum limit the normal coordinates
should be treated as operators, and in the second-quantized representation the
interaction Hamiltonian is
\begin{equation}\label{V-H5}
 \hat H_{ev}=\sum_{\alpha\beta}\sum_q\lambda^q_{\alpha\beta}(a_q+a_q^\dag)
 d^{\dag}_\alpha d_\beta.
\end{equation}
This Hamiltonian is similar to the usual electron-phonon Hamiltonian, but the
vibrations are like localized phonons and $q$ is an index labeling them, not the
wave-vector. We include both diagonal coupling, which describes a change of the
electrostatic energy with the distance between atoms, and the off-diagonal coupling,
which describes the dependence of the matrix elements $t_{\alpha\beta}$ over the
distance between atoms.

\begin{figure}[b]
\begin{center}
\vskip 0.35cm\epsfxsize=0.6\hsize
 \epsfbox{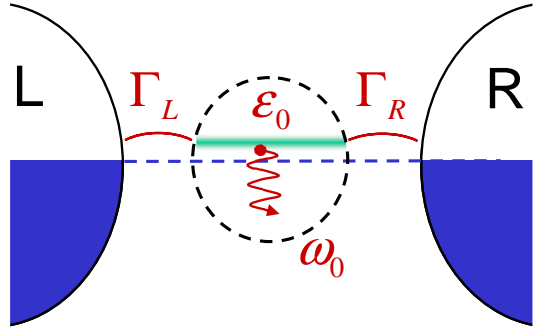}
\caption{(Color) Single-level electron-vibron model.} \label{fig-5-1a}
\end{center}
\end{figure}

The full Hamiltonian
\begin{equation}\label{H}
 \hat H=\hat H^0_S+\hat H_V+\hat H_L+\hat H_R+\hat H_T
\end{equation}
is the sum of the noninteracting Hamiltonian $\hat H^0_S$, the Hamiltonians of the
leads $\hat H_{R(L)}$, the tunneling Hamiltonian $\hat H_T$ describing the
system-to-lead coupling, the vibron Hamiltonian $\hat H_V$ including electron-vibron
interaction and coupling of vibrations to the environment (describing dissipation of
vibrons).

Vibrons and the electron-vibron coupling are described by the Hamiltonian
($\hbar=1$)
\begin{equation}\label{H_V}
 \hat H_{V}=\sum_q\omega_qa_q^\dag a_q
 +\sum_{\alpha\beta}\sum_q\lambda^q_{\alpha\beta}(a_q+a_q^\dag)
 d^{\dag}_\alpha d_\beta +\hat H_{env}.
\end{equation}
The first term represents free vibrons with the energy $\hbar\omega_q$. The second
term is the electron-vibron interaction. The rest part $\hat H_{env}$ describes
dissipation of vibrons due to interaction with other degrees of freedom, we do not
consider the details in this chapter.

The Hamiltonians of the right (R) and left (L) leads read as usual
\begin{equation}
 \hat H_{i=L(R)}=\sum_{k\sigma}(\epsilon_{ik\sigma}+e\varphi_i)
 c^{\dag}_{ik\sigma}c_{ik\sigma},
\end{equation}
$\varphi_i$ are the electrical potentials of the leads. Finally, the tunneling
Hamiltonian
\begin{equation}\label{H_T}
 \hat H_T=\sum_{i=L,R}\sum_{k\sigma,\alpha}\left(V_{ik\sigma,\alpha}
 c^{\dag}_{ik\sigma}d_\alpha+V^*_{ik\sigma,\alpha}d^{\dag}_\alpha c_{ik\sigma}\right)
\end{equation}
describes the hopping between the leads and the molecule. A direct hopping between
two leads is neglected.

The simplest example of the considered model is a single-level model
(Fig.~\ref{fig-5-1a}) with the Hamiltonian
\begin{equation}
  \hat H=\tilde\epsilon_0d^{\dag}d+\omega_0a^{\dag}a+\lambda\left(a^{\dag}+a\right)d^{\dag}d
  +\sum_{ik}\left[\tilde\epsilon_{ik}
  c^{\dag}_{ik}c_{ik}+V_{ik}c^{\dag}_{ik}d+h.c.\right],
\end{equation}
where the first and the second terms describe free electron state and free vibron,
the third term is electron-vibron interaction, and the rest is the Hamiltonian of
the leads and tunneling coupling ($i=L,R$ is the lead index).

\begin{figure}[b]
\begin{center}
\vskip 0.35cm \epsfxsize=0.6\hsize
\epsfbox{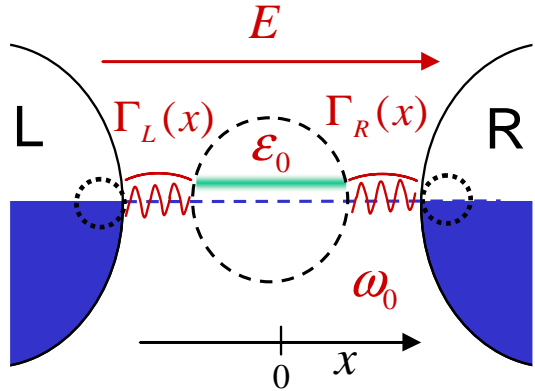}
\caption{(Color) A center-of-mass vibration.} \label{fig-5-1b}
\end{center}
\end{figure}

The other important case is a center-of-mass motion of molecules between the leads
(Fig.~\ref{fig-5-1b}). Here not the internal overlap integrals, but the coupling to
the leads $V_{ik\sigma,\alpha}(x)$ is fluctuating. This model is easily reduced to
the general model (\ref{H_V}), if we consider additionaly two not flexible states in
the left and right leads (two atoms most close to a system), to which the central
system is coupled (shown by the dotted circles).

Tunneling Hamiltonian includes $x$-dependent matrix elements, considered in linear
approximation
\begin{equation}
  H_T=\sum_{i=L,R}\sum_{k\sigma,\alpha}\left(V_{ik\sigma,\alpha}{(\hat x)}
  c^{\dag}_{ik\sigma}d_\alpha+h.c.\right),
  \end{equation}
\begin{equation}
V_{L,R}(x)=V_{0}{e^{\mp\hat x/L}}\approx V_{0} \left(1{\mp\frac{\hat x}{L}}\right).
\end{equation}

Consider now a single-level molecule ($\alpha\equiv 0$) and extend our system,
including two additional states from the left ($\alpha\equiv l$) and right
($\alpha\equiv r$) sides of a molecule, which are coupled to the central state
through $x$-dependent matrix elements, and to the leads in a usual way through
$\Gamma_{L(R)}$. Then the Hamiltonian is of linear electron-vibron type
\begin{align} \nonumber
  \hat H_{M+V}= & \sum_{\alpha=l,0,r}\left(\epsilon_\alpha+e\varphi_\alpha\right)
  d^{\dag}_\alpha d_\alpha+
  t_l(d_l^\dag d_0+h.c.)+t_r(d_r^\dag d_0+h.c.)+ \\
& +{\omega_0}a^\dag a+(a+a^\dag)\left(\lambda_0 d_0^\dag d_0-\lambda_l(d_l^\dag
  d_0+h.c.)+\lambda_r(d_r^\dag d_0+h.c.)\right).
\end{align}

\paragraph*{(iii) Local polaron and canonical transformation}

Now let us start to consider the situation, when the electron-vibron interaction is
strong. For an isolated system with the Hamiltonian, including only diagonal terms,
\begin{equation}
 \hat H_{S+V}=\sum_{\alpha}\tilde\epsilon_{\alpha}d^{\dag}_\alpha d_\alpha
 +\sum_q\omega_qa_q^\dag a_q
 +\sum_{\alpha}\sum_q\lambda^q_{\alpha}(a_q+a_q^\dag)
 d^{\dag}_\alpha d_\alpha,
\end{equation}
the problem can be solved exactly. This solution, as well as the method of the
solution (canonical transformation), plays an important role in the theory of
electron-vibron systems, and we consider it in detail.

Let's start from the simplest case. The single-level electron-vibron model is
described by the Hamiltonian
\begin{equation}\label{V-P-H1}
  \hat H_{S+V}=\tilde\epsilon_0d^{\dag}d+\omega_0a^{\dag}a+\lambda\left(a^{\dag}+a\right)d^{\dag}d,
\end{equation}
where the first and the second terms describe free electron state and free vibron,
and the third term is the electron-vibron interaction.

This Hamiltonian is diagonalized by the canonical transformation (called
"Lang-Firsov" or "polaron") \cite{Lang63jetp,Hewson74jjap,Mahan90book}
\begin{equation}\label{V-P-LF1}
  \bar H=\hat S^{-1} \hat H \hat S,
\end{equation}
with
\begin{equation}\label{V-P-LF2}
  \hat S=\exp\left[-\frac{\lambda}{\omega_0}\left(a^\dag-a\right)d^\dag d\right],
\end{equation}
the Hamiltonian (\ref{V-P-H1}) is transformed as
\begin{equation}\label{V-P-H2}
  \bar H_{S+V}=\hat S^{-1} \hat H_{S+V} \hat S=
  \tilde\epsilon_0\bar d^{\dag}\bar d+\omega_0\bar a^{\dag}\bar a+
  \lambda\left(\bar a^{\dag}+\bar a\right)\bar d^{\dag}\bar d,
\end{equation}
it has the same form as (\ref{V-P-H1}) with new operators, it is a trivial
consequence of the general property
\begin{equation}
  \hat S^{-1} \left(\hat f_1 \hat f_2 \hat f_3...\right) \hat S=
  (\hat S^{-1}\hat f_1\hat S)(\hat S^{-1}\hat f_2\hat S)(\hat S^{-1}\hat f_3\hat S)...=
  \bar f_1 \bar f_2 \bar f_3 ...
\end{equation}
and new single-particle operators are
\begin{eqnarray}
& \bar a=\hat S^{-1} a \hat S=a-\frac{\lambda}{\omega_0}d^\dag d, \\ & \bar
a^\dag=\hat S^{-1} a^\dag \hat S=a^\dag-\frac{\lambda}{\omega_0}d^\dag d, \\ & \bar
d=\hat S^{-1} d \hat
S=\exp\left[-\frac{\lambda}{\omega_0}\left(a^\dag-a\right)\right] d,
\label{V-P-dbar1} \\ & \bar d^\dag=\hat S^{-1}d^\dag\hat
S=\exp\left[\frac{\lambda}{\omega_0}\left(a^\dag-a\right)\right] d^\dag.
\label{V-P-dbar2}
\end{eqnarray}

Substituting these expressions into (\ref{V-P-H2}) we get finally
\begin{equation}\label{V-P-H3}
  \bar H_{S+V}=\left(\tilde\epsilon_0-\frac{\lambda^2}{\omega_0}\right)d^{\dag}d+\omega_0a^{\dag}a.
\end{equation}

We see that the electron-vibron Hamiltonian (\ref{V-P-H1}) is equivalent to the
free-particle Hamiltonian (\ref{V-P-H3}). This equivalence means that any quantum
state $|\bar\psi_\lambda\rangle$, obtained as a solution of the Hamiltonian
(\ref{V-P-H3}) is one-to-one equivalent to the state $|\psi_\lambda\rangle$ as a
solution of the initial Hamiltonian (\ref{V-P-H1}), with the same matrix elements
for any operator
\begin{equation}\label{V-P-ME1}
  \langle\bar\psi_\lambda|\bar f|\bar\psi_\lambda\rangle=\langle\psi_\lambda|\hat
  f|\psi_\lambda\rangle,
\end{equation}
\begin{equation}\label{V-P-LF3}
  \bar f=\hat S^{-1} \hat f \hat S,
\end{equation}
\begin{equation}\label{V-P-LF4}
  |\bar\psi_\lambda\rangle=\hat S^{-1}|\psi_\lambda\rangle.
\end{equation}

It follows immediately that the eigenstates of the free-particle Hamiltonian are
\begin{equation}\label{V-P-ES}
  |\bar\psi_{nm}\rangle=
  |n=0,1;m=0,1,2,...\rangle=(d^\dag)^{n}\frac{(a^\dag)^{m}}{\sqrt{m!}}|0\rangle,
\end{equation}
and the eigen-energies are
\begin{equation}\label{V-P-EE}
  E(n,m)=\left(\tilde\epsilon_0-\frac{\lambda^2}{\omega_0}\right)n+\omega_0m.
\end{equation}

The eigenstates of the {\em initial} Hamiltonian (\ref{V-P-H1}) are
\begin{equation}\label{V-P-ES2}
  |\psi_{nm}\rangle=\hat S|\bar\psi_{nm}\rangle=
  e^{-\frac{\lambda}{\omega_0}\left(a^\dag-a\right)d^\dag d}
  (d^\dag)^{n}\frac{(a^\dag)^{m}}{\sqrt{m!}}|0\rangle,
\end{equation}
with the same quantum numbers $(n,m)$ and the same energies (\ref{V-P-EE}). This
representation of the eigenstates demonstrates clearly the collective nature of the
excitations, but it is inconvenient for practical calculations.

Now let us consider the polaron transformation (\ref{V-P-LF1})-(\ref{V-P-LF2})
applied to the tunneling Hamiltonian
\begin{equation}
 \hat H_T=\sum_{i=L,R}\sum_{k\sigma}\left(V_{ik\sigma}
 c^{\dag}_{ik\sigma}d+V^*_{ik\sigma}d^{\dag} c_{ik\sigma}\right)
\end{equation}
The electron operators in the left and right leads $c_{ik\sigma}$ are not changed by
this operation, but the dot operators $d_\alpha$, $d^{\dag}_\alpha$ are changed in
accordance with (\ref{V-P-dbar1}) and (\ref{V-P-dbar2}). So that transformed
Hamiltonian is
\begin{equation}\label{V-SPA-HTbar}
 \bar H_T=\sum_{i=L,R}\sum_{k\sigma}\left(V_{ik\sigma}
 e^{-\frac{\lambda}{\omega_0}\left(a^\dag-a\right)}
 c^{\dag}_{ik\sigma}d+V^*_{ik\sigma}
 e^{\frac{\lambda}{\omega_0}\left(a^\dag-a\right)}
 d^{\dag} c_{ik\sigma}\right).
\end{equation}

Now we see clear the problem: while the new dot Hamiltonian (\ref{V-P-H3}) is very
simple and exactly solvable, the new tunneling Hamiltonian (\ref{V-SPA-HTbar}) is
complicated. Moreover, instead of one linear electron-vibron interaction term, the
exponent in (\ref{V-SPA-HTbar}) produces all powers of vibronic operators. Actually,
we simply remove the complexity from one place to the other. This approach works
well, if the tunneling can be considered as a perturbation, we consider it in the
next section. In the general case the problem is quite difficult, but in the
single-particle approximation it can be solved exactly
\cite{Glazman88jetp,Wingreen88prl,Wingreen89prb,Jonson89prb}.

To conclude, after the canonical transformation we have two equivalent models: (1)
the initial model (\ref{V-P-H1}) with the eigenstates (\ref{V-P-ES2}); and (2) the
{\em fictional} free-particle model (\ref{V-P-H3}) with the eigenstates
(\ref{V-P-ES}). We shall call this second model {\em polaron representation}. The
relation between the models is established by (\ref{V-P-ME1})-(\ref{V-P-LF4}). It is
also clear from the Hamiltonian (\ref{V-P-H2}), that the operators $\bar d^\dag$,
$\bar d$, $\bar a^\dag$, and $\bar a$ describe the initial electrons and vibrons in
the fictional model.

\paragraph*{(iv) Inelastic tunneling in the single-particle approximation}

In this section we consider a special case of a {\em single particle transmission}
through an electron-vibron system. It means that we consider a system coupled to the
leads, but without electrons in the leads. This can be considered equivalently as the limit
of large electron level energy $\epsilon_0$ (far from the Fermi surface in the leads).

The inelastic {\em transmission matrix $T(\epsilon',\epsilon)$} describes the probability
that an electron with energy $\epsilon$, incident from one lead, is transmitted with the
energy $\epsilon'$ into a second lead. The transmission function can be defined as the
total transmission probability
\begin{equation}
  T(\epsilon)=\int T(\epsilon',\epsilon)d\epsilon'.
\end{equation}

For a noninteracting single-level system the transmission matrix is
\begin{equation}
  T^0(\epsilon',\epsilon)=\frac{\Gamma_R(\epsilon)\Gamma_L(\epsilon)\delta(\epsilon-\epsilon')}
  {(\epsilon-\epsilon_0-\Lambda(\epsilon))^2+(\Gamma(\epsilon)/2)^2},
\end{equation}
where $\Gamma(\epsilon)=\Gamma_L(\epsilon)+\Gamma_R(\epsilon)$ is the level-width
function, and $\Lambda(\epsilon)$ is the real part of the self-energy.

\begin{figure}[b]
\begin{center}
\epsfxsize=0.6\hsize \epsfbox{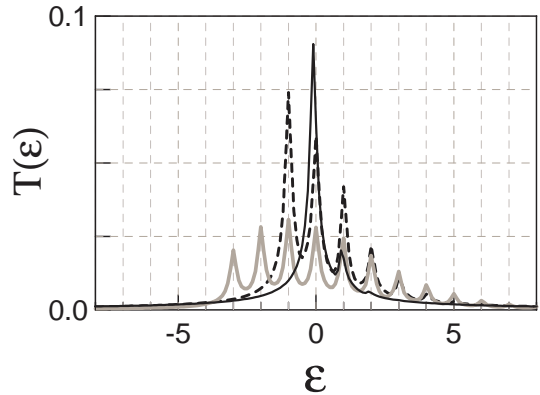}
\caption{Transmission function as a function of energy at different electron-vibron
coupling: $g=0.1$ (thin solid line), $g=1$ (dashed line), and $g=3$ (thick solid line),
at $\Gamma=0.1$.}
\label{V-SPA-T1}
\end{center}
\end{figure}

We can do some general conclusions, based on the form of the tunneling
Hamiltonian (\ref{V-SPA-HTbar}). Expanding the exponent in the same way as before, we get
\begin{equation}\label{V-SPA-HTbar2}
 \bar H_T=\sum_{i=L,R}\sum_{k\sigma}\left(V_{ik\sigma}
 c^{\dag}_{ik\sigma}d\left[\alpha_0+\sum_{m=1}^\infty\alpha_m\left((a^\dag)^m+a^m\right)
 \right]+h.c.\right),
\end{equation}
with the coefficients
\begin{equation}
  \alpha_m=\left(-\frac{\lambda}{\omega_0}\right)^{m}
  \frac{e^{-(\lambda/\omega_0)^2/2}}{m!}.
\end{equation}
This complex Hamiltonian has very clear interpretation, the tunneling of one electron
from the right to the left lead is accompanied by the excitation of vibrons. The energy
conservation implies that
\begin{equation}
  \epsilon-\epsilon'=\pm m\omega_0,
\end{equation}
so that the inelastic tunneling with emission or absorption of vibrons is possible.

\begin{figure}
\begin{center}
\epsfxsize=0.6\hsize \epsfbox{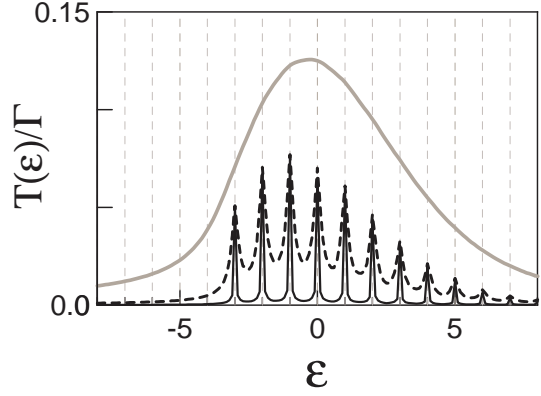}
\caption{Transmission function as a function of energy at different coupling to the
leads: $\Gamma=0.01$ (thin solid line), $\Gamma=0.1$ (dashed line), and $\Gamma=1$ (thick
solid line), at $g=3$.} \label{V-SPA-T2}
\end{center}
\end{figure}

The exact solution is possible in the wide-band
limit.~\cite{Glazman88jetp,Wingreen88prl,Wingreen89prb,Jonson89prb}

It is convenient to introduce the dimensionless electron-vibron coupling constant
\begin{equation}\label{V-SPA-g}
  g=\left(\frac{\lambda}{\omega_0}\right)^2.
\end{equation}

At zero temperature the solution is
\begin{align}\nonumber
  T & (\epsilon',\epsilon)=\Gamma_L\Gamma_Re^{-2g}\sum_{m=0}^\infty
  \frac{g^m}{m!}\delta(\epsilon-\epsilon'-m\omega_0) \\
& \times\left|\sum_{j=0}^m(-1)^j\frac{m!}{j!(m-j)!}
  \sum_{l=0}^\infty\frac{g^l}{l!}\frac{1}{\epsilon-\epsilon_0+g\omega_0-(j+l)\omega_0+i\Gamma/2}
  \right|^2,
\end{align}
the total transmission function $T(\epsilon)$ is trivially obtain by integration over
$\epsilon'$. The representative results are presented in Figs. \ref{V-SPA-T1} and
\ref{V-SPA-T2}.

At finite temperature the general expression is too cumbersome, and we present here only
the expression for the total transmission function
\begin{align}\nonumber
  T & (\epsilon)=\frac{\Gamma_L\Gamma_R}{\Gamma}e^{\displaystyle -g(1+2n_\omega)}\int_{-\infty}^\infty dt \\
& \times\exp\left(-\frac{\Gamma}{2}|t|+i(\epsilon-\epsilon_0+g\omega_0)t-
  g\left[(1+n_\omega)e^{-i\omega_0t}+n_\omega e^{i\omega_0t}\right]
  \right),
\end{align}
where $n_\omega$ is the equilibrium number of vibrons.

\paragraph*{(v) Master equation}

When the system is weakly coupled to the leads, the polaron representation
(\ref{V-P-H3}), (\ref{V-SPA-HTbar}) is a convenient starting point. Here we consider
how the sequential tunneling is modified by vibrons.

The master equation for the probability $p(n,m,t)$ to find the system in one of the
polaron eigenstates (\ref{V-P-ES}) can be written as
\begin{equation}\label{V-ST-ME}
  \frac{dp(n,m)}{dt}=\sum_{n'm'}\Gamma^{nn'}_{mm'}p(n',m')-
  \sum_{n'm'}\Gamma^{n'n}_{m'm}p(n,m)+I^V[p],
\end{equation}
where the first term describes tunneling transition {\em into the state} $|n,m
\rangle$, and the second term -- tunneling transition {\em out of the state} $|n,m
\rangle$, $I^V[p]$ is the vibron scattering integral describing the relaxation to
equilibrium. The transition rates $\Gamma^{nn'}_{mm'}$ should be found from the
Hamiltonian (\ref{V-SPA-HTbar}).

Taking into account all possible single-electron tunneling processes, we obtain the
incoming tunneling rate
\begin{align}\nonumber
  \Gamma^{10}_{mm'} =
  \frac{2\pi}{\hbar} & \sum_{ik\sigma} f_i^0(E_{ik\sigma})
  \left|\left\langle i\bar k,1,m\left|\bar H_{T}\right|ik,0,m'\right\rangle\right|^2
  \delta(E_{0m'}+E_{ik\sigma}-E_{1m}) \\ \nonumber
  =\frac{2\pi}{\hbar}\sum_{ik\sigma} & f_i^0(E_{ik\sigma})
  \left|V_{ik\sigma}\right|^2
  \left|\left\langle m\left|e^{\frac{\lambda}{\omega_0}\left(a^\dag-a\right)}
  \right|m'\right\rangle\right|^2\delta(E_{0m'}+E_{ik\sigma}-E_{1m})\\
& =\sum_{i=L,R}\Gamma_{i}(E_{1m}-E_{0m'})\left|M_{mm'}\right|^2
  f_i^0(E_{1m}-E_{0m'}),
\end{align}
where
\begin{equation}\label{V-ST-FC}
  M_{mm'}=\left\langle m\left|e^{\frac{\lambda}{\omega_0}\left(a^\dag-a\right)}
  \right|m'\right\rangle
\end{equation}
is the Franck-Condon matrix element.
We use usual short-hand notations: $|ik,n,m\rangle$ is the state with occupied
$k$-state in the $i-$th lead, $n$ electrons, and $m$ vibrons, while $|i\bar
k,n,m\rangle$ is the state with unoccupied $k$-state in the $i-$th lead, $E_{nm}$ is
the polaron energy (\ref{V-P-EE}).

Similarly, the outgoing rate is
\begin{align}
  \Gamma^{01}_{mm'} =
  \sum_{i=L,R}\Gamma_{i}(E_{1m'}-E_{0m})\left|M_{mm'}\right|^2
  \left(1-f_i^0(E_{1m'}-E_{0m})\right).
\end{align}

The current (from the left or right lead to the system) is
\begin{equation}\label{V-J1}
  J_{i=L,R}(t)=e\sum_{mm'}\left(\Gamma^{10}_{imm'}p(0,m')-
  \Gamma^{01}_{imm'}p(1,m')\right).
\end{equation}

The system of equations (\ref{V-ST-ME})-(\ref{V-J1}) solves the transport problem in
the sequential tunneling regime.

\paragraph*{(v) Franck-Condon blockade}

Now let us consider some details of the tunneling at small and large values of the
electro-vibron coupling parameter $g=\left(\frac{\lambda}{\omega_0}\right)^2$.

The matrix element (\ref{V-ST-FC}) can be calculated analytically, it is symmetric
in $m-m'$ and for $m<m'$ is
\begin{equation}\label{V-ST-FC2}
  M_{m<m'}=\sum_{l=0}^{m}\frac{(-g)^l\sqrt{m!m'!}e^{-g/2}g^{(m'-m)/2}}
  {l!(m-l)!(l+m'-m)!}.
\end{equation}

The lowest order elements are
\begin{eqnarray}
M_{0m}=e^{-g/2}\frac{g^{m/2}}{\sqrt{m!}}, \\ M_{11}=(1-g)e^{-g/2}, \\
M_{12}=\sqrt{2g}\left(1-\frac{g}{2}\right)e^{-g/2}...
\end{eqnarray}

\begin{figure}[t]
\begin{center}
\epsfxsize=0.6\hsize
\epsfbox{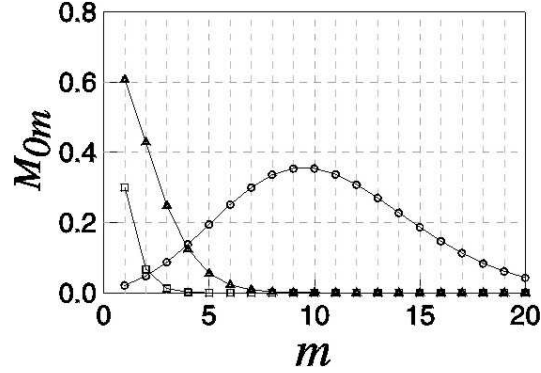}
\caption{Franck-Condon matrix elements $M_{0m}$ for weak ($g=0.1$, squares),
intermediate ($g=1$, triangles), and strong ($g=10$, circles) electron-vibron
interaction. Lines are the guides for eyes.}
\label{V-ST-M}
\end{center}
\end{figure}

The characteristic feature of these matrix elements is so-called Franck-Condon
blockade \cite{Koch05prl,Koch06prb}, illustrated in Fig.~\ref{V-ST-M} for the matrix
element $M_{0m}$. From the picture, as well as from the analytical formulas, it is
clear, that in the case of strong electron-vibron interaction the tunneling with
small change of the vibron quantum number is suppressed exponentially, and only the
tunneling through high-energy states is possible, which is also suppressed at low
bias voltage and low temperature. Thus, the electron transport through a system
(linear conductance) is very small.

\begin{figure}[t]
\begin{center}
\epsfxsize=0.6\hsize
\epsfbox{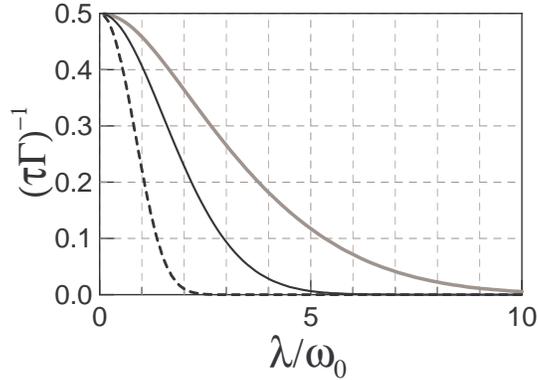}
\caption{The inverse life-time $(\tau\Gamma)^{-1}$ as a function of
$\lambda/\omega_0$ at optimal electron level position $\epsilon_0=\lambda^2/2\omega_0$
for neutral state (thin solid line), and for the charged state (dashed line), and
for the neutral state at other level position $\epsilon_0=\lambda^2/4\omega_0$ (thick solid line).}
\label{V-ST-M2}
\end{center}
\end{figure}

There are several interesting manifestations of the Franck-Condon blockade.

The {\em life-time} of the state $|n,m\rangle$ is determined by the sum of the rates
of all possible processes which change this state in the assumption that all other
states are empty
\begin{equation}\label{V-ST-Time}
  \tau^{-1}_{nm}=\sum_{n'm'}\Gamma^{n'n}_{m'm}.
\end{equation}

As an example, let us calculate the life-time of the neutral state
$|0,0\rangle$, which has the energy higher than the charged ground state
$|1,0\rangle$.
\begin{equation}\label{V-ST-T1}
  \tau^{-1}_{00}=\sum_{n'm'}\Gamma^{n'0}_{m'0}=
  \sum_{m}
  \sum_{i=L,R}\Gamma_{i}(E_{1m}-E_{00})\left|M_{m0}\right|^2
  f_i^0(E_{1m}-E_{00}).
\end{equation}

In the wide-band limit we obtain the simple analytical expression
\begin{equation}\label{V-ST-T2}
  \tau^{-1}_{00}=\Gamma\sum_{m}e^{-g}\frac{g^{m}}{m!}
  f^0\left(\tilde\epsilon_0-\frac{\lambda^2}{\omega_0}+\omega_0m\right).
\end{equation}
The corresponding expression for the life-time of the charged state (which can be
excited by thermal fluctuations) is
\begin{equation}\label{V-ST-T3}
  \tau^{-1}_{10}=\Gamma\sum_{m}e^{-g}\frac{g^{m}}{m!}
  f^0\left(-\tilde\epsilon_0+\frac{\lambda^2}{\omega_0}+\omega_0m\right).
\end{equation}

The result of the calculation is shown in Fig.~\ref{V-ST-M2}, it is clear seen that
the tunneling from the state $|0,0\rangle$ to the charged state and from the state
$|1,0\rangle$ to the neutral state is exponentially suppressed in comparison with
the bare tunneling rate $\Gamma$ at large values of the electron-vibron interaction
constant $\lambda$. This {\em polaron memory effect} can be used to create
nano-memory and nano-switches. At finite voltage the switching between two states is
easy accessible through the excited vibron states. It can be used to switch between
memory states \cite{Ryndyk08preprint}.

The other direct manifestation of the Franck-Condon blockade, -- suppression of the
linear conductance, was considered in Refs.\,\cite{Koch05prl,Koch06prb}.

\section{Nonequilibrium Green function theory of transport}

\subsection{Standard transport model: a nanosystem between ideal leads}

First of all, we formulate a standard discrete-level model to describe nanoscale
interacting quantum systems (quantum dot, system of quantum dots, molecule, below
"nanosystem", "central system", or simply "system") coupled to free conduction
electrons in the leads. We include the Coulomb interaction with the help of the
Anderson-Hubbard Hamiltonan to be able to describe correlation effects, such as
Coulomb blockade and Kondo effect, which could dominate at low temperatures. At high
temperatures or weak interaction the self-consistent mean-field effects are well
reproduced by the same model. Furthermore, electrons are coupled to vibrational
modes, below we use the electron-vibron model introduced previously.

\paragraph*{(i) The model Hamiltonian}

The full Hamiltonian is the sum of the free system Hamiltonian $\hat H_S^{(0)}$, the
inter-system electron-electron interaction Hamiltonian $\hat H_C$, the vibron Hamiltonian
$\hat H_V$ including the electron-vibron interaction and coupling of vibrations to the
environment (dissipation of vibrons), the Hamiltonians of the leads $\hat H_{R(L)}$, and
the tunneling Hamiltonian $\hat H_T$ describing the system-to-lead coupling
\begin{equation}\label{Model-H}
 \hat H=\hat H_S+\hat H_C+\hat H_V+\hat H_L+\hat H_R+\hat H_T.
\end{equation}

An isolated noninteracting nanosystem is described as a set of discrete states
$|\alpha\rangle$ with energies $\epsilon_\alpha$ and inter-orbital overlap integrals
$t_{\alpha\beta}$ by the following model Hamiltonian:
\begin{equation}\label{Model-H_M}
 \hat H^{(0)}_S=\sum_\alpha\left(\epsilon_\alpha+e\varphi_\alpha(t)\right)
 d^{\dag}_\alpha d_\alpha +
 \sum_{\alpha\neq\beta}t_{\alpha\beta}
 d^{\dag}_\alpha d_\beta,
\end{equation}
where $d^{\dag}_\alpha$,$d_\alpha$ are creation and annihilation operators in the states
$|\alpha\rangle$, and $\varphi_\alpha(t)$ is the effective (self-consistent) electrical
potential. The index $\alpha$ is used to mark single-electron states (e.g. atomic
orbitals) including the spin degree of freedom. In the eigenstate (molecular orbital)
representation the second term is absent and the Hamiltonian is diagonal.

For molecular transport the parameters of a model are to be determined by {\em ab initio}
methods or considered as semi-empirical. This is a compromise, which allows us to
consider complex molecules with a relatively simple model.

The Hamiltonians of the right (R) and left (L) leads are
\begin{equation}\label{Model-H_L}
 \hat H_{i=L(R)}=\sum_{k\sigma}(\epsilon_{ik\sigma}+e\varphi_i(t))
 c^{\dag}_{ik\sigma}c_{ik\sigma},
\end{equation}
$\varphi_i(t)$ are the electrical potentials of the leads, the index $k$ is the wave
vector, but can be considered as representing an other conserved quantum number, $\sigma$
is the spin index, but can be considered as a generalized channel number, describing e.g.
different bands or subbands in semiconductors. Alternatively, the tight-binding model can
be used also for the leads, then (\ref{Model-H_L}) should be considered as a result of the
Fourier transformation. The leads are assumed to be noninteracting and equilibrium.

The tunneling Hamiltonian
\begin{equation}\label{Model-H_T}
 \hat H_T=\sum_{i=L,R}\sum_{k\sigma,\alpha}\left(V_{ik\sigma,\alpha}
 c^{\dag}_{ik\sigma}d_\alpha+V^*_{ik\sigma,\alpha}d_\alpha^{\dag}c_{ik\sigma}\right)
\end{equation}
describes the hopping between the leads and the system. The direct hopping between two
leads is neglected (relatively weak molecule-to-lead coupling case). Note, that the
direct hoping between equilibrium leads can be easy taken into account as an additional
independent current channel.

The Coulomb interaction inside a system is described by the Anderson-Hubbard Hamiltonian
\begin{equation}\label{Model-H_C}
  \hat H_C=\frac{1}{2}\sum_{\alpha\neq\beta}U_{\alpha\beta}\hat n_\alpha \hat n_\beta.
\end{equation}
This Hamiltonian is used usually only for the short-range part of Coulomb interaction.
The long-range interactions can be better introduced through the self-consistent
electrical potential $\varphi_\alpha$, which is determined by the Poison equation with
the average electron density.

Vibrations and the electron-vibron coupling are described by the
Hamiltonian
\begin{equation}\label{Model-H_V}
 \hat H_{V}=\sum_q\hbar\omega_qa_q^\dag a_q
 +\sum_{\alpha\beta}\sum_q\lambda^q_{\alpha\beta}(a_q+a_q^\dag)
 d^{\dag}_\alpha d_\beta +\hat H_{e}.
\end{equation}
Here vibrations are considered as localized phonons and $q$ is the index labeling them,
not the wave-vector. The first term describes free vibrons with the energy
$\hbar\omega_q$. The second term represents the electron-vibron interaction. The third
term describes the coupling to the environment and the dissipation of vibrons. We include
both diagonal coupling, which originates from a change of the electrostatic energy with
the distance between atoms, and the off-diagonal coupling, which can be obtained from the
dependence of the matrix elements $t_{\alpha\beta}$ over the distance between atoms.

\paragraph*{(ii) Nonequilibrium current and charge}

To connect the microscopic description of a system with the macroscopic (electrodynamic)
equations and calculate the observables, we need the expressions for the nonequilibrium
electrical charge of the system and the current between the system and the leads.

The charge in a nonequilibrium state is given by ($Q_0$ is the background charge)
\begin{equation}\label{Q1}
  Q_S(t)=e\sum_\alpha\left\langle d^{\dag}_{\alpha}d_{\alpha}\right\rangle-Q_0.
\end{equation}

To calculate the current we find the time evolution of the particle number operator
\mbox{$\hat N_S=\sum_\alpha d^{\dag}_{\alpha}d_{\alpha}$} due to tunneling from the left
($i=L$) or right ($i=R$) contact.

The current {\em from} the left ($i=L$) or right ($i=R$) contact {\em to} the
nanosystem is determined by (note, that we consider $e$ as the charge of the
electron (negative) or the hole (positive))
\begin{equation}
  J_i(t)=-e\left\langle\left(\frac{dN_S}{dt}\right)_i\right\rangle=-\frac{ie}{\hbar}
  \left\langle[H_T^{(i)},N_S]\right\rangle,
\end{equation}
where
\begin{equation}
  H_T^{(i)}=\sum_{k\sigma,\alpha}\left(V_{ik\sigma,\alpha}
  c^{\dag}_{ik\sigma}d_\alpha +V^*_{ik\sigma,\alpha}d_\alpha^{\dag}c_{ik\sigma}\right)
\end{equation}
is the Hamiltonian of the coupling to the corresponding contact. The current is
determined by this only part of the full Hamiltonian (\ref{H}), because all other terms
commute with $\hat N_S$.

Applying the commutation relation
\begin{align}
\left[d_{\alpha},d^{\dag}_{\beta}d_{\beta}\right]=
d_{\alpha}d^{\dag}_{\beta}d_{\beta}-d^{\dag}_{\beta}d_{\beta}d_{\alpha}= &
d_{\alpha}d^{\dag}_{\beta}d_{\beta}+d^{\dag}_{\beta}d_{\alpha}d_{\beta}= \nonumber \\
(d_{\alpha}d^{\dag}_{\beta}+ \delta_{\alpha\beta}-d_{\alpha}d^{\dag}_{\beta}) &
d_{\beta}= \delta_{\alpha\beta}d_{\alpha},
\end{align}
one obtains finally
\begin{equation}\label{J1}
  J_i(t)=\frac{ie}{\hbar}\sum_{k\sigma,\alpha}\left[V_{ik\sigma,\alpha}
  \left\langle c^{\dag}_{ik\sigma}d_{\alpha} \right\rangle-
  V^*_{ik\sigma,\alpha}\left\langle d^{\dag}_{\alpha}c_{ik\sigma} \right\rangle
  \right].
\end{equation}

\paragraph*{(iii) Density matrix and NGF}

The averages of the operators in Eqs. (\ref{Q1}) and (\ref{J1}) are the elements of the
density matrix in the single-particle space
\begin{eqnarray}\label{Model-DM}
 & \rho_{\alpha\alpha}(t)=\left\langle d^{\dag}_{\alpha}(t)d_{\alpha}(t)\right\rangle, \\
 & \rho_{\alpha,ik\sigma}(t)=\left\langle c^{\dag}_{ik\sigma}(t)d_{\alpha}(t) \right\rangle.
\end{eqnarray}

It is possible, also, to express it as a two-time Green function at equal times
\begin{equation}
\displaystyle Q_S(t)=e\sum_\alpha {\rho_{\alpha\alpha}(t)} =-ie\sum_\alpha
{G^<_{\alpha\alpha}(t,t)},
\end{equation}
\begin{align}
  J_i(t)=\frac{2e}{\hbar}{\rm Im} \left(\sum_{k\sigma,\alpha} V_{ik\sigma,\alpha}
  {\rho_{\alpha,ik\sigma}(t)}\right)
   =\frac{2e}{\hbar}{\rm Re}\left(\sum_{k\sigma,\alpha} V_{ik\sigma,\alpha}
  {G^<_{\alpha,ik\sigma}(t,t)}\right), \label{Model-J}
\end{align}
where we define the system-to-lead lesser Green function
\begin{equation}
G^<_{\alpha,ik\sigma}(t_1,t_2)=i\left\langle
c^\dag_{ik\sigma}(t_2)d_\alpha(t_1)\right\rangle,
\end{equation}
while nonequilibrium charge distribution of the molecule is determined by the system
lesser function
\begin{equation}
 G^<_{\alpha\beta}(t_1,t_2)=i\left\langle
 d^{\dag}_{\beta}(t_2)d_{\alpha}(t_1)\right\rangle.
\end{equation}

One can ask: what is the advantage to use the more complex two-time Green functions
instead of density matrices? There are several reasons. First of all, NGF give, as we
shall see below, a clear description of both density of states and distribution of
particles over this states. Then, the equations of motion including interactions and the
influence of environment can be obtained with the help of a diagrammatic technique, and
(very important) all diagrammatic results of {\em equilibrium} theory can be easily
incorporated. Retardation effects are conveniently taken into account by two-time Green
functions. And, ... finally, one can always go back to the density matrix when necessary.

It is important to note, that the {\em single-particle} density matrix
(\ref{Model-DM}) should not be mixed up with the density matrix in the basis of {\em
many-body eigenstates}.

In these review we consider different methods. The density matrix can be determined
from the master equation. For Green functions the EOM method or Keldysh method can
be applied. Traditionally, the density matrix is used in the case of very weak
system-to-lead coupling, while the NGF methods are more successful in the
description of strong and intermediate coupling to the leads. The convenience of one
or other method is determined essentially by the type of interaction. Our aim is to
combine the advantages of both methods.

\newpage
\subsection{Nonequilibrium Green functions: definition and properties}

In the previous section we found, that the current through a system (as well as other
observables) can be expressed through nonequilibrium Green functions. Here we give the
definitions of retarded, advanced, lesser, and greater Green functions and consider some
simple examples. We also introduce a very important concept of the Schwinger-Keldysh
closed-time contour, and define contour Green functions. This section is a little bit
technical, but we need these definitions in the next sections.

\subsubsection{ Spectral - retarded ($G^R$) and advanced ($G^A$) functions}

\paragraph*{(i) Definition}

Retarded Green function for fermions is defined as
\begin{equation}\label{GR}
  G^R_{\alpha\beta}(t_1,t_2)=-i\theta(t_1-t_2)\left\langle\left[c_\alpha(t_1),c_\beta^\dagger(t_2)
  \right]_+\right\rangle,
\end{equation}
where $c^\dagger_\alpha(t)$, $c_\alpha(t)$ are creation and annihilation
time-dependent (Heisenberg) operators, $[c,d]_+=cd+dc$ is the anti-commutator,
$\langle...\rangle$ denotes averaging over equilibrium state.

We use notations $\alpha$, $\beta$, ... to denote single-particle quantum states, the
other possible notation is more convenient for bulk systems
\begin{equation}
  G^R(x_1,x_2)=-i\theta(t_1-t_2)\left\langle\left[c(x_1),c^\dagger(x_2)
  \right]_+\right\rangle,
\end{equation}
where $x\equiv{\bf r},t,\sigma,...$ or $x\equiv{\bf k},t,\sigma,...$, etc. Some other
types of notations can be found in the literature, they are equivalent to (\ref{GR}).

The advanced function for fermions is defined as
\begin{equation}
  G^A_{\alpha\beta}(t_1,t_2)=i\theta(t_2-t_1)\left\langle\left[c_\alpha(t_1),c_\beta^\dagger(t_2)
  \right]_+\right\rangle.
\end{equation}

Finally, retarded and advanced functions for {\em bosons} can be defined as
\begin{equation}
  \tilde{G}^R_{\alpha\beta}(t_1,t_2)=-i\theta(t_1-t_2)\left\langle\left[a_\alpha(t_1),a_\beta^\dagger(t_2)
  \right]_-\right\rangle,
\end{equation}
\begin{equation}
  \tilde{G}^A_{\alpha\beta}(t_1,t_2)=i\theta(t_2-t_1)\left\langle\left[a_\alpha(t_1),a_\beta^\dagger(t_2)
  \right]_-\right\rangle,
\end{equation}
where $a^\dagger_\alpha(t)$, $a_\alpha(t)$ are creation and annihilation boson
operators, $[a,b]_-=ab-ba$ is the commutator.

\paragraph*{(ii) Discussion of averaging}

The average value of any operator $\hat O$ can be written as $\langle\hat
O\rangle=\langle t|\hat O^S|t\rangle$ in the Schr\"odinger representation or $\langle\hat
O\rangle=\langle 0|\hat O^H(t)|0\rangle$ in the Heisenberg representation, where
$|0\rangle$ is some initial state. This initial state is in principle arbitrary, but in
many-particle problems it is convenient to take this state as an equilibrium state,
consequently without time-dependent perturbation we obtain usual equilibrium Green
functions.

In accordance with this definition the Heisenberg operators $c_\alpha(t)$,
$c_\beta^\dagger(t)$, etc. are equal to the time-independent Schr\"odinger operators
at some initial time $t_0$: $c_\alpha(t_0)=c_\alpha$, etc. Density matrix of the
system is assumed to be equilibrium at this time $\hat\rho(t_0)=\hat\rho_{eq}$.
Usually we can take $t_0=0$ for simplicity, but if we want to use $t_0\neq 0$ the
transformation to Heisenberg operators should be written as
\begin{equation}\label{H t0}
  \hat f^H(t)=e^{i\hat H(t-t_0)}\hat f^S e^{-i\hat H(t-t_0)}.
\end{equation}

In fact, the initial conditions are not important because of dissipation (the memory
about the initial state is completely lost after the relaxation time). However, in
some pathological cases, for example for free noninteracting particles, the initial
state determines the state at all times. Note also, that the initial conditions can be
more convenient formulated for Green functions itself, instead of corresponding
initial conditions for operators or wave functions.

Nevertheless, thermal averaging is widely used and we define it here explicitly. If we
introduce the basis of exact time-independent many-particle states $|n\rangle$ with
energies $E_n$, the averaging over equilibrium state can be written as
\begin{equation}\label{Z}
  \langle\hat O\rangle=\frac{1}{Z}\sum_n e^{-E_n/T}\left\langle n\left|\hat
  O^H(t)\right|n\right\rangle, \ \ \ Z=\sum_n e^{-E_n/T}.
\end{equation}

In the following when we use notations like $\left\langle \hat O\right\rangle$ or
$\left\langle \Psi\left|\hat O(t)\right|\Psi\right\rangle$, we assume the averaging
with density matrix (density operator) $\hat\rho$
\begin{equation}
\left\langle \hat O\right\rangle=Sp\left(\hat\rho\hat O\right),
\end{equation}
for equilibrium density matrix and Heisenberg operators it is equivalent to (\ref{Z}).

\paragraph*{(iii) Free-particle retarded function for fermions}

Now consider the simplest possible example -- retarded Green function for free
particles (fermions).

The free-particle Hamiltonian has equivalent form if one uses Schr\"odinger or
Heisenberg operators
\begin{equation}
  \hat H=\sum_\alpha\epsilon_\alpha c^\dagger_\alpha c_\alpha=
  \sum_\alpha\epsilon_\alpha c^\dagger_\alpha(t) c_\alpha(t),
\end{equation}
because (here we assume $t_0=0$)
\begin{align}
  c^\dagger_\alpha(t) c_\alpha(t)=e^{i\hat Ht} & c^\dagger_\alpha e^{-i\hat Ht}
  e^{i\hat Ht}c_\alpha e^{-i\hat Ht}
  \nonumber \\
& =e^{i\hat Ht}c^\dagger_\alpha c_\alpha e^{-i\hat Ht}=c^\dagger_\alpha c_\alpha,
\end{align}
where we used that $c^\dagger_\alpha c_\alpha$ is commutative with the Hamiltonian
$\hat H=\sum_\alpha\epsilon_\alpha c^\dagger_\alpha c_\alpha$.

From the definitions (\ref{GR}) and (\ref{Z})
%
\begin{align}
 & \left\langle\left[c_\alpha(t_1),c_\beta^\dagger(t_2)\right]_+\right\rangle=
  \left\langle c_\alpha(t_1)c_\beta^\dagger(t_2)+c_\beta^\dagger(t_2)c_\alpha(t_1)\right\rangle=
  \nonumber \\
 & =\left\langle
  e^{i\hat Ht_1}c_\alpha(t_1)e^{-i\hat Ht_1}e^{i\hat Ht_2}c_\beta^\dagger(t_2)e^{-i\hat Ht_2}+
  e^{i\hat Ht_2}c_\beta^\dagger(t_2)e^{-i\hat Ht_2}e^{i\hat Ht_1}c_\alpha(t_1)e^{-i\hat Ht_1}
  \right\rangle= \nonumber \\
 & =e^{i\epsilon_\beta t_2-i\epsilon_\alpha t_1}\left\langle c_\alpha c_\beta^\dagger+
  c_\beta^\dagger c_\alpha\right\rangle=e^{-i\epsilon_\alpha(t_1-t_2)}\delta_{\alpha\beta},
\end{align}
%
%
\begin{align}\label{GR free}
& G^R_{\alpha\beta}(t_1,t_2)=-i\theta(t_1-t_2)
  \left\langle\left[c_\alpha(t_1),c_\beta^\dagger(t_2)\right]_+\right\rangle \nonumber \\
& = -i\theta(t_1-t_2)e^{-i\epsilon_\alpha(t_1-t_2)}\delta_{\alpha\beta},
\end{align}
where we used some obvious properties of the creation and annihilation operators and
commutation relations.

We consider also the other method, based on the equations of motion for operators.
From Liuville -- von Neuman equation we find (all $c$-operators are Heisenberg
operators in the formula below, $(t)$ is omitted for shortness)

\begin{align}
&
i\frac{dc_\alpha(t)}{dt}\!=\!\left[c_\alpha(t),H\right]_-\!=\!\sum_\beta\epsilon_\beta\left[c_\alpha,
c^\dag_\beta c_\beta\right]_-\! \nonumber
\\
& =\!\sum_\beta\epsilon_\beta\left(c_\alpha c^\dag_\beta c_\beta-c^\dag_\beta c_\beta
c_\alpha\right)\!=\!\sum_\beta\epsilon_\beta\left(c_\alpha c^\dag_\beta c_\beta
+c^\dag_\beta c_\alpha c_\beta\right)\! \nonumber
\\
& =\!\sum_\beta\epsilon_\beta\left(c_\alpha c^\dag_\beta +c^\dag_\beta
c_\alpha\right)c_\beta\!=\!\sum_\beta\epsilon_\beta\delta_{\alpha\beta}c_\beta\!=\!\epsilon_\alpha
c_\alpha(t),
\end{align}
so that Heisenberg operators for free fermions are
\begin{equation}
  c_\alpha(t)=e^{-i\epsilon_\alpha t}c_\alpha(0),\ \ \
  c^\dagger_\alpha(t)=e^{i\epsilon_\alpha t}c^\dagger_\alpha(0).
\end{equation}

Substituting these expressions into (\ref{GR}) we obtain again (\ref{GR free}). Note
also that if we take $t_0\neq 0$, then Heisenberg operators for free fermions are
\begin{equation}
  c_\alpha(t)=e^{-i\epsilon_\alpha (t-t_0)}c_\alpha(t_0),\ \ \
  c^\dagger_\alpha(t)=e^{i\epsilon_\alpha (t-t_0)}c^\dagger_\alpha(t_0),
\end{equation}
but the result for the Green functions is just the same, because
\begin{align}
&  \left\langle\left[c_\alpha(t_1),c_\beta^\dagger(t_2)\right]_+\right\rangle=
  \left\langle c_\alpha(t_1)c_\beta^\dagger(t_2)+
  c_\beta^\dagger(t_2)c_\alpha(t_1)\right\rangle= \nonumber \\
& =e^{i\epsilon_\beta (t_2-t_0)-i\epsilon_\alpha (t_1-t_0)}\left\langle c_\alpha
  c_\beta^\dagger+c_\beta^\dagger
  c_\alpha\right\rangle=e^{-i\epsilon_\alpha(t_1-t_2)}\delta_{\alpha\beta}.
\end{align}

It is interesting to make Fourie-transform of this function. In equilibrium two-time
function $G^R_{\alpha\beta}(t_1,t_2)$ is a function of the time difference only, so
that we define transform over time difference $(t_1-t_2)$
\begin{equation}\label{fourie GR}
  G^R(\epsilon)=\int_0^{\infty}G^R(t_1-t_2)e^{i(\epsilon+i0)(t_1-t_2)}d(t_1-t_2),
\end{equation}
we add infinitely small positive complex part to $\epsilon$ to make this integral well
defined in the upper limit (this is necessary for free particles without dissipation
because function (\ref{GR free}) oscillates at large times $\tau=t_1-t_2$ and the
integral (\ref{fourie GR}) can not be calculated without $i0$ term. Then we obtain
\begin{equation}
  G^R_{\alpha\beta}(\epsilon)=\frac{\delta_{\alpha\beta}}{\epsilon-\epsilon_\alpha+i0}.
\end{equation}

More generally, transformation (\ref{fourie GR}) can be considered as the Laplas
transformation with complex argument $z=\epsilon+i\eta$.

For advanced function
\begin{equation}\label{GA free}
  G^A_{\alpha\beta}(t_1,t_2)=i\theta(t_2-t_1)e^{-i\epsilon_\alpha(t_1-t_2)}\delta_{\alpha\beta},
\end{equation}
the Fourier transform is given by
\begin{equation}\label{fourie GA}
  G^A(\epsilon)=\int_{-\infty}^0G^A(t_1-t_2)e^{i(\epsilon-i0)(t_1-t_2)}d(t_1-t_2),
\end{equation}
with other sign of the term $i0$.

\paragraph*{(iv) Spectral function}

Finally, we introduce the important combination of retarded and advanced functions
known as {\em spectral} or {\em spectral weight} function
\begin{equation}
  A_{\alpha\beta}(\epsilon)=
  i\left(G^R_{\alpha\beta}(\epsilon)-G^A_{\alpha\beta}(\epsilon)\right),
\end{equation}
in equilibrium case Fourie-transformed retarded and advanced functions are complex
conjugate $G^A(\epsilon)=\left(G^R(\epsilon)\right)^*$, and
$A_{\alpha\beta}(\epsilon)=-2{\rm Im}G^R_{\alpha\beta}(\epsilon)$.

For free fermions the spectral function is
\begin{equation}
  A_{\alpha\beta}(\epsilon)=
  -2{\rm Im}\left(\frac{\delta_{\alpha\beta}}{\epsilon-\epsilon_\alpha+i0}\right)=
  2\pi\delta(\epsilon-\epsilon_\alpha)\delta_{\alpha\beta}.
\end{equation}
The result is transparent -- the function $A_{\alpha\beta}(\epsilon)$ is nonzero only
at particle eigen-energies, so that
\begin{equation}
  \rho(\epsilon)=\frac{1}{2\pi}{\rm Sp}A_{\alpha\beta}(\epsilon)=
  \frac{1}{2\pi}\sum_\alpha A_{\alpha\alpha}(\epsilon)=
  \sum_\alpha\delta(\epsilon-\epsilon_\alpha)
\end{equation}
is the usual energy density of states. Note that the imaginary part $i0$ is necessary
to obtain this result, thus it is not only mathematical trick, but reflects the
physical sense of the retarded Green function.

If we introduce finite relaxation time
\begin{equation}
  G^R_{\alpha\beta}(\tau)=-i\theta(\tau)e^{-i\epsilon_\alpha\tau-\gamma\tau}
  \delta_{\alpha\beta},
\end{equation}
then the spectral function has familiar Lorentzian form
\begin{equation}
  A_{\alpha\beta}(\epsilon)=\frac{2\gamma\delta_{\alpha\beta}}
  {(\epsilon-\epsilon_\alpha)^2+\gamma^2}.
\end{equation}

Finally, spectral function has a special property, so-called {\em sum rule}, namely
\begin{equation}
  \int_{-\infty}^\infty A_{\alpha\beta}(\epsilon)\frac{d\epsilon}{2\pi}=
  \delta_{\alpha\beta}.
\end{equation}

\subsubsection{Kinetic - lesser ($G^<$) and greater ($G^>$) functions}

\paragraph*{(i) Definition}

Spectral functions, described before, determine single-particle properties of the
system, such as quasiparticle energy, broadening of the levels (life-time), and
density of states. These functions can be modified in nonequilibrium state, but most
important {\em kinetic} properties, such as distribution function, charge, and
current, are determined by lesser Green function
\begin{equation}
  G^<_{\alpha\beta}(t_1,t_2)=i\left\langle
  c_\beta^\dagger(t_2)c_\alpha(t_1)\right\rangle.
\end{equation}

Indeed, density matrix is the same as equal-time lesser function
\begin{equation}
  \rho_{\alpha\beta}(t)=
  \left\langle c_\beta^\dagger(t)c_\alpha(t)\right\rangle=-iG^<_{\alpha\beta}(t,t).
\end{equation}
the number of particles in state $|\alpha\rangle$ (distribution function) is
\begin{equation}
  n_\alpha(t)=
  \left\langle c_\alpha^\dagger(t)c_\alpha(t)\right\rangle=-iG^<_{\alpha\alpha}(t,t),
\end{equation}
the tunneling current is
\begin{align}
  J(t)=\frac{ie}{\hbar}\sum_{kq} &
  \left[V_{qk}\left\langle c^{\dag}_{q}(t)c_{k}(t) \right\rangle-
  V^*_{qk}\left\langle c^{\dag}_{k}(t)c_{q}(t) \right\rangle
  \right] \nonumber \\
& = \frac{2e}{\hbar}{\rm Re}\left(\sum_{kq} V_{qk}{G^<_{kq}(t,t)}\right).
\end{align}

In addition to the lesser the other (greater) function is used
\begin{equation}
  G^>_{\alpha\beta}(t_1,t_2)=-i\left\langle
  c_\alpha(t_1)c_\beta^\dagger(t_2)\right\rangle.
\end{equation}

For bosons lesser and greater functions are defined as
\begin{equation}
  \tilde G^<_{\alpha\beta}(t_1,t_2)=-i\left\langle a_\beta^\dagger(t_2)a_\alpha(t_1)\right\rangle,
\end{equation}
\begin{equation}
  \tilde G^>_{\alpha\beta}(t_1,t_2)=-i\left\langle
  a_\alpha(t_1)a_\beta^\dagger(t_2)\right\rangle.
\end{equation}

The name "lesser" originates from the time-ordered Green function, the main function
in equilibrium theory, which can be calculated by diagrammatic technique
\begin{equation}
  G_{\alpha\beta}(t_1,t_2)=-i\left\langle T\left(
  c_\alpha(t_1)c_\beta^\dagger(t_2)\right)\right\rangle,
\end{equation}
%
\begin{equation}
G_{\alpha\beta}(t_1,t_2)=\left\{\begin{array}{lll} -i\left\langle
  c_\alpha(t_1)c_\beta^\dagger(t_2)\right\rangle & {\rm if\ \ \ } t_1>t_2 &
{\Rightarrow\ \ G_{\alpha\beta}\equiv G^>_{\alpha\beta}}, \\[0.4cm] i\left\langle
  c_\beta^\dagger(t_2)c_\alpha(t_1)\right\rangle & {\rm if\ \ \ } t_1<t_2 & {\Rightarrow\ \
G_{\alpha\beta}\equiv G^<_{\alpha\beta}},
\end{array}\right.
\end{equation}
%
here additional sing minus appears for interchanging of fermionic
creation-annihilation operators. Lesser means that $t_1<t_2$.

From the definitions it is clear that the retarded function can be combined from
lesser and greater functions
\begin{equation}\label{GR through lesser}
  G^R_{\alpha\beta}(t_1,t_2)=\theta(t_1-t_2)\left[G^>_{\alpha\beta}(t_1,t_2)-
  G^<_{\alpha\beta}(t_1,t_2)\right].
\end{equation}

\paragraph*{(ii) Free-particle lesser function for fermions}

Now let us consider again free fermions. Heisenberg operators for free fermions are
($t_0=0$)
\begin{equation}
  c_\alpha(t)=e^{-i\epsilon_\alpha t}c_\alpha(0),\ \ \
  c^\dagger_\alpha(t)=e^{i\epsilon_\alpha t}c^\dagger_\alpha(0).
\end{equation}

Lesser function is
\begin{align}\label{lesser free}
   G^<_{\alpha\beta}(t_1,t_2)= & i\left\langle
   c_\beta^\dagger(t_2)c_\alpha(t_1)\right\rangle
  =ie^{i\epsilon_\beta t_2-i\epsilon_\alpha t_1}\left\langle c_\beta^\dagger c_\alpha\right\rangle
  \nonumber \\
& =ie^{-i\epsilon_\alpha(t_1-t_2)}f^0(\epsilon_\alpha)\delta_{\alpha\beta},
\end{align}
one sees that contrary to the retarded function, the lesser function is proportional
to the distribution function, in equilibrium this is Fermi distribution function
\begin{equation}
f^0(\epsilon)=\frac{1}{e^{\frac{\epsilon-\mu}{T}}+1}.
\end{equation}

It is interesting to compare this answer with the result for {\em nonthermal} initial
conditions. Assume that initial state is described by the density matrix
$\rho^0_{\alpha\beta}=\left\langle c_\beta^\dagger c_\alpha\right\rangle$, now with
nonzero off-diagonal elements. Time dependence of the density matrix is given by
\begin{equation}
  \rho_{\alpha\beta}(t)=e^{i(\epsilon_\beta-\epsilon_\alpha)t}\rho^0_{\alpha\beta}.
\end{equation}
We obtain the well known result that off-diagonal elements oscillate in time.

Now define Fourier-transform for lesser function ($\tau=t_1-t_2$)
\begin{equation}\label{fourie lesser}
  G^<(\epsilon)=\int_{-\infty}^{\infty}G^<(\tau)e^{i\left[\epsilon+i0{\rm sign}(\tau)\right]\tau}d\tau,
\end{equation}
note that here we use Fourie-transform with complicated term $i0{\rm sign}(\tau)$,
which makes this transformation consistent with previously introduced transformations
(\ref{fourie GR}) for retarded ($\tau>0$) and (\ref{fourie GA}) advanced ($\tau<0$)
functions.

Applying this transformation to (\ref{lesser free}) we obtain
\begin{align}
  G^<_{\alpha\beta}(\epsilon)= & if^0(\epsilon_\alpha)\delta_{\alpha\beta}\int_{-\infty}^{\infty}
  e^{+i\left[\epsilon-\epsilon_\alpha+i0{\rm sign}(\tau)\right]\tau}d\tau \nonumber \\
& =2\pi if^0(\epsilon_\alpha)\delta(\epsilon-\epsilon_\alpha)\delta_{\alpha\beta}.
\end{align}

For free fermion greater function one obtaines
\begin{equation}
  G^>_{\alpha\beta}(t_1,t_2)=
  -ie^{-i\epsilon_\alpha(t_1-t_2)}(1-f^0(\epsilon_\alpha))\delta_{\alpha\beta},
\end{equation}
\begin{equation}
  G^>_{\alpha\beta}(\epsilon)=
  -2\pi i(1-f^0(\epsilon_\alpha))\delta(\epsilon-\epsilon_\alpha)\delta_{\alpha\beta}.
\end{equation}

\paragraph*{(iii) Equilibrium case. Fluctuation-dissipation theorem}

Now we want to consider some general properties of interacting systems. In equilibrium
the lesser function is not independent and is simply related to the spectral function
by the relation
\begin{equation}\label{fdt}
  G^<_{\alpha\beta}(\epsilon)=iA_{\alpha\beta}(\epsilon)f^0(\epsilon).
\end{equation}
This relation is important because establish equilibrium initial condition for
nonequilibrium lesser function, and propose useful Ansatz if equilibrium distribution
function $f^0(\epsilon)$ is replaced by some unknown nonequilibrium function.

Here we prove this relation using {\em Lehmann representation} -- quite useful method
in the theory of Green functions. The idea of the method is to use exact many-particle
eigenstates $|n\rangle$, even if they are not explicitly known.

Consider first the greater function. Using states $|n\rangle$ we represent this
function as
%
\begin{eqnarray}
 & \displaystyle
  G^>_{\alpha\beta}(t_1,t_2)=-i\left\langle c_\alpha(t_1)c_\beta^\dagger(t_2)\right\rangle=
  -\frac{i}{Z}\sum_n\left\langle n\left|e^{-{\hat H}/{T}}c_\alpha(t_1)c_\beta^\dagger(t_2)
  \right|n\right\rangle= \nonumber \\
  & \displaystyle =-\frac{i}{Z}\sum_{nm}e^{-{E_n}/{T}}\langle n|c_\alpha|m\rangle
  \langle m|c_\beta^\dagger|n\rangle e^{i(E_n-E_m)(t_1-t_2)}.
\end{eqnarray}

In Fourie representation
\begin{equation}
  G^>_{\alpha\beta}(\epsilon)=-\frac{2\pi i}{Z}\sum_{nm}e^{-{E_n}/{T}}\langle n|c_\alpha|m\rangle
  \langle m|c_\beta^\dagger|n\rangle \delta(E_n-E_m+\epsilon).
\end{equation}

Similarly, for the lesser function we find
\begin{equation}\label{lehmann 2}
  G^<_{\alpha\beta}(\epsilon)=\frac{2\pi i}{Z}\sum_{nm}e^{-{E_m}/{T}}\langle n|c_\beta^\dagger|m\rangle
  \langle m|c_\alpha|n\rangle \delta(E_m-E_n+\epsilon).
\end{equation}

Now we can use these expressions to obtain some general properties of Green functions
without explicit calculation of the matrix elements. Exchanging indices $n$ and $m$ in
the expression (\ref{lehmann 2})  and taking into account that $E_m=E_n-\epsilon$
because of delta-function, we see that
\begin{equation}
  G^>_{\alpha\beta}(\epsilon)=-e^{-\epsilon/T}G^<_{\alpha\beta}(\epsilon).
\end{equation}

From this expression and relation (\ref{GR through lesser}), which can be written as
\begin{equation}\label{A through lesser}
  A_{\alpha\beta}(\epsilon)=i\left[G^>_{\alpha\beta}(\epsilon)-
  G^<_{\alpha\beta}(\epsilon)\right]
\end{equation}
we derive (\ref{fdt}).

\subsubsection{Interaction representation}

In the previous lectures we found that nonequilibrium Green functions can be quite
easy calculated for free particles, and equations of motion for one-particle Green
functions (the functions which are the averages of two creation-annihilation
operators) can be formulated if we add interactions and time-dependent perturbations,
but these equations include high-order Green functions (the averages of three, four,
and larger number of operators). The equations can be truncated and formulated in
terms of one-particle Green functions in some simple approximations. However,
systematic approach is needed to proceed with perturbation expansion and
self-consistent methods (all together is known as {\em diagrammatic approach}). The
main idea of the diagrammatic approach is to start from some "simple" Hamiltonian
(usually for free particles) and, treating interactions and external fields as a
perturbation, formulate perturbation expansion, and summarize all most important terms
(diagrams) {\em in all orders of perturbation theory}. The result of such procedure
gives, in principle, {\em nonperturbative} description (ordinary mean-field theory is
the simplest example). The starting point of the method is so-called {\em interaction
representation}.

Let us consider the full Hamiltonian $\hat H$ as the sum of a {\em free-particle}
time-independent part $\hat H_0$ and (possibly time-dependent) perturbation $\hat V(t)$
(note that this "perturbation" should not be necessarily small)
\begin{equation}
  \hat H=\hat H_0+\hat V(t).
\end{equation}

We define new operators in {\em interaction representation} by
\begin{equation}\label{f^I}
  \hat f^I(t)=e^{i\hat H_0t}\hat f^Se^{-i\hat H_0t},
\end{equation}
where $\hat f^S$ is the time-independent Schr\"odinger operator. This is equivalent
to the time-dependent Heisenberg operator, defined by the part $\hat H_0$ of the
Hamiltonian. For a free-particle Hamiltonian $\hat H_0$ the operators $\hat f^I(t)$
can be calculated exactly.

A new wave function corresponding to (\ref{f^I}) is
\begin{equation}\label{Psi^I}
  \Psi^I(t)=e^{i\hat H_0t}\Psi^S(t).
\end{equation}
It is easy to see that transformation (\ref{f^I}), (\ref{Psi^I}) is unitary
transformation and conserves the average value of any operator
\begin{equation}
  \langle\Psi^S|\hat f^S|\Psi^S\rangle=\langle\Psi^I|\hat f^I|\Psi^I\rangle.
\end{equation}

Substituting (\ref{Psi^I}) into the ordinary Schr\"odinger equation, we derive the
equation
\begin{equation}\label{eq I}
  i\frac{\partial\Psi^I}{\partial t}=\hat V^I(t)\Psi^I,
\end{equation}
where $\hat V^I(t)=e^{i\hat H_0t}\hat V^S(t)e^{-i\hat H_0t}$ is in the intreraction
representation.

Equation (\ref{eq I}) seems to be quite simple, however the operator nature of $\hat
V$ makes this problem nontrivial. Indeed, consider a small time-step $\Delta t$. Then
\begin{equation}
  \Psi(t+\Delta t)=\left[1-i\hat V^S(t)\Delta t\right]\Psi(t)=
  \exp^{-i\hat V^S(t)\Delta t}\Psi(t),
\end{equation}
linear in $\Delta t$ term can be transformed into the exponent if we understand the
exponential function of the operator in the usual way
\begin{equation}\label{exp_exp}
  \exp^{\hat A}=1+\hat A +\frac{1}{2!}\hat A^2+...+\frac{1}{n!}\hat A^n+... ,
\end{equation}
and assume that only linear term should be taken at $\Delta t\rightarrow 0$.

If we now repeat this procedure at times $t_i$ with step $\Delta t$, we obtain finally
\begin{equation}\label{Psi S}
  \Psi^I(t)=\hat S(t,t_0)\Psi^I(t_0),
\end{equation}
with
\begin{equation}\label{S prod}
  \hat S(t,t_0)=\prod_{t_i=t_0}^{t}\exp\left(-i\hat V^I(t_i)\Delta t\right),
\end{equation}
this product, however, is not simply $\displaystyle\exp\left(-i\int_{t_0}^{t}\hat
V^I(t')dt'\right)$ in the limit $\Delta t\rightarrow 0$, because operators $\hat
V^I(t')$ are not commutative at different times, and for two noncommutative operators
$\hat A$ and $\hat B$ $e^{\hat A+\hat B}\neq e^{\hat A}e^{\hat B}$.

In the product (\ref{S prod}) operators at earlier times should be applied first,
before operators at later times. In the limit $\Delta t\rightarrow 0$ we obtain
\begin{equation}\label{S int}
  \hat S(t,t_0)=T\exp\left(-i\int_{t_0}^{t}\hat V^I(t')dt'\right),
\end{equation}
where $T$ is the time-ordering operator ("-" for fermionic operators)
\begin{equation}
  T\left(\hat A(t_1)\hat B(t_2)\right)=\left\{\begin{array}{ll}
  \hat A(t_1)\hat B(t_2) & {\rm if\ \ \ } t_1>t_2, \\[0.4cm]
  \pm\hat B(t_2)\hat A(t_1) & {\rm if\ \ \ } t_1<t_2.
\end{array}\right.
\end{equation}

Of cause, expression (\ref{S int}) is defined only in the sense of expansion
(\ref{exp_exp}). Consider for example the second-order term in the time-ordered
expansion.
%
\begin{align}
  T\left[\int_{t_0}^{t}\hat V^I(t')dt'\right]^2 & =
  T\left[\int_{t_0}^{t}\hat V^I(t')dt'\int_{t_0}^{t}\hat V^I(t'')dt''\right]= \nonumber \\
  & =\int_{t_0}^{t}dt'\int_{t_0}^{t'}dt''\hat V^I(t')\hat V^I(t'')+
  \int_{t_0}^{t}dt''\int_{t_0}^{t''}dt'\hat V^I(t'')\hat V^I(t').
\end{align}
%
If we exchange $t'$ and $t''$ in the second integral, we see finally that
\begin{equation}
  T\left[\int_{t_0}^{t}\hat V^I(t')dt'\right]^2=
  2\int_{t_0}^{t}dt'\int_{t_0}^{t'}dt''\hat V^I(t')\hat V^I(t'').
\end{equation}

\paragraph*{(i) Properties of $\hat S(t,t_0)$}

$\hat S$ is the unitary operator and
\begin{equation}
  \hat S^{-1}(t,t_0)=\hat S^\dagger(t,t_0)=
  \tilde T\exp\left(i\int_{t_0}^{t}\hat V^I(t')dt'\right),
\end{equation}
where $\tilde T$ is time-anti-ordering operator. Some other important properties are
\begin{eqnarray}
  & \hat S^{-1}(t,t_0)=\hat S(t_0,t), \\[0.2cm]
  & \hat S(t_3,t_2)\hat S(t_2,t_1)=\hat S(t_3,t_1), \\[0.2cm]
  & \hat S^{-1}(t_2,t_1)\hat S^{-1}(t_3,t_2)=\hat S^{-1}(t_3,t_1).
\end{eqnarray}

Finally, we need the expression of a Heisenberg operator, defined by the full
Hamiltonian \mbox{$\hat H=\hat H_0+\hat V(t)$}, through an operator in the interaction
representation. The transformation, corresponding to (\ref{Psi S}), is given by
\begin{equation}\label{f S}
  \hat f^H(t)=e^{-i\hat H_0t_0}\hat S^{-1}(t,t_0)\hat f^I(t)\hat S(t,t_0)e^{i\hat
  H_0t_0},
\end{equation}
and the state $\Psi^I(t_0)$ is related to the Heisenberg time-independent wave
function by
\begin{equation}
\Psi^I(t_0)\equiv e^{i\hat H_0t_0}\Psi^S(t_0)=e^{i\hat H_0t_0}\Psi^H,
\end{equation}
in accordance with our previous discussion of averaging we assume that at time
\mbox{$t=t_0$} Heisenberg operators coincide with time-independent Schr\"odinger
operators \mbox{$\hat f^H(t_0)=\hat f^S$}, and Schr\"odinger wave function coincides
at the same time with Heisenberg time-independent wave function
\mbox{$\Psi^S(t_0)=\Psi^H$}. To avoid these additional exponents in (\ref{f S}) we
can redefine the transformation to the interaction representation as
\begin{equation}
  \hat f^I(t)=e^{i\hat H_0(t-t_0)}\hat f^Se^{-i\hat H_0(t-t_0)},
\end{equation}
in accordance with the transformation (\ref{H t0}) for time-independent Hamiltonian.
Previously we showed that free-particle Green functions are not dependent on $t_0$
for equilibrium initial condition, if we want to consider some nontrivial initial
conditions, it is easier to formulate these conditions directly for Green functions.
Thus below we shall use relations
\begin{equation}\label{f S2}
  \hat f^H(t)=\hat S^{-1}(t,t_0)\hat f^I(t)\hat S(t,t_0),
\end{equation}
and
\begin{equation}
\Psi^I(t_0)\equiv \Psi^S(t_0)=\Psi^H.
\end{equation}

\paragraph*{(ii) Green functions in the interaction representation}

Consider, for example, the lesser function
\begin{equation}
  G^<_{\alpha\beta}(t_1,t_2)=i\left\langle
  c_\beta^\dagger(t_2)c_\alpha(t_1)\right\rangle=i\left\langle\Psi^H\left|
  c_\beta^\dagger(t_2)c_\alpha(t_1)\right|\Psi^H\right\rangle,
\end{equation}
$c$-operators here are Heisenberg operators and they should be replaced by operators
$c^I(t)\equiv\tilde c(t)$ in the interaction representation:
%
\begin{equation}
  G^<_{\alpha\beta}(t_1,t_2)=i\left\langle\Psi^H\left|
  \hat S^{-1}(t_2,t_0)\tilde c_\beta^\dagger(t_2)
  \hat S(t_2,t_0)\hat S^{-1}(t_1,t_0)\tilde c_\alpha(t_1)
  \hat S(t_1,t_0)\right|\Psi^H\right\rangle.
\end{equation}
%
Using properties of $\hat S$ operators, we rewrite this expression as
\begin{equation}\label{lesserTC}
  G^<_{\alpha\beta}(t_1,t_2)=i\left\langle \hat S(t_0,t_2)\tilde c_\beta^\dagger(t_2)
  \hat S(t_2,t_1)\tilde c_\alpha(t_1)\hat S(t_1,t_0)\right\rangle.
\end{equation}

\subsubsection{Schwinger-Keldysh time contour and contour functions}

\paragraph*{(i) Closed time-path integration}

Now let us introduce one useful trick, so-called {\em closed time-path contour of
integration}. First, note that the expression of the type
\begin{align}
  \hat f^H(t)=\hat S^{-1}(t,t_0)\hat f^I(t)\hat S(t,t_0) 
  =\tilde Te^{i\int_{t_0}^{t}\hat V^I(t')dt'}\hat f^I(t)
  Te^{-i\int_{t_0}^{t}\hat V^I(t')dt'},
\end{align}
can be written as
\begin{equation}
  \hat f^H(t)=T_{C_t}\exp\left(-i\int_{C_t}\hat V^I(t')dt'\right)\hat f^I(t),
\end{equation}
where the integral is taken along closed time contour from $t_0$ to $t$ and then back
from $t$ to $t_0$
\begin{equation}
  \int_{C_t}dt'=\int_{t_0}^{t}dt'+\int_{t}^{t_0}dt',
\end{equation}
contour time-ordering operator $T_{C_t}$ works along the contour $C_t$, it means that
for times $t^\rightarrow$ it is usual time-ordering operator $T$, and for times
$t^\leftarrow$ it is anti-time-ordering operator $\tilde T$. Symbolically
\begin{equation}
  T_{C_t}\int_{C_t}dt'=T\int_\rightarrow dt'+\tilde T\int_\leftarrow dt'.
\end{equation}

Consider now the application of this closed time-path contour to calculation of Green
functions. It is convenient to start from the time-ordered function at $t_2>t_1$
\begin{equation}\label{TC}
  \left\langle T\left(\hat B(t_2)\hat A(t_1)\right)\right\rangle=
  \left\langle \hat S(t_0,t_2)\tilde B(t_2)\hat S(t_2,t_1)\tilde A(t_1)
  \hat S(t_1,t_0)\right\rangle,
\end{equation}
here $\hat A(t)$ and $\hat B(t)$ are Heisenberg operators, $\tilde A(t)$ and $\tilde
B(t)$ are operators in the interaction representation, in the case of fermionic
operators the additional minus should be added for any permutation of two operators.

Using the properties of the $\hat S$-operator, we transform this expression as
%
\begin{align}
  \left\langle \hat S(t_0,t_2)\tilde B(t_2)\hat S(t_2,t_1)\tilde A(t_1)
  \hat S(t_1,t_0)\right\rangle=
  \left\langle \hat S^{-1}(t_2,t_0)\tilde B(t_2)\hat S(t_2,t_1)\tilde A(t_1)
  \hat S(t_1,t_0)\right\rangle= \nonumber \\
  =\left\langle \hat S^{-1}(\infty,t_0)\hat S(\infty,t_2)\tilde B(t_2)\hat S(t_2,t_1)\tilde A(t_1)
  \hat S(t_1,t_0)\right\rangle=
  \left\langle \hat S^{-1} T\left(\tilde B(t_2)\tilde A(t_1)\hat
  S\right)\right\rangle,
\end{align}
%
where we defined operator
\begin{equation}
  \hat S=\hat S(\infty,t_0).
\end{equation}

Using contour integration, it can be written as
\begin{equation}\label{TCC}
  \left\langle T\left(\hat B(t_2)\hat A(t_1)\right)\right\rangle=
  \left\langle T_C\left(\hat S_C\tilde B(t^\rightarrow_2)\tilde
  A(t^\rightarrow_1)\right)\right\rangle,
\end{equation}

\begin{equation}
  \hat S_C=T_C\exp\left(-i\int_C\hat V^I(t')dt'\right),
\end{equation}
contour $C$ goes from $t_0$ trough $t_1$ and $t_2$, and back to $t_0$. If $t_2>t_1$ it
is obvious that contour ordering along $C^\rightarrow$ gives the terms from $\hat
S(t_1,t_0)$ to $\hat B(t_2)$ in (\ref{TC}). The integral over the back path
$C^\leftarrow$ gives
\begin{align}
& T_C\exp\left(-i\int_\leftarrow\hat V^I(t')dt'\right)=
  \tilde T\exp\left(-i\int_{t_2}^{t_0}\hat V^I(t')dt'\right)= \nonumber \\
&  =\tilde T\exp\left(i\int_{t_0}^{t_2}\hat V^I(t')dt'\right)=
  \hat S^{-1}(t_2,t_0)=\hat S(t_0,t_2).
\end{align}

For $t_2<t_1$ the operators in (\ref{TC}) are reordered by $T$-operator and we again
obtain (\ref{TCC}).

The lesser and greater functions are not time-ordered and arguments of the operators
are not affected by time-ordering operator. Nevertheless we can write such functions
in the same form (\ref{TCC}). The trick is to use one time argument from the forward
contour and the other from the backward contour, for example
\begin{equation}\label{LCC}
  \left\langle \hat B(t_2)\hat A(t_1)\right\rangle=
  \left\langle T_C\left(\hat S_C\tilde B(t^\leftarrow_2)\tilde
  A(t^\rightarrow_1)\right)\right\rangle,
\end{equation}
here the time $t_1$ is always before $t_2$.

\paragraph*{(ii) Contour (contour-ordered) Green function}

Now we are able to define {\em contour} or {\em contour-ordered} Green function -- the
useful tool of Keldysh diagrammatic technique. The definition is similar to the
previous one
\begin{equation}
  G^C_{\alpha\beta}(\tau_1,\tau_2)=-i\left\langle T_C\left(
  c_\alpha(\tau_1)c_\beta^\dagger(\tau_2)\right)\right\rangle,
\end{equation}
where, however, $\tau_1$ and $\tau_2$ are contour times. This function includes all
nonequilibrium Green functions introduced before. Indeed, depending on contour
position of times we obtain lesser, greater, or time-ordered functions (below we give
different notations used in the literature)
%
\begin{equation}
G^C_{\alpha\beta}(\tau_1,\tau_2)=\left\{\begin{array}{lllll}
  \tau_1,\tau_2\in C^\rightarrow:
  & -i\left\langle T c_\alpha(t_1)c_\beta^\dagger(t_2)\right\rangle &
  \Longrightarrow\ \ G^{--} & or & G^T(t_1,t_2), \\[0.4cm]
  \tau_1\in C^\leftarrow, \tau_2\in C^\rightarrow:
  & -i\left\langle c_\alpha(t_1)c_\beta^\dagger(t_2)\right\rangle &
  \Longrightarrow\ \ G^{+-} & or & G^>(t_1,t_2), \\[0.4cm]
  \tau_1\in C^\rightarrow, \tau_2\in C^\leftarrow:
  & i\left\langle c_\beta^\dagger(t_2)c_\alpha(t_1)\right\rangle &
  \Longrightarrow\ \ G^{-+} & or & G^<(t_1,t_2), \\[0.4cm]
  \tau_1,\tau_2\in C^\leftarrow:
  & -i\left\langle\tilde T c_\alpha(t_1)c_\beta^\dagger(t_2)\right\rangle &
  \Longrightarrow\ \ G^{++} & or & G^{\tilde T}(t_1,t_2).
\end{array}\right.
\end{equation}

These four functions are not independent, from definitions it follows that
\begin{equation}
  G^<+G^>=G^{T}+G^{\tilde T},
\end{equation}
and anti-hermitian relations
\begin{eqnarray}
G^{T}_{\alpha\beta}(t_1,t_2)=-{G^T}^*_{\beta\alpha}(t_2,t_1), \\
G^<_{\alpha\beta}(t_1,t_2)=-{G^<}^*_{\beta\alpha}(t_2,t_1), \\
G^>_{\alpha\beta}(t_1,t_2)=-{G^>}^*_{\beta\alpha}(t_2,t_1).
\end{eqnarray}

It is more convenient to use retarded and advanced functions instead of time-ordered
functions. There is a number of ways to express $G^R$ and $G^A$ through above defined
functions
\begin{equation}
  G^R=\theta(t_1-t_2)\left[G^>-G^<\right]=G^{T}-G^<=G^>-G^{\tilde T},
\end{equation}
\begin{equation}
  G^A=\theta(t_2-t_1)\left[G^<-G^>\right]=G^T-G^>=G^<-G^{\tilde T}.
\end{equation}

\paragraph*{(iii) Contour Green function in the interaction representation}

In the interaction representation one should repeat the calculations performed before
and given the expressions (\ref{lesserTC}), (\ref{TC}), and then replace usual times
by contour times $\tau$, so we obtain
\begin{align}
&  \left\langle T_C\left(c_\alpha(\tau_1)c_\beta^\dagger(\tau_2)\right)\right\rangle
&  =\left\langle T_C\left(\hat S(\tau_0,\tau_2)\tilde c_\beta^\dagger(\tau_2)\hat
   S(\tau_2,\tau_1)
    \tilde c_\alpha(\tau_1)\hat S(\tau_1,\tau_0)\right)\right\rangle.\label{TCtau}
\end{align}

Using contour integration, it can be written as
\begin{align}\label{TCCtau}
G^C_{\alpha\beta}(\tau_1,\tau_2)=
  -i\left\langle T_C\left(c_\alpha(\tau_1)c_\beta^\dagger(\tau_2)\right)\right\rangle 
=-i\left\langle T_C\left(\hat S_C\tilde c_\alpha(\tau_1)\tilde
  c_\beta^\dagger(\tau_2)\right)\right\rangle,
\end{align}
\begin{equation}
  \hat S_C=T_C\exp\left(-i\int_C\hat V^I(t')dt'\right).
\end{equation}

\subsection{Current through a nanosystem: Meir-Wingreen-Jauho formula}

Now we consider the central point of the NGF transport theory through nanosystems -
the Meir-Wingreen-Jauho current formula \cite{Meir92prl,Jauho94prb,Haug96book}. This
important expression shows that the current can be calculated, if the spectral and
kinetic Green functions of the central system are known, and it is exact in the case
of noninteracting leads. The details of the derivation can be found in the above
cited papers, so we only briefly outline it.

\paragraph*{(i) Derivation by the NGF method}

In the absence of interactions in the leads (besides the tunneling) one can derive the
following exact expression for the lead-system function:
\begin{equation}
  G^<_{\alpha,ik\sigma}(\epsilon)=\sum_\beta V^*_{ik\sigma,\beta}\left[
  G^R_{\alpha\beta}(\epsilon)g^<_{ik\sigma}(\epsilon)+
  G^<_{\alpha\beta}(\epsilon)g^A_{ik\sigma}(\epsilon)\right],
\end{equation}
where $g^<_{ik\sigma}(\epsilon)$ and $g^A_{ik\sigma}(\epsilon)$ are Green functions
of {\em isolated} leads, Substituting it into (\ref{Model-J}), we obtain for the
current
%
\begin{equation}
  J_{i}(t)=\frac{2e}{\hbar}\int\frac{d\epsilon}{2\pi}{\rm Re}\left[\sum_{k\sigma,\alpha\beta}
  V_{ik\sigma,\alpha}V^*_{ik\sigma,\beta}\left[
  G^R_{\alpha\beta}(\epsilon)g^<_{ik\sigma}(\epsilon)+
  G^<_{\alpha\beta}(\epsilon)g^A_{ik\sigma}(\epsilon)\right]
  \right].
\end{equation}

For equilibrium right or left lead Green functions we obtain directly
\begin{align}
&\displaystyle g^<_{k\sigma}(t_1-t_2) =  i\left\langle
 c^{\dag}_{k\sigma}(t_2)c_{k\sigma}(t_1)\right\rangle=
 if^0_\sigma(\epsilon_{k\sigma})e^{-i(\epsilon_{k\sigma}+e\varphi)(t_1-t_2)}, \\
&\displaystyle g^R_{k\sigma}(t_1-t_2) =
 -i\theta(t_1-t_2)\left\langle\left[c_{k\sigma}(t_1),c^{\dag}_{k\sigma}(t_2)\right]_+\right\rangle=
 -i\theta(t_1-t_2)e^{-i(\epsilon_{k\sigma}+e\varphi)(t_1-t_2)},  \\
&\displaystyle g^A_{k\sigma}(t_1-t_2) =
 i\theta(t_2-t_1)\left\langle\left[c_{k\sigma}(t_1),c^{\dag}_{k\sigma}(t_2)\right]_+\right\rangle=
 i\theta(t_2-t_1)e^{-i(\epsilon_{k\sigma}+e\varphi)(t_1-t_2)},
\end{align}
or after the Fourier transform
\begin{align}
 \displaystyle g^<_{k\sigma}(\epsilon) & =
 \int g^<_{k\sigma}(t_1-t_2)e^{i\epsilon(t_1-t_2)}d(t_1-t_2)
 =2\pi i f^0_\sigma(\epsilon_{k\sigma})\delta(\epsilon-\epsilon_{k\sigma}-e\varphi), \\
 \displaystyle g^>_{k\sigma}(\epsilon) & =-2\pi i
 [1-f^0_\sigma(\epsilon_{k\sigma})]\delta(\epsilon-\epsilon_{k\sigma}-e\varphi), \\
 \displaystyle g^R_{k\sigma}(\epsilon) & =\frac{1}{\epsilon-\epsilon_{k\sigma}-e\varphi+i0}, \\
 \displaystyle g^A_{k\sigma}(\epsilon) & =\frac{1}{\epsilon-\epsilon_{k\sigma}-e\varphi-i0},
 \end{align}
\begin{equation}
f^0_\sigma(\epsilon)=\frac{1}{\exp\left(\frac{\epsilon-\mu_\sigma}{T}\right)+1}.
\end{equation}

Using the level-width function (below without {\em spin polarization} of the leads)
\begin{align}
& {\Gamma}_{i=L(R)}(\epsilon)\equiv\Gamma_{i\alpha\beta}(\epsilon) =2\pi\sum_{k\sigma}
  V_{ik\sigma,\beta}V^*_{ik\sigma,\alpha}\delta(\epsilon-\epsilon_{ik\sigma}) 
  =2\pi\sum_\sigma\rho_{i\sigma}(\epsilon)
  V_{i\sigma,\beta}(\epsilon)V^*_{i\sigma,\alpha}(\epsilon),
\end{align}
and changing the momentum summation to the energy integration
\mbox{$\displaystyle\sum_{k}\,\Rightarrow\,\int \rho(\epsilon_k)d\epsilon_{k}$},
we obtain the following expression for the current
%
\begin{equation}
  J_{i=L,R}=\frac{ie}{\hbar}\int\frac{d\epsilon}{2\pi}{\rm Tr}\left\{
  {\Gamma}_i(\epsilon-e\varphi_i)\left({\bf G}^<(\epsilon)+f^0_i(\epsilon-e\varphi_i)
  \left[{\bf G}^R(\epsilon)-{\bf G}^A(\epsilon)\right]\right)\right\},
\end{equation}
%
where $f^0_i$ is the equilibrium Fermi distribution function with chemical potential
$\mu_i$. Thus, we obtain the well-known Meir-Wingreen formula. Note, that we use
explicitly the electrical potential of the leads in this expression. It is important to
mention, that at finite voltage the arguments of the left and right level-width functions
are changed in a different way, which means, in particular, that the known condition of
proportional coupling ${\Gamma}_L=\lambda{\Gamma}_R$ can be fulfilled only in the
wide-band limit, when both functions are energy independent.

\paragraph*{(ii) Different forms of the MWJ formula}

In a stationary state $J_R=-J_L=J$ and one can use the symmetric form
%
\begin{equation}\label{J_sym}
\begin{array}{c}\displaystyle
  J=\frac{ie}{2\hbar}\int\frac{d\epsilon}{2\pi}{\rm Tr}{\Big\{}\Big[
  {\Gamma}_L(\epsilon-e\varphi_L)-{\Gamma}_R(\epsilon-e\varphi_R)\Big]
  {\bf G}^<(\epsilon)+ \\[0.3cm]
  \displaystyle \left.
  +\Big[{\Gamma}_L(\epsilon-e\varphi_L)f^0_L(\epsilon-e\varphi_L)-
  {\Gamma}_R(\epsilon-e\varphi_R)f^0_R(\epsilon-e\varphi_R)\right]
  \left[{\bf G}^R(\epsilon)-{\bf G}^A(\epsilon)\Big]\right\}.
\end{array}
\end{equation}

For the proportional coupling ${\Gamma}_L(\epsilon)=\lambda{\Gamma}_R(\epsilon)$ in
{\em linear response} ($\varphi_i$ dependence of $\Gamma_i$ is ignored!)
%
\begin{equation}
  J=\frac{2e}{\hbar}\int\frac{d\epsilon}{4\pi}\left[f^0_{L}(\epsilon-e\varphi_L)-
  f^0_{R}(\epsilon-e\varphi_R)\right]
  {\rm Tr}\left(\frac{{\Gamma}_L(\epsilon){\Gamma}_R(\epsilon)}
  {{\Gamma}_L(\epsilon)+{\Gamma}_R(\epsilon)}
  {\bf A}(\epsilon)\right).
\end{equation}
%
${\bf A}=i({\bf G}^R-{\bf G}^A)$ is the spectral function. This expression is valid for
{\em nonlinear response} if the energy dependence of $\Gamma$ can be neglected (wide
band limit).

\paragraph*{(iii) Noninteracting case}

Finally, in the noninteracting case it is possible to obtain the usual
Landauer-B\"uttikier formula with the transmission function
\begin{equation}\label{NGF-LB-Tr}
  T(\epsilon)={\rm Tr}\left[{\Gamma}_L(\epsilon-e\varphi_L){\bf G}^R(\epsilon)
  {\Gamma}_R(\epsilon-e\varphi_R){\bf G}^A(\epsilon)\right].
\end{equation}
This expression is equivalent to the one derived earlier by the single-particle
Green function method.

We should stress once more that this formula is valid for finite voltage. Therefore, the
voltage dependence of the level-width functions is important.

\subsection{Nonequilibrium equation of motion method}

Now we start to consider the case of interacting nanosystems. Although the MWJ
current formula is exact, the problem to find the Green functions of the central
region is sometimes highly nontrivial. At the present time there are several
techniques developed to solve this problem.

Nonequilibrium equation of motion (NEOM) method is the simplest approximate
approach. In spite of its simplicity, it is very useful in many cases, and is very
convenient for numerical implementation. In this section we consider only a general
formulation, some particular examples are considered further.

We start from the general definition of a Green function as the average of two
Heisenberg operators $\hat A(t)$ and $\hat B(t)$, denoted as
$$\left\langle\!\left\langle\hat  A(t_1),\hat
B(t_2)\right\rangle\!\right\rangle^{R,A,<}.$$
The particular definitions of the averages for spectral and kinetic functions are
\begin{equation}
  \Big\langle\!\!\Big\langle\hat A(t_1),\hat B(t_2)\Big\rangle\!\!\Big\rangle^R=
  -i\theta(t_1-t_2)\Big\langle\Big[\hat A(t_1),\hat B(t_2)\Big]_\mp\Big\rangle,
\end{equation}
where upper sing here and below is for boson functions, lower sing for fermions,
\begin{equation}
  \Big\langle\!\!\Big\langle \hat A(t_1),\hat B(t_2)\Big\rangle\!\!\Big\rangle^<=
  -i\Big\langle \hat A(t_1),\hat B(t_2)\Big\rangle.
\end{equation}

The equations of motion for NGF are obtained from the Heisenberg equation of motion
for operators
\begin{equation}\label{HeisenbergEq}
i\frac{\partial\hat  A}{\partial t}=\left[\hat A,\hat H\right]_-=\hat A\hat H-\hat H
\hat A,
\end{equation}
for any Heisenberg operator $\hat A(t)$. Here and below all Hamiltonians are {\em
time-independent}. We consider the {\em stationary problem}.

\paragraph*{(i) Spectral (retarded and advanced) functions}

Let us start from a retarded function
\begin{equation}
  \Big\langle\!\!\Big\langle\hat A(t_1),\hat B(t_2)\Big\rangle\!\!\Big\rangle^R=
  -i\theta(t_1-t_2)\Big\langle\Big[\hat A(t_1),\hat B(t_2)\Big]_\mp\Big\rangle.
\end{equation}
Taking the time derivative we obtain
\begin{align}\label{R5}
  i\frac{\partial}{\partial t_1}\Big\langle\!\!\Big\langle\hat A(t_1),\hat B(t_2)\Big\rangle\!\!\Big\rangle^R
   = \delta(t_1-t_2)\Big\langle\Big[\hat A(t_1),\hat B(t_1)\Big]_\mp\Big\rangle
   +\Big\langle\!\!\Big\langle\Big[\hat A(t_1),\hat H\Big]_-,\hat B(t_2)\Big\rangle\!\!\Big\rangle^R,
\end{align}
where the first term originates from the time-derivative of the $\theta$-function,
and the equation (\ref{HeisenbergEq}) is used in the second term.

In the stationary case the Fourier transform can be used
\begin{equation}
  (\epsilon+i\eta)\Big\langle\!\!\Big\langle \hat A,\hat B\Big\rangle\!\!\Big\rangle^R_\epsilon=
  \Big\langle\Big[\hat A,\hat B\Big]_\mp\Big\rangle
  +\Big\langle\!\!\Big\langle\Big[\hat A,\hat H\Big]_-,\hat B\Big\rangle\!\!\Big\rangle^R_\epsilon.
\end{equation}

Now let us assume that the Hamiltonian can be divided into "free particle" and
"interaction" parts $\hat H=\hat H_0+\hat H_1$, and $[\hat A,\hat
H_0]_-=\hat\epsilon_0 \hat A$. (The simple example. For the free particle
Hamiltonian $\hat H_0=\sum_\beta\epsilon_\beta d^{\dag}_\beta d_\beta$ and the
operator \mbox{$\hat A=d^{\dag}_\alpha$} one has $[\hat A,\hat
H_0]_-=\sum_\beta\epsilon_\beta[d^{\dag}_\alpha,d^{\dag}_\beta
d_\beta]_-=\epsilon_\alpha d^{\dag}_\alpha$, $\hat\epsilon_0=\epsilon_\alpha$ is
simply a number. In general, $\hat\epsilon_0$ is some time-independent operator). So
that
\begin{equation}\label{R7}
  (\epsilon+i\eta-\hat\epsilon_0)\Big\langle\!\!\Big\langle \hat A,\hat B\Big\rangle\!\!\Big\rangle^R_\epsilon=
  \Big\langle\Big[\hat A,\hat B\Big]_\mp\Big\rangle
  +\Big\langle\!\!\Big\langle\Big[\hat A,\hat H_1\Big]_-,\hat
  B\Big\rangle\!\!\Big\rangle^R_\epsilon,
\end{equation}
the second term includes interaction and can not be easy simplified.

It is convenient now to introduce the "free particle" function $\hat g^R_\epsilon$
as a solution of the equation
\begin{equation}
  (\epsilon+i\eta-\hat\epsilon_0)\hat g^R_\epsilon=1.
\end{equation}

Now we multiply the right and left parts of (\ref{R7}) by $\hat g^R_\epsilon$. Using
the function $\hat g^R(t)=\int \hat g^R_\epsilon e^{-i\epsilon
t}\frac{d\epsilon}{2\pi}$ we can write the time-dependent solution of (\ref{R5}) as
\begin{align}\label{Rtd}
  \Big\langle\!\!\Big\langle \hat A(t_1), \hat B(t_2)\Big\rangle\!\!\Big\rangle^R= &
  \hat g^R(t_1-t_2)\Big\langle\Big[\hat A(t_1),\hat B(t_1)\Big]_\mp\Big\rangle \nonumber \\
& +\int \hat g^R(t_1-t')
  \Big\langle\!\!\Big\langle\Big[\hat A(t'),\hat H_1\Big]_-,\hat B(t_2)\Big\rangle\!\!\Big\rangle^Rdt'.
\end{align}

\paragraph*{(ii) EOM on the Schwinger-Keldysh contour}

The calculation of the lesser functions by the EOM technique requires some care. To
demonstrate it let us compare the EOM for retarded and lesser functions of free
particles.

The equation for $g^R_{\alpha\beta}$ is (assuming the diagonal matrix
$\tilde\epsilon_{\alpha\beta}$)
\begin{equation}\label{gR}
\left(\epsilon+i\eta-\tilde\epsilon_\alpha\right)g^R_{\alpha\beta}=
\delta_{\alpha\beta},
\end{equation}
from which the free-particle Green function is easily obtained.

At the same time for the lesser function we have the equation
\begin{equation}\label{gL}
\left(\epsilon-\tilde\epsilon_\alpha\right)g^<_{\alpha\beta}=0,
\end{equation}
from which, however, the free-particle lesser function $g^<_{\alpha\beta}=2\pi
f_0(\epsilon)\delta(\epsilon-\epsilon_\alpha)\delta_{\alpha\beta}$ can not be
obtained.

The problem can be generally resolved by using the EOM on the Schwinger-Keldysh time
contour. Contour-ordered Green function is defined as
\begin{equation}
  \Big\langle\!\!\Big\langle \hat A(\tau_1),\hat B(\tau_2)\Big\rangle\!\!\Big\rangle^C=
  -i\Big\langle T_c\Big( \hat A(\tau_1),\hat B(\tau_2)\Big)\Big\rangle,
\end{equation}
where $\hat A(\tau_1)$ and $\hat B(\tau_2)$ are two Heisenberg operators, defined
along the contour.

Taking the time derivative we obtain the equation
\begin{align}
  i\frac{\partial}{\partial \tau_1}\Big\langle\!\!\Big\langle \hat A(\tau_1),\hat B(\tau_2)\Big\rangle\!\!\Big\rangle^C=
  \delta^c(\tau_1-\tau_2)\Big\langle\Big[\hat A(\tau_1),\hat B(\tau_1)\Big]_\mp\Big\rangle 
  +\Big\langle\!\!\Big\langle\Big[\hat A(\tau_1),\hat H\Big]_-,\hat B(\tau_2)\Big\rangle\!\!\Big\rangle^C,
\end{align}
in the stationary case this equation can be formally solved if one applies the
Fourier transform along the contour, or perturbation expansion in the interaction
representation (Niu et al. 1999).  Using the free particle solution $\hat
g^C(\tau_1-\tau_2)$ we can write the time-dependent solution as
\begin{align}
  \Big\langle\!\!\Big\langle \hat A(\tau_1),\hat B(\tau_2)\Big\rangle\!\!\Big\rangle^C= &
  \hat g^C(\tau_1-\tau_2)\Big\langle\Big[\hat A(\tau_1),\hat B(\tau_1)\Big]_\mp\Big\rangle \nonumber \\
& +\int \hat g^C(\tau_1-\tau')
  \Big\langle\!\!\Big\langle\Big[\hat A(\tau'),\hat H_1\Big]_-,\hat B(\tau_2)\Big\rangle\!\!\Big\rangle^C d\tau'.
\end{align}

\paragraph*{(iii) Kinetic (lesser) function}

Applying now the Langreth rules (see the next section for details), which shows,
that from
\begin{equation}
  C(\tau_1,\tau_2)=\int_CA(\tau_1,\tau_3)B(\tau_3,\tau_2)d\tau_3
\end{equation}
it follows
\begin{eqnarray}
 & C^R(t_1,t_2)=\int A^R(t_1,t_3)B^R(t_3,t_2)dt_3, \\
 & C^<(t_1,t_2)=\int\left(A^R(t_1,t_3)B^R(t_3,t_2)+A^<(t_1,t_3)B^A(t_3,t_2)\right)dt_3,
\end{eqnarray}
we get (\ref{Rtd}) for the retarded function, and
\begin{align}\label{Ltd}
  \Big\langle\!\!\Big\langle \hat A(t_1),\hat B(t_2)\Big\rangle\!\!\Big\rangle^<= & \
  \hat g^<(t_1-t_2)  \Big\langle\Big[\hat A(t_1),\hat B(t_1)\Big]_\mp\Big\rangle \nonumber \\
& +\int \hat g^R(t_1-t')
  \Big\langle\!\!\Big\langle\Big[\hat A(t'),\hat H_1\Big]_-,\hat B(t_2)\Big\rangle\!\!\Big\rangle^< dt' \nonumber \\
& +\int \hat g^<(t_1-t')
  \Big\langle\!\!\Big\langle\Big[\hat A(t'),\hat H_1\Big]_-,\hat B(t_2)\Big\rangle\!\!\Big\rangle^A dt'
\end{align}
for the lesser function. And the Fourier transform is
\begin{align}\label{Lfd}
  \Big\langle\!\!\Big\langle \hat A,\hat B\Big\rangle\!\!\Big\rangle^<_\epsilon=
  \hat g^<_\epsilon\Big\langle\Big[\hat A,\hat B\Big]_\mp\Big\rangle
  +\hat g^R_\epsilon
  \Big\langle\!\!\Big\langle\Big[\hat A,\hat H_1\Big]_-,\hat B\Big\rangle\!\!\Big\rangle^<_\epsilon 
  + \hat g^<_\epsilon
  \Big\langle\!\!\Big\langle\Big[\hat A,\hat H_1\Big]_-,\hat B\Big\rangle\!\!\Big\rangle^A_\epsilon.
\end{align}

\newpage
\subsection{Kadanoff-Baym-Keldysh method}

Now we review briefly the other approach. Kadanoff-Baym-Keldysh (KBK) method
systematically extends the equilibrium many-body theory to the nonequilibrium case.
Potentially, it is the most powerful approach. Below we give a simple introduction
into the method, which is currently actively developed.

\paragraph*{(i) Perturbation expansion and diagrammatic rules for contour functions}

We found that Green functions can be written in the interaction representation with a
help of the $\hat S$-operator. For example, time-ordered fermionic Green function is
\begin{align}\label{GT}
  G^T_{\alpha\beta}(t_1,t_2)= &
  -i\left\langle T\left(c_\alpha(t_1)c^\dagger_\beta(t_2)\right)\right\rangle 
  =-i\left\langle \hat S^{-1} T\left(\tilde c_\alpha(t_1)\tilde c^\dagger_\beta(t_2)\hat
  S\right)\right\rangle,
\end{align}
using "usual" $\hat S$-operator
\begin{equation}
  \hat S=\hat S(\infty,t_0)=T\exp\left(-i\int_{t_0}^{\infty}\hat V^I(t')dt'\right),
\end{equation}
or
\begin{equation}\label{GTC}
  G^T_{\alpha\beta}(t_1,t_2)=
  -i\left\langle T_C\left(\tilde c_\alpha(t^\rightarrow_1)
  \tilde c^\dagger_\beta(t^\rightarrow_2)\hat
  S_C\right)\right\rangle,
\end{equation}
using "contour" $\hat S_C$-operator
\begin{equation}
  \hat S_C=T_C\exp\left(-i\int_C\hat V^I(t')dt'\right).
\end{equation}

We first consider the zero temperature case, when one can set $t_0=-\infty$,
\begin{equation}
  \hat S=\hat S(\infty,-\infty)=T\exp\left(-i\int_{-\infty}^{\infty}\hat V^I(t')dt'\right),
\end{equation}
and assume that interaction is switched on and switched off at $t\rightarrow +\infty$
{\em adiabatically}. This condition is necessary to prevent excitation of the system
from its ground state. The other necessary condition is that the perturbation is
time-independent in the Schr\"odinger representation. In this case if the initial
state $|\Psi(t=-\infty)\rangle=|\Psi_0\rangle$ is the ground state (of free
particles), then the final state $|\Psi(t=+\infty)\rangle=\hat
S|\Psi^0\rangle=e^{i\theta}|\Psi^0\rangle$ is also the ground state, only the phase
can be changed. Now, using the average value of the $\hat S$-operator
\begin{equation}
  \langle\hat S\rangle=\langle\Psi^0|\hat S|\Psi^0\rangle=
  e^{i\theta}\langle\Psi^0|\Psi^0\rangle=e^{i\theta},
\end{equation}
we obtain
\begin{equation}
  \hat S|\Psi^0\rangle=\langle\hat S\rangle|\Psi^0\rangle,
\end{equation}
and
\begin{equation}
  \langle\Psi^0|\hat S^{-1}=\frac{\langle\Psi^0|}{\langle\hat S\rangle}.
\end{equation}

So that (\ref{GT}) can be written as
\begin{equation}\label{GT2}
  G^T_{\alpha\beta}(t_1,t_2)=
  -i\frac{\left\langle T\left(\tilde c_\alpha(t_1)\tilde c^\dagger_\beta(t_2)\hat
  S\right)\right\rangle}{\langle\hat S\rangle}.
\end{equation}

Now we can expand the exponent (note that $S$-operator is defined only in the sense of
this expansion)
\begin{align}
  \hat S= & T\exp\left(-i\int_{-\infty}^{\infty}\hat V^I(t')dt'\right) 
  =T\sum_{n=0}^{\infty}\frac{(-i)^n}{n!}\int_{-\infty}^\infty dt'_1...
  \int_{-\infty}^\infty dt'_n\ \hat V^I(t'_1)...\hat V^I(t'_n),
\end{align}
and numerator and denominator of the expression (\ref{GT2}) are
\begin{align}
& \left\langle T\left(\tilde c_\alpha(t_1)\tilde c^\dagger_\beta(t_2)\hat
  S\right)\right\rangle= 
  \sum_{n=0}^{\infty}\frac{(-i)^n}{n!}\int_{-\infty}^\infty dt'_1...
  \int_{-\infty}^\infty dt'_n\left\langle T
  \tilde c_\alpha(t_1)\tilde c^\dagger_\beta(t_2)
  \hat V^I(t'_1)...\hat V^I(t'_n)\right\rangle,
\end{align}
\begin{equation}
  \langle\hat S\rangle=\sum_{n=0}^{\infty}\frac{(-i)^n}{n!}\int_{-\infty}^\infty dt'_1...
  \int_{-\infty}^\infty dt'_n\left\langle T\hat V^I(t'_1)...\hat
  V^I(t'_n)\right\rangle.
\end{equation}

These expressions are used to produce the perturbation series.

The main quantity to be calculated is the contour Green function
\begin{equation}
  G(1,2)\equiv G^C_{\alpha\beta}(\tau_1,\tau_2)=-i\left\langle T_C\left(
  c_\alpha(\tau_1)c_\beta^\dagger(\tau_2)\right)\right\rangle,
\end{equation}
where $\tau_1$ and $\tau_2$ are contour times. Here $1_c\equiv\alpha,\tau_1$.

The general diagrammatic rules for contour Green functions are exactly the same as
in the usual zero-temperature technique (we call it standard rules). The
correspondence between diagrams and analytical expressions is established in the
following way.

\begin{enumerate}

\item Open bare electron line is $iG_0(1,2)$.

\item Closed bare electron line is $n_0(1)\equiv n^{(0)}_\alpha(\tau_1)$.

\item Bare interaction line is $-iv(1,2)$.

\item Self-energy is $-i\Sigma(1,2)$.

\item Integration over internal vertices, and other standard rules.

\end{enumerate}

\paragraph*{(ii) Langreth rules}

Although the basic equations and diagrammatic rules are formulated for contour Green
functions, the solution of these equation and final results are much more
transparent when represented by real-time spectral and kinetic functions.

As in the ordinary diagrammatic technique, the important role is played by the
integration (summation) over space and contour-time arguments of Green functions,
which is denoted as
\begin{equation}
  \int d1_c\equiv\sum_\alpha\int_C d\tau_1.
\end{equation}
After application of the Langreth rules, for real-time functions these integrals
become
\begin{equation}
  \int d1\equiv\sum_\alpha\int_{-\infty}^\infty dt_1.
\end{equation}

The Langreth rules show, for example, that from
\begin{equation}
  C(\tau_1,\tau_2)=\int_CA(\tau_1,\tau_3)B(\tau_3,\tau_2)d\tau_3
\end{equation}
it follows
\begin{eqnarray}
&  C^R(t_1,t_2)=\int A^R(t_1,t_3)B^R(t_3,t_2)dt_3,  \\ &
C^<(t_1,t_2)=\int\big(A^R(t_1,t_3)B^<(t_3,t_2)+A^<(t_1,t_3)B^A(t_3,t_2)\big)dt_3.
\end{eqnarray}

The other important rules are: from
\begin{equation}
  C(\tau_1,\tau_2)=A(\tau_1,\tau_2)B(\tau_1,\tau_2)
\end{equation}
it follows
\begin{eqnarray}
&
C^R(t_1,t_2)=A^R(t_1,t_2)B^R(t_1,t_2)+A^R(t_1,t_2)B^<(t_1,t_2)+A^<(t_1,t_2)B^R(t_1,t_2),
\\ &  C^<(t_1,t_2)=A^<(t_1,t_2)B^<(t_1,t_2),
\end{eqnarray}
and from
\begin{equation}
  C(\tau_1,\tau_2)=A(\tau_1,\tau_2)B(\tau_2,\tau_1)
\end{equation}
it follows
\begin{eqnarray}
&  C^R(t_1,t_2)=A^R(t_1,t_2)B^<(t_2,t_1)+A^<(t_1,t_2)B^A(t_2,t_1), \\ &
C^<(t_1,t_2)=A^<(t_1,t_2)B^>(t_2,t_1).
\end{eqnarray}

\paragraph*{(iii) First-order self-energy and polarization operator}

\begin{figure}[t]
\begin{center}
\epsfxsize=0.5\hsize \epsfbox{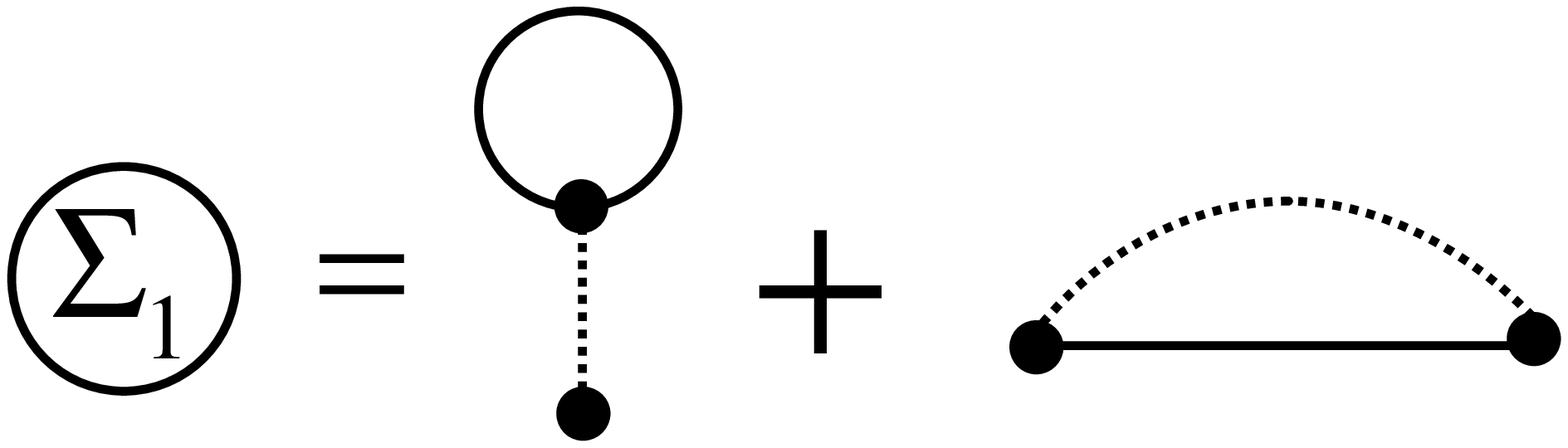} \caption{Diagrammatic representation of
the first-order self-energy.} \label{NQT-9-x1}
\end{center}
\end{figure}

Consider, as an example, the first order expression for the self-energy, shown in
Fig.~\ref{NQT-9-x1}. Following the diagrammatic rules, we find
\begin{equation}
 \Sigma_1(1,2)=\delta(1-2)\int v(1,3)n_0(3)d3+iv(1,2)G_0(1,2),
\end{equation}
where the first term is the Hartree contribution, which can be included into the
unperturbed Green function $G_0(1,2)$. This expression is actually symbolic, and
translation from contour (Keldysh-time) to real-time functions is necessary. Using
the Langreth rules, one obtains
\begin{align}
 \Sigma_1^R(1,2)= & \delta(1^+-2)\int v^R(1,3)n_0(3,3)d3 +iv^R(1,2)G_0^R(1,2) \nonumber \\ &
 +iv^<(1,2)G_0^R(1,2)+iv^R(1,2)G_0^<(1,2),
\end{align}
\begin{equation}
 \Sigma_1^<(1,2)=iv^<(1,2)G_0^<(1,2).
\end{equation}
There is no Hartree term for lesser function, because the times $\tau_1$ and
$\tau_2$ are always at the different branches of the Keldysh contour, and the
$\delta$-function $\delta(\tau_1-\tau_2)$ is zero.

In the stationary case and using explicit matrix indices, we have, finally
($\tau=t_1-t_2$!, not to mix with the Keldysh time)
\begin{eqnarray}
& \Sigma_{\alpha\beta}^{R(1)}(\tau)=\delta(\tau^+)\delta_{\alpha\beta}\sum_\gamma
  \tilde v_{\alpha\gamma}^R(0)n^{(0)}_{\gamma} \nonumber \\ &
  +iv_{\alpha\beta}^R(\tau)G_{\alpha\beta}^{R(0)}\tau)
  +iv_{\alpha\beta}^<(\tau)G_{\alpha\beta}^{R(0)}(\tau)
  +iv_{\alpha\beta}^R(\tau)G_{\alpha\beta}^{<(0)}(\tau),
  \\[0.3cm]
& \Sigma_
{\alpha\beta}^{<(1)}(\tau)=iv_{\alpha\beta}^<(\tau)G_{\alpha\beta}^{<(0)}(\tau),
\end{eqnarray}
and we define the Fourier transform of the bare interaction
\begin{equation}
\tilde v_{\alpha\gamma}^R(0)=\int v_{\alpha\gamma}^R(\tau)d\tau.
\end{equation}

Finally, the Fourier transforms are
\begin{align}
& \Sigma_{\alpha\beta}^{R(1)}(\epsilon)= \delta_{\alpha\beta}\sum_\gamma
  \tilde v_{\alpha\gamma}^R(0)n^{(0)}_{\gamma} \nonumber \\
& +i\int\frac{d\epsilon'}{2\pi}\left[
  v_{\alpha\beta}^R(\epsilon')G_{\alpha\beta}^{R(0)}(\epsilon-\epsilon')
  +v_{\alpha\beta}^<(\epsilon')G_{\alpha\beta}^{R(0)}(\epsilon-\epsilon')+
  v_{\alpha\beta}^R(\epsilon')G_{\alpha\beta}^{<(0)}(\epsilon-\epsilon')
  \right],
\end{align}
\begin{equation}
  \Sigma_ {\alpha\beta}^{<(1)}(\epsilon)=i\int\frac{d\epsilon'}{2\pi}
  v_{\alpha\beta}^<(\epsilon')G_{\alpha\beta}^{<(0)}(\epsilon-\epsilon').
\end{equation}

\begin{figure}[t]
\begin{center}
\epsfxsize=0.4\hsize \epsfbox{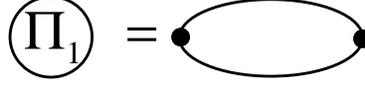} \caption{Diagrammatic representation of
the first-order polarization operator.} \label{NQT-9-x2}
\end{center}
\end{figure}

The second important function is the polarization operator ("self-energy for
interaction"), showing in Fig.~\ref{NQT-9-x2}. Following the diagrammatic rules, we
find
\begin{equation}
 \Pi_1(1,2)=-iG_0(1,2)G_0(2,1),
\end{equation}
note the order of times in this expression.

Using the Langreth rules,
\begin{equation}
 \Pi_1^R(1,2)=iG_0^R(1,2)G_0^<(2,1)+iG_0^<(1,2)G_0^A(2,1),
\end{equation}
\begin{equation}
 \Pi_1^<(1,2)=iG_0^<(1,2)G_0^>(2,1).
\end{equation}

And in the stationary case, restoring the matrix indices
\begin{equation}\label{PiR1}
 \Pi^{R(1)}_{\alpha\beta}(\epsilon)=-i\left[
 G^{R(0)}_{\alpha\beta}(\tau)G^{<(0)}_{\beta\alpha}(-\tau)
 +G^{<(0)}_{\alpha\beta}(\tau)G^{A(0)}_{\beta\alpha}(-\tau)
 \right],
\end{equation}
\begin{equation}\label{PiK1}
 \Pi^{<(1)}_{\alpha\beta}(\epsilon)=-i
 G^{<(0)}_{\alpha\beta}(\tau)G^{>(0)}_{\beta\alpha}(-\tau).
\end{equation}

In the Fourier representation
\begin{equation}\label{PiRF}
 \Pi^{R(1)}_{\alpha\beta}(\tau)=-i\int\frac{d\epsilon'}{2\pi}\left[
 G^{R(0)}_{\alpha\beta}(\epsilon')G^{<(0)}_{\beta\alpha}(\epsilon'-\epsilon)
 + G^{<(0)}_{\alpha\beta}(\epsilon')G^{A(0)}_{\beta\alpha}(\epsilon'-\epsilon)
 \right],
\end{equation}
\begin{equation}\label{PiKF}
 \Pi^{<(1)}_{\alpha\beta}(\tau)=-i\int\frac{d\epsilon'}{2\pi}
 G^{<(0)}_{\alpha\beta}(\epsilon')G^{>(0)}_{\beta\alpha}(\epsilon'-\epsilon).
\end{equation}

These expressions are quite general and can be used for both electron-electron and
electron-vibron interaction.

For Coulomb interaction the bare interaction is is $v(1,2)\equiv
U_{\alpha\beta}\delta(\tau_1^+-\tau_2)$, so that
\begin{eqnarray}
& v^R(1,2)\equiv U_{\alpha\beta}\delta(t^+_1-t_2), \\ & v^<(1,2)=0.
\end{eqnarray}

\begin{figure}[t]
\begin{center}
\epsfxsize=0.5\hsize\hskip -1.5cm \epsfbox{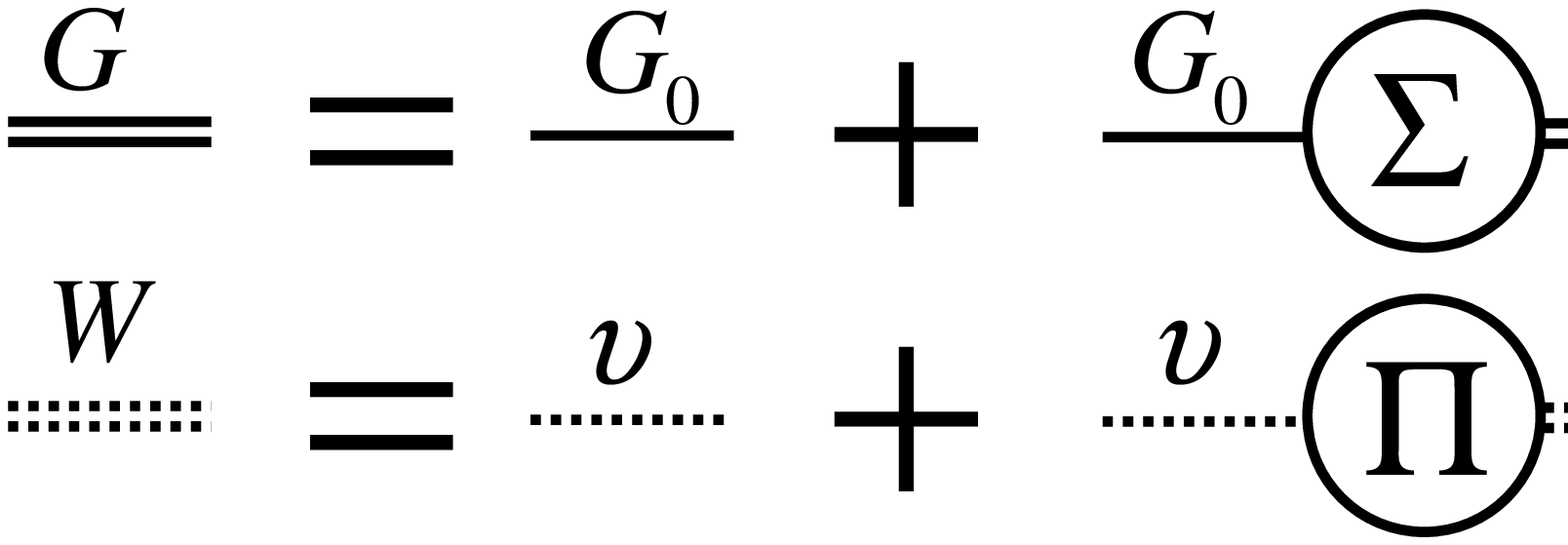} \caption{Diagrammatic
representation of the Dyson equations.} \label{NQT-9-1}
\end{center}
\end{figure}

\begin{figure}[b]
\begin{center}
\epsfxsize=0.5\hsize\hskip -2cm \epsfbox{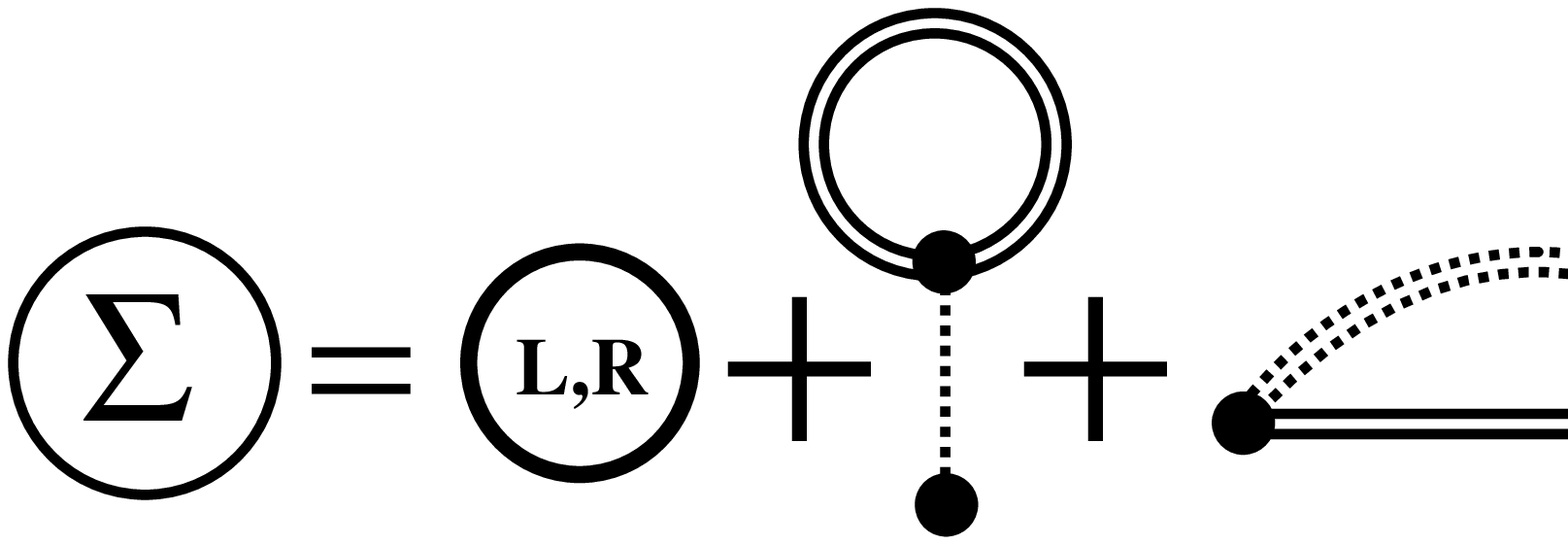} \caption{Diagrammatic
representation of the full self-energy.} \label{NQT-9-2}
\end{center}
\end{figure}
\begin{figure}
\begin{center}
\epsfxsize=0.4\hsize \epsfbox{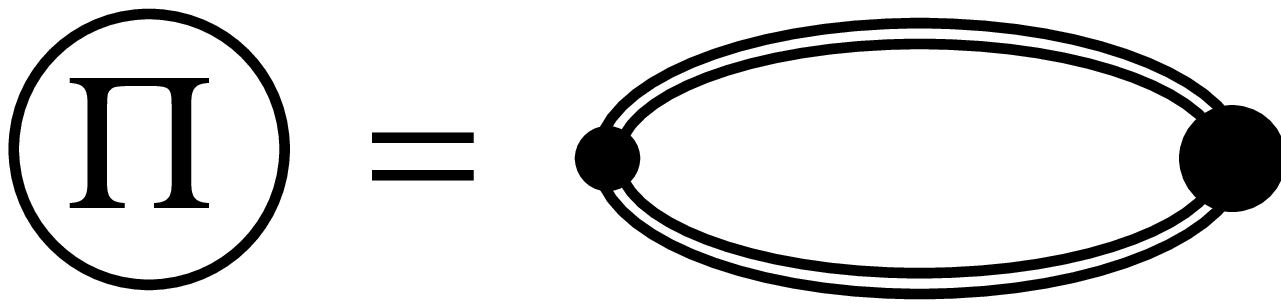} \caption{Diagrammatic representation of
the full polarization operator.} \label{NQT-9-3}
\end{center}
\end{figure}
\begin{figure}
\begin{center}
\epsfxsize=0.5\hsize \epsfbox{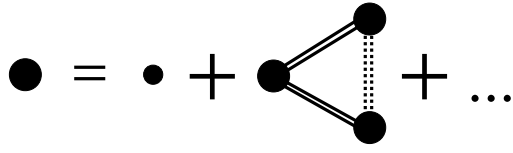} \caption{Diagrammatic representation of
the vertex function.} \label{NQT-9-4}
\end{center}
\end{figure}

\paragraph*{(iv) Self-consistent equations}

The diagrams can be partially summed in all orders of perturbation theory. The
resulting equations are known as Dyson equations for the dressed Green function
$G(1,2)$ and the effective interaction $W(1,2)$ (Fig.~\ref{NQT-9-1}). Analytically
these equations are written as
\begin{equation}\label{Hedin1}
 G(1,2)=G_0(1,2)+\int\!\!\int G_0(1,3)\Sigma(3,4)G(4,2)d3d4,
\end{equation}
\begin{equation}
 W(1,2)=v(1,2)+\int\!\!\int v(1,3)\Pi(3,4)W(4,2)d3d4.
\end{equation}

In the perturbative approach the first order (or higher order) expressions for the
self-energy and the polarization operator are used. The other possibility is to
summarize further the diagrams and obtain the self-consistent approximations
(Figs.~\ref{NQT-9-2},\ref{NQT-9-3}), which include, however, a new unknown function,
called vertex function. We shall write these expressions analytically, including the
Hartree-Fock part into unperturbed Green function $G_0(1,2)$.
\begin{equation}
 \Sigma'(1,2)=i\int\!\!\int W(1,3)G(1,4)\Gamma(3;4,2)d3d4,
\end{equation}
\begin{equation}\label{Hedin4}
 \Pi(1,2)=-i\int\!\!\int G(1,3)G(4,1)\Gamma(2;3,4)d3d4.
\end{equation}

The equation for the vertex function can not be closed diagrammatically
(Fig.~\ref{NQT-9-4}). Nevertheless, it is possible to write close set of equations
({\em Hedin's equations}), which are exact equations for full Green functions
written through a functional derivative. Hedin's equations are equations
(\ref{Hedin1})-(\ref{Hedin4}) and the equation for the vertex function
\begin{equation}
 \Gamma(1;2,3)=\delta(1,2)\delta(1,3)+
 \int\!\!\int\!\!\int\!\!\int G(4,6)G(7,5)\Gamma(1;6,7)\frac{\delta\Sigma(2,3)}{\delta G(4,5)}d4d5d6d7.
\end{equation}

\section{Applications}

\subsection{Coulomb blockade}

In Section~{\bf 2} we have seen that Coulomb blockade phenomena mediated by
electron-electron interactions on a quantum dot can be dealt with in a
straightforward way by using master equation (ME) approaches, which are based on
Fermi's Golden Rule.
\cite{Beenakker91prb,Averin91prb,vanHouten92inbook,vonDelft01pr,Bonet02prb,Hettler03prl,
vanderWiel02rmp,Muralidharan06prb} However, due to its intrinsic perturbative
character in the lead-dot coupling, ME techniques cannot cover the whole interaction
range from weak-coupling (Coulomb blockade), intermediate coupling (Kondo physics),
up to strong coupling (Fabry-Perot physics). It is thus of methodological and
practical interest to develop schemes which allow, in a systematic way, to describe
the three mentioned regimes also in out-of-equilibrium situations. As stated in the
introduction, we believe that Green function techniques are such a tool; in this
section we will show how a non-equilibrium treatment of the Hubbard-Anderson model
together with appropriate approximations allow us to reproduce the well-known
Coulomb blockade  stability diagrams obtained with the master equation approach (see
also Section~{\bf 2}). For the sake of simplicity we will deal with the problem of
single and double-site dots in the CB regime, although the method can be
straightforwardly extended to multi-level systems. Our purpose is to study the
problem of a two site donor/acceptor molecule in the CB regime within the NEGF as a
first step to deal with the phenomenology of a rigid multilevel island. The nuclear
dynamics (vibrations) always present in molecular junctions could be then modularly
included in this theory. Our method can be calibrated on the well-studied double
quantum dot problem \cite{vanderWiel02rmp,Pedersen07prb} and could be possibly
integrated in the density functional theory based approaches to molecular
conductance. The Kondo regime would require a separate treatment involving more
complex decoupling schemes and will be thus left out of this review, for some new
results see Ref.\,\cite{Swirkowicz03prb} (EOM method) and
Refs.\,\cite{Thygesen07jcp,Thygesen08prl,Thygesen08prb} (the self-consistent GW
approximation).

The {\it linear conductance} properties  of a single site junction (\SSJ) with
Coulomb interactions (Anderson impurity model),  have been extensively studied by
means of the EOM approach in the cases related to CB
\cite{Lacroix81jphysf,Meir91prl} and the Kondo effect. \cite{Meir93prl} Later the
same method was applied to some two-site models.
\cite{Niu95prb,Pals96jpcm,Lamba00prb,Bulka04prb} Multi-level systems were started to
be considered only recently. \cite{Palacios97prb,Yi02prb}  For out-of-equilibrium
situations (finite applied bias), there are some methodological unclarified issues
for calculating correlation functions using EOM techniques.
\cite{Niu99jpcm,Swirkowicz03prb,Bulka04prb} We have developed  an EOM-based method
which allows to deal with the finite-bias case in a self-consistent way.
\cite{Song07prb}

\subsubsection{Nonequilibrium EOM formalism}

\paragraph*{(i) The Anderson-Hubbard Hamiltonian}

We consider the following model Hamiltonian (which can be called the multi-level
Anderson impurity model, the Hubbard model, or the quantum cluster model)
\begin{equation}\label{H1}
  \hat H=\sum_{\alpha\beta}\tilde\epsilon_{\alpha\beta} d^{\dag}_\alpha d_\beta+
  \frac{1}{2}\sum_{\alpha\beta}U_{\alpha\beta}\hat n_\alpha \hat n_\beta  
  +\sum_{ik\sigma}\tilde\epsilon_{ik\sigma}c^{\dag}_{ik\sigma}c_{ik\sigma}+
  \sum_{ik\sigma,\alpha}\left(V_{ik\sigma,\alpha}c^{\dag}_{ik\sigma}d_\alpha+h.c.\right),
\end{equation}
electrical potentials are included into the energies
\mbox{$\tilde\epsilon_{ik\sigma}=\epsilon_{ik\sigma}+e\varphi_i(t)$} and
$\tilde\epsilon_{\alpha\alpha}=\epsilon_{\alpha\alpha}+e\varphi_\alpha(t)$.

This model is quite universal, describing a variety of correlated electron systems
coupled to the leads: the Anderson impurity model, the  multilevel quantum dot with
diagonal noninteracting Hamiltonian $\tilde\epsilon_{\alpha\beta}$, a system
(cluster) of several quantum dots, when the off-diagonal matrix elements of
$\tilde\epsilon_{\alpha\beta}$ describe hopping between individual dots, and,
finally, the 1D and 2D quantum point contacts.

\paragraph*{(ii) EOM for Heisenberg operators}

Using the Hamiltonian (\ref{H1}) one derives
\begin{equation}
i\frac{\partial c_{ik\sigma}}{\partial t}=\left[c_{ik\sigma},\hat H\right]_-=
\tilde\epsilon_{ik\sigma}c_{ik\sigma}+\sum_\alpha V_{ik\sigma,\alpha}d_\alpha,
\end{equation}
\begin{equation}
i\frac{\partial c^\dag_{ik\sigma}}{\partial t}=
-\tilde\epsilon_{ik\sigma}c^\dag_{ik\sigma}-\sum_\alpha
V^*_{ik\sigma,\alpha}d^\dag_\alpha,
\end{equation}
\begin{equation}\label{Heisenberg3}
i\frac{\partial d_\alpha}{\partial t}=
\sum_\beta\tilde\epsilon_{\alpha\beta}d_\beta+\sum_{\beta\neq\alpha}U_{\alpha\beta}\hat
n_\beta d_\alpha +\sum_{ik\sigma} V^*_{ik\sigma,\alpha}c_{ik\sigma},
\end{equation}
\begin{equation}
i\frac{\partial d^\dag_\alpha}{\partial t}=
-\sum_\beta\tilde\epsilon_{\alpha\beta}d^\dag_\beta-
\sum_{\beta\neq\alpha}U_{\alpha\beta}\hat n_\beta d^\dag_\alpha -\sum_{ik\sigma}
V_{ik\sigma,\alpha}c^\dag_{ik\sigma},
\end{equation}
\begin{align}
i\frac{\partial \hat n_\gamma}{\partial t}=
\sum_{ik\sigma} & \left[-V_{ik\sigma,\gamma}c^\dag_{ik\sigma}d_\gamma+
V^*_{ik\sigma,\gamma}d^\dag_\gamma c_{ik\sigma}\right] \nonumber \\
& +\sum_\beta\tilde\epsilon_{\gamma\beta}d^\dag_\gamma d_\beta-
\sum_\alpha\tilde\epsilon_{\alpha\gamma}d^\dag_\alpha d_\gamma.
\end{align}

These equations look like a set of ordinary differential equations, but are, in
fact, much more complex. The first reason is, that there are the equations for {\em
operators}, and special algebra should be used to solve it. Secondly, the number of
$c_{ik\sigma}$ operators is infinite! Because of that, the above equations are not
all sufficient, but are widely used to obtain the equations for Green functions.

\paragraph*{(iii) Spectral (retarded and advanced) functions}

Now we follow the general NEOM method described in the Section~{\bf 3}. Using
(\ref{Heisenberg3}), we get the equation for
\mbox{$G^R_{\alpha\beta}=-i\left\langle\left[d_\alpha,d_\beta^\dagger\right]_+\right\rangle_\epsilon$}
\begin{align}\label{R1}
(\epsilon+i\eta)G^R_{\alpha\beta}-\sum_\gamma\tilde\epsilon_{\alpha\gamma}
G^R_{\gamma\beta} & =
\delta_{\alpha\beta}+\sum_{\gamma\neq\alpha}U_{\alpha\gamma}G^{(2)R}_{\alpha\gamma,\beta}
+\sum_{ik\sigma}V^*_{ik\sigma,\alpha}G^R_{ik\sigma,\beta}
\end{align}
which includes two new functions: $G^{(2)R}_{\alpha\gamma,\beta}$ and
$G^R_{ik\sigma,\beta}$.

The equation for $G^R_{ik\sigma,\beta}$ is closed (includes only the function
$G^R_{\alpha\beta}$ introduced before)
\begin{equation}
(\epsilon+i\eta-\tilde\epsilon_{ik\sigma})G^R_{ik\sigma,\beta}= \sum_\delta
V_{ik\sigma,\delta}G^R_{\delta\beta}.
\end{equation}

The equation for $$G^{(2)R}_{\alpha\gamma,\beta}(t_1-t_2)=
-i\theta(t_1-t_2)\left\langle\left[d_\alpha(t_1)\hat
n_\gamma(t_1),d_\beta^\dagger(t_2) \right]_+\right\rangle$$ is more complicated
\begin{align}
\label{G2_1}
(&\epsilon+i\eta)G^{(2)R}_{\alpha\gamma,\beta}\!-\!\sum_\delta\tilde\epsilon_{\alpha\delta}
G^{(2)R}_{\delta\gamma,\beta}=n_\gamma\delta_{\alpha\beta}+
(\delta_{\alpha\beta}-\rho_{\alpha\beta})\delta_{\beta\gamma} \nonumber \\
\!+\! & \sum_\delta U_{\alpha\delta}\left\langle\!\!\left\langle\hat n_\delta d_\alpha \hat n_\gamma;
d^\dag_\beta\right\rangle\!\!\right\rangle^R
+\sum_{ik\sigma}V^*_{ik\sigma,\alpha}\left\langle\!\!\left\langle c_{ik\sigma}n_\gamma;
d^\dag_\beta\right\rangle\!\!\right\rangle^R+  \nonumber \\
+ &\! \sum_{ik\sigma}V^*_{ik\sigma,\gamma}\left\langle\!\!\left\langle d_\alpha d^\dag_\gamma c_{ik\sigma};
d^\dag_\beta\right\rangle\!\!\right\rangle^R
\!\!-\!\sum_{ik\sigma}V_{ik\sigma,\gamma}\left\langle\!\!\left\langle d_\alpha c^\dag_{ik\sigma}d_\gamma;
d^\dag_\beta\right\rangle\!\!\right\rangle^R \nonumber \\
+ &\! \sum_\delta \tilde\epsilon_{\gamma\delta}\left\langle\!\!\left\langle d_\alpha d^\dag_\gamma d_\delta;
d^\dag_\beta\right\rangle\!\!\right\rangle^R
\!\!-\!\sum_\delta \tilde\epsilon_{\delta\gamma}\left\langle\!\!\left\langle d_\alpha d^\dag_\delta d_\gamma;
d^\dag_\beta\right\rangle\!\!\right\rangle^R.
\end{align}

The equation (\ref{G2_1}) is not closed again and produces new Green functions of
higher order. And so on. These sequence of equations can not be closed in the
general case and should be truncated at some point. Below we consider some possible
approximations. The other important point is, that average populations and lesser
Green functions should be calculated self-consistently. In equilibrium (linear
response) these functions are easy related to the spectral functions. But at finite
voltage it should be calculated independently.

\paragraph*{(iv) Kinetic (lesser) function}

Following the same way, as for the retarded functions (using only the definitions of
NGF and Heisenberg equations of motion) one derives instead of
(\ref{R1})-(\ref{G2_1})
\begin{equation}\label{L1}
\epsilon G^<_{\alpha\beta}-\sum_\gamma\tilde\epsilon_{\alpha\gamma}
G^<_{\gamma\beta}=
\sum_{\gamma\neq\alpha}U_{\alpha\gamma}G^{(2)<}_{\alpha\gamma,\beta}+
\sum_{ik\sigma}V^*_{ik\sigma,\alpha}G^<_{ik\sigma,\beta},
\end{equation}
\begin{equation}\label{L2}
(\epsilon-\tilde\epsilon_{ik\sigma})G^<_{ik\sigma,\beta}= \sum_\delta
V_{ik\sigma,\delta}G^<_{\delta\beta},
\end{equation}
\begin{align}
\label{L3}
\epsilon G^{(2)<}_{\alpha\gamma,\beta}-\sum_\delta\tilde\epsilon_{\alpha\delta}
G^{(2)<}_{\delta\gamma,\beta}=
\sum_{\delta\neq\alpha}U_{\alpha\delta}\left\langle\!\!\left\langle\hat n_\delta d_\alpha \hat n_\gamma;
d^\dag_\beta\right\rangle\!\!\right\rangle^<+ \nonumber \\
+\sum_{ik\sigma}V^*_{ik\sigma,\alpha}\left\langle\!\!\left\langle c_{ik\sigma}n_\gamma;
d^\dag_\beta\right\rangle\!\!\right\rangle^<
+\sum_{ik\sigma}V^*_{ik\sigma,\gamma}\left\langle\!\!\left\langle d_\alpha d^\dag_\gamma c_{ik\sigma};
d^\dag_\beta\right\rangle\!\!\right\rangle^< \nonumber \\
-\sum_{ik\sigma}V_{ik\sigma,\gamma}\left\langle\!\!\left\langle d_\alpha c^\dag_{ik\sigma}d_\gamma;
d^\dag_\beta\right\rangle\!\!\right\rangle^< \nonumber \\
+\sum_\delta \tilde\epsilon_{\gamma\delta}\left\langle\!\!\left\langle d_\alpha d^\dag_\gamma d_\delta;
d^\dag_\beta\right\rangle\!\!\right\rangle^<
\!\!-\!\sum_\delta \tilde\epsilon_{\delta\gamma}\left\langle\!\!\left\langle d_\alpha d^\dag_\delta d_\gamma;
d^\dag_\beta\right\rangle\!\!\right\rangle^<.
\end{align}

To find $G^<_{ik\sigma,\beta}$ we should divide the right parts by
$(\epsilon-\tilde\epsilon_{ik\sigma})$, which is not well defined at
$\epsilon=\tilde\epsilon_{ik\sigma}$. In the section~{\bf 3} we considered the
general prescription to avoid this problem, we use the equation (\ref{Lfd}), and
instead of (\ref{L2})we obtain
\begin{equation}\label{L2A}
G^<_{ik\sigma,\beta}= g^R_{ik\sigma}\sum_\delta
V_{ik\sigma,\delta}G^<_{\delta\beta}+ g^<_{ik\sigma}\sum_\delta
V_{ik\sigma,\delta}G^A_{\delta\beta}.
\end{equation}
The equations (\ref{L1}) and (\ref{L3}) can be used without modifications because
they include the imaginary parts (dissipation) from the lead terms.

At this point we stop the general consideration, and introduce a powerful
\emph{Ansatz} for the NGF which is related both to the equation-of-motion (EOM)
method and to the Dyson equation approach.~\cite{Song07prb} From the knowledge of
the Green functions we then calculate the transport observables. For clarity, we
first describe our method in the more familiar problem of a single site junction,
which is the well-known Anderson impurity model. Then we apply it to a double
quantum dot. The equations obtained below by the heuristic mapping method can be
obtained straightforward from the general NEOM equations derived in this section
using the same approximations as in the mapping method.

\subsubsection{Anderson impurity model (single site)}

The Anderson impurity model is used to describe the Coulomb interaction on a single
site:
\beann H=H_{\textrm{D}}+\sum_{\alpha}(H_{\alpha}+H_{\alpha \textrm{D}}),
\eeann where \beann
&&H^{\phantom{\dagger}}_{\textrm{D}}=\sum_{\sigma}\epsilon^{\phantom{\dagger}}_{\sigma}
d^{\dagger}_{\sigma}d^{\phantom{\dagger}}_{\sigma} +\frac{1}{2}U
n_{\sigma}n_{\bar{\sigma}},\\
&&H^{\phantom{\dagger}}_{\alpha}=\sum_{k,\sigma}\epsilon^{\alpha}_{k,\sigma}
c^{\dagger}_{\alpha,k,\sigma}c^{\phantom{\dagger}}_{\alpha,k,\sigma},\\
&&H^{\phantom{\dagger}}_{\alpha \textrm{D}
}=\sum_{k,\sigma}\left(V^{\phantom{\dagger}}_{\alpha,k,\sigma}
c^{\dagger}_{\alpha,k,\sigma}d^{\phantom{\dagger}}_{\sigma}
+V^{*\phantom{\dagger}}_{\alpha,k,\sigma}d^{\dagger}_{\sigma}c^{\phantom{\dagger}}_{\alpha,k,\sigma}\right),
\eeann
where $d$ and $c$ are the operators for electrons on the dot and on the left
($\alpha=\textrm{L}$) and the right ($\alpha=\textrm{R}$) lead, $U$ is the Coulomb
interaction parameter, $\epsilon_{\sigma}$ is the $\sigma$ level of the quantum dot,
while $\epsilon^{\alpha}_{k,\sigma}$ is the spin $\sigma$ level of lead $\alpha$ in
$k$ space, $\sigma=\uparrow,\downarrow$. With the help of the EOM and the truncation
approximation, we can get a closed set of equations for the retarded and advanced
GFs $G^{\textrm{r}/\textrm{a}}_{\sigma,\tau}$~\cite{Meir91prl,Haug96book}
\begin{subequations}
\begin{align}
\label{eq-GF-SS-1} &(\omega-\epsilon_{\sigma}-\Sigma^{\textrm{r}/\textrm{a}}_{\sigma})
G^{\textrm{r}/\textrm{a}}_{\sigma,\tau}
=\delta_{\sigma,\tau}+UG^{(2)\textrm{r}/\textrm{a}}_{\sigma,\tau},\\ \label{eq-GF-SS-2}
&(\omega-\epsilon_{\sigma}-U-\Sigma^{\textrm{r}/\textrm{a}}_{\sigma})
G^{(2)\textrm{r}/\textrm{a}}_{\sigma,\tau} =\langle n_{\bar{\sigma}}\rangle
\delta_{\sigma,\tau},
\end{align}
\end{subequations}
where $G^{\textrm{r}/\textrm{a}}_{\sigma,\tau}=\langle\langle d_{\sigma}
|d^{\dagger}_{\tau}\rangle\rangle^{\textrm{r}/\textrm{a}}$,
$G^{(2)\textrm{r}/\textrm{a}}_{\sigma,\tau}=\langle\langle
n_{\bar{\sigma}}d_{\sigma}|d^{\dagger}_{\tau}\rangle\rangle^{\textrm{r}/\textrm{a}}$ and
\bea\label{eq-SS-self-energy} \Sigma^{\textrm{r}/\textrm{a}}_{\sigma}(\omega)
=\Sigma^{\textrm{r}/\textrm{a}}_{\textrm{L},\sigma}
+\Sigma^{\textrm{r}/\textrm{a}}_{\textrm{R},\sigma}
=\sum_{\alpha,k}{\frac{|V_{\alpha,k,\sigma}|^{2}} {\omega-\epsilon^{\alpha}_{k,\sigma}\pm
\textrm{i}0^{+}}} \eea are the electron self-energies.

\paragraph*{(i) Mapping on retarded Green functions}

For retarded GFs, from the EOM method, and with the help of Eqs.~(\ref{eq-GF-SS-1})
and (\ref{eq-GF-SS-2}), we can get
\beann
G^{\textrm{r}}&=&G^{\textrm{r}}_{0}+G^{\textrm{r}}_{0}UG^{(2)\textrm{r}} 
=G^{\textrm{r}}_{0}+G^{\textrm{r}}_{0}\Sigma^{\textrm{EOM}}G^{(1)\textrm{r}},
\eeann
where $G^{\textrm{r}}$ is single-particle GF matrix \beann
G^{\textrm{r}}=\begin{pmatrix}
G^{\textrm{r}}_{\uparrow,\uparrow} & G^{\textrm{r}}_{\uparrow,\downarrow} \\
G^{\textrm{r}}_{\downarrow,\uparrow} &
G^{\textrm{r}}_{\downarrow,\downarrow}
\end{pmatrix},
\eeann
and
$G^{(1)\textrm{r}}_{\sigma,\tau}=G^{(2)\textrm{r}}_{\sigma,\tau}/\langle
n_{\bar{\sigma}}\rangle$. $G^{\textrm{r}}_{0}$ describes the
single-particle spectrum without Coulomb interaction, but
including the effects from the electrodes.
$\Sigma^{\textrm{EOM}}_{\sigma,\tau}=U\langle
n_{\bar{\sigma}}\rangle$ is the Hartree-like self-energy of our
model. Since there is only Coulomb interaction on the site with
the levels $\epsilon_{\sigma}$, the Fock-like self-energy is
vanishing.

Alternatively, by means of the Dyson equation and the second-order
truncation approximation, taking Hartree-like self-energies
$\Sigma^{\textrm{H}}_{\sigma,\tau}=U\langle
n_{\bar{\sigma}}\rangle~(=\Sigma^{\textrm{EOM}}_{\sigma,\tau})$, we can
also get the retarded GFs as follows
\bea
\label{eq-Dyson-SS-1}
G^{\textrm{r}}=G^{\textrm{r}}_{0}+G^{\textrm{r}}_{0}\Sigma^{\textrm{H}}G^{\textrm{r}}_{1},
\eea where
$G^{\textrm{r}}_{1}=G^{\textrm{r}}_{0}+G^{\textrm{r}}_{0}\Sigma^{\textrm{H}}G^{\textrm{r}}_{0}$
is the first-order truncation GF.

Within the level of the second-order truncation approximation, we
see that there is a map between the EOM results and the Dyson
results:
\begin{subequations}
\label{eq-mapping-SS}
\begin{align}
\label{eq-a}
&&G^{\textrm{r}}~=~G^{\textrm{r}}_{0}~+~G^{\textrm{r}}_{0}~
\Sigma^{\textrm{H}}~G^{(1)\textrm{r}}\hspace{1cm}\textrm{(EOM)},\\
&&\updownarrow\hspace{3.1cm}\updownarrow\hspace{2.7cm}\nonumber\\
\label{eq-b} &&G^{\textrm{r}}~=~G^{\textrm{r}}_{0}~+~
G^{\textrm{r}}_{0}~\Sigma^{\textrm{H}}~G^{\textrm{r}}_{1}~\hspace{1cm}\textrm{(Dyson)}.
\end{align}
\end{subequations}
Eqs.~(\ref{eq-mapping-SS}) prompts a way to include further
many-particle effects into the Dyson equation, Eq.~(\ref{eq-b}),
by replacing the \textit{Dyson-first-order} retarded Green
function $G^{\textrm{r}}_{1}$ with the EOM $G^{(1)r}$.
Then one obtains already the correct results to describe CB while
keeping the Hartree-like self-energy.

\paragraph*{(ii) Mapping on contour and lesser Green functions}

Introducing now the contour GF $\check{G}$, we can get the Dyson equation as
follows~\cite{Kadanoff62book,Keldysh64,Rammer86RMP,Haug96book} \bea
\label{eq-Dyson-SS-2} \check{G}=\check{G}_{0}+\check{G}_{0}\check{\Sigma}\check{G},
\eea where $\check{\Sigma}$ is the self-energy matrix. \cite{Haug96book}

According to the approximation for the retarded GF in
Eq.~(\ref{eq-Dyson-SS-1}), we take the second-order truncation on
Eq.~(\ref{eq-Dyson-SS-2}), and then get \beann
\check{G}=\check{G}_{0}+\check{G}_{0}\check{\Sigma}^{\textrm{H}}\check{G}_{1},
\eeann where
$\check{G}_{1}=\check{G}_{0}+\check{G}_{0}\check{\Sigma}^{\textrm{H}}\check{G}_{0}$
is the first-order contour GF, and $\check{G}_{0}$ has already
included the lead broadening effects.

Similar to the mapping in Eq.~(\ref{eq-mapping-SS}), we perform an
\textit{Ansatz} consisting in substituting the
\textit{Dyson-first-order} $G^{\textrm{r}/\textrm{a}/<}_{1}$ with
the EOM one $G^{(1)\textrm{r}/\textrm{a}/<}$ to consider
more many-particle correlations, while the EOM
self-energy is used for the \textit{Dyson} equation for
consistency: \bea \label{eq:mapping}
\begin{matrix}
\check{G} & = & \check{G}_{0} & + &
\check{G}_{0}&\check{\Sigma}^{\textrm{H}}&\check{G}_{1}
& & & \textrm{(Dyson)},\\
\updownarrow& & & & & &\uparrow & & &\\
\check{G}& &&&&&\check{G}^{(1)} & & & \textrm{(EOM)}.
\end{matrix}
\eea Then, using the Langreth theorem \cite{Haug96book} we get the lesser GF,
\bea \label{eq:Gkeldysh-SS-new}
G^{<}&=&G^{<}_{0}+G^{\textrm{r}}_{0}\Sigma^{\textrm{H},\textrm{r}}G^{(1)<}
+G^{<}_{0}\Sigma^{\textrm{H},\textrm{a}}G^{(1)\textrm{a}}\nonumber\\
&=&G^{<}_{0}+G^{\textrm{r}}_{0}UG^{(2)<}
+G^{<}_{0}UG^{(2)\textrm{a}} \eea where
$G^{\textrm{r}/\textrm{a}/<}_{0}$ are GFs for $U=0$, \textit{but}
including the lead broadening effects, \textit{i.e.} \beann
&&G^{<}_{0}=g^{<}_{0}+g^{\textrm{r}}_{0}\Sigma^{<}G^{\textrm{a}}_{0}
+g^{<}_{0}\Sigma^{\textrm{a}}G^{\textrm{a}}_{0}
+g^{\textrm{r}}_{0}\Sigma^{\textrm{r}}G^{<}_{0},\\
&&G^{\textrm{r}/\textrm{a}}_{0}=g^{\textrm{r}/\textrm{a}}_{0}
+g^{\textrm{r}/\textrm{a}}_{0}\Sigma^{\textrm{r}/\textrm{a}}G^{\textrm{r}/\textrm{a}}_{0}
, \eeann
with $g^{\textrm{r}/\textrm{a}/<}_{0}$ the free electron GF, and
\beann
\Sigma^{\textrm{r}/\textrm{a}/<}=
\begin{pmatrix}
\Sigma^{\textrm{r}/\textrm{a}/<}_{\uparrow} & 0 \\
0 & \Sigma^{\textrm{r}/\textrm{a}/<}_{\downarrow}
\end{pmatrix},
\eeann
$\Sigma^{<}_{\sigma}=\textrm{i}\sum_{\alpha}{\Gamma_{\alpha}
f_{\alpha}(\omega)}$, and
$\Gamma_{\alpha}=\textrm{i}(\Sigma^{\textrm{r}}_{\alpha}-\Sigma^{\textrm{a}}_{\alpha})$,
$f_{\alpha}(\omega)=f(\omega-\mu_{\alpha})$, $f$ is the
equilibrium Fermi function and $\mu_{\alpha}$ is the
electro-chemical potential in lead $\alpha$;
$\Sigma^{\textrm{r}/\textrm{a}}_{\alpha}$ are the
retarded/advanced electron self-energies from
Eq.~(\ref{eq-SS-self-energy}) and
$G^{(1)\textrm{r}/\textrm{a}/<}_{\sigma,\tau}=G^{(2)\textrm{r}/\textrm{a}/<}_{\sigma,\tau}
/\langle n_{\bar{\sigma}}\rangle$. Performing the same
\textit{Ansatz} on the double-particle GF, from
Eq.~(\ref{eq-GF-SS-2}) we can get \bea \label{GF2-SS}
G^{(2)<}=G^{(2)\textrm{r}}\Sigma^{(2)<}G^{(2)\textrm{a}}, \eea
with $\Sigma^{(2)<}_{\sigma}=\Sigma^{<}_{\sigma}/\langle
n_{\bar{\sigma}}\rangle$.

The lesser GFs in Eq.~(\ref{eq:Gkeldysh-SS-new}) can also be obtained directly from
the general formula~\cite{Haug96book}
\beann {G}^{<}={G}^{<}_{0}+{G}^{\textrm{r}}_{0} {\Sigma}^{\textrm{r}}{G}^{<}
+{G}^{\textrm{r}}_{0}{\Sigma}^{<}{G}^{\textrm{a}}
+{G}^{<}_{0}{\Sigma}^{\textrm{a}}{G}^{\textrm{a}}, \eeann with the help of the
\textit{Ansatz} in Eq.~(\ref{eq:mapping}). It should be noted that
Eq.~(\ref{eq:Gkeldysh-SS-new}) is very different from the lesser GF formula,
\bea\label{widely-used-LGF} G^{<}=G^{\textrm{r}}\Sigma^{<}G^{\textrm{a}},\eea
with the self-energy $\Sigma^{<}$ containing \textit{only} contributions from the
electrodes. The equation~(\ref{widely-used-LGF}) is widely used for both
first-principle~\cite{Brandbyge02prb,Rocha06prb,Pecchia04repprogphys} and model
Hamiltonian calculations.~\cite{Pals96jpcm}

\begin{figure}[t]
\begin{center}
\psfig{file=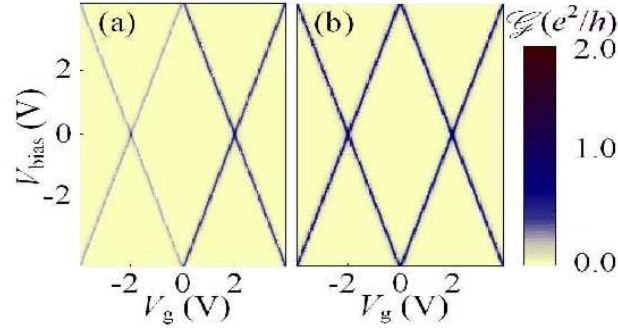,width=0.7\columnwidth}
\caption{\label{stab-diagram-DS}(Color) The stability diagram of a SSJ with
$\epsilon_{\sigma}=2.0~$eV, \mbox{$U=4.0~$eV}, \mbox{$\Gamma_{\textrm{L}}
=\Gamma_{\textrm{R}}=0.05~$eV}. (a) The uncorrect result obtained by means of the
widely used formula in Eq.~(\ref{widely-used-LGF}) for the lesser GF is not
symmetric for levels $\epsilon_{\sigma}$ and $\epsilon_{\sigma}+U$. (b) Results
obtained by means of our \textit{Ansatz} in Eq.~(\ref{eq:Gkeldysh-SS-new}) shows
correctly symmetric for levels $\epsilon_{\sigma}$ and $\epsilon_{\sigma}+U$.}
\end{center}
\end{figure}

The numerical calculation results of conductance dependence on the bias and gate
voltages by the two different NGF Eqs.~(\ref{eq:Gkeldysh-SS-new}) and
(\ref{widely-used-LGF}) are shown in Fig.~\ref{stab-diagram-DS}. As we can see in
the left panel, the adoption of Eq.~(\ref{widely-used-LGF}) results in an
incorrectly symmetry-breaking in the gate potential. This wrong behavior is
corrected in the right panel where Eq.~(\ref{eq:Gkeldysh-SS-new}) has been used.

Note, that the expressions for the retarded and lesser functions, described above,
can be obtained in a more formal way by the EOM method formulated on the Keldysh
contour.

\paragraph*{(iii) Comparison with the master equation result}

In the single site model with two (spin-up and spin-down) levels it is possible to make
the direct comparison between our \emph{Ansatz} and the master equation methods.
For the latter, we used the well known master equations for quantum dots
\cite{Beenakker91prb,Averin91prb}.

\begin{figure}
\begin{center}
\psfig{file=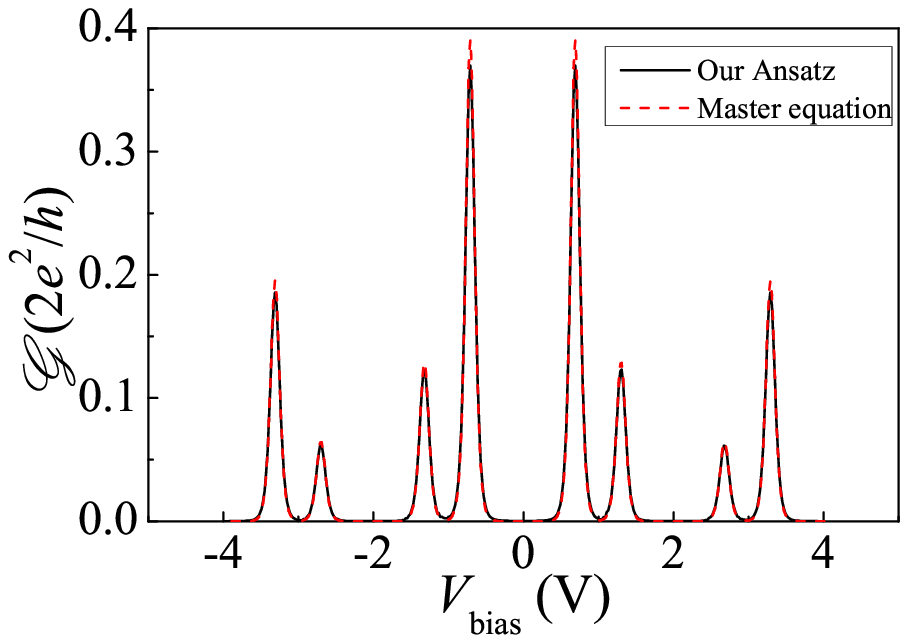, width=0.65\columnwidth}
\caption{(Color) The
comparison of the master equation method and our \emph{Ansatz} for the differential
conductance of the two level model with \mbox{$\epsilon_{\uparrow}=-0.35~$eV},
\mbox{$\epsilon_{\downarrow}=-0.65~$eV}, \mbox{$U=1.0~$eV},
\mbox{$V_{\textrm{g}}=1.0~$V},
\mbox{$\Gamma_{\textrm{L}}=\Gamma_{\textrm{R}}=0.05~$eV}.} \label{fig-1x}
\end{center}
\end{figure}

\begin{figure}
\begin{center}
\psfig{file=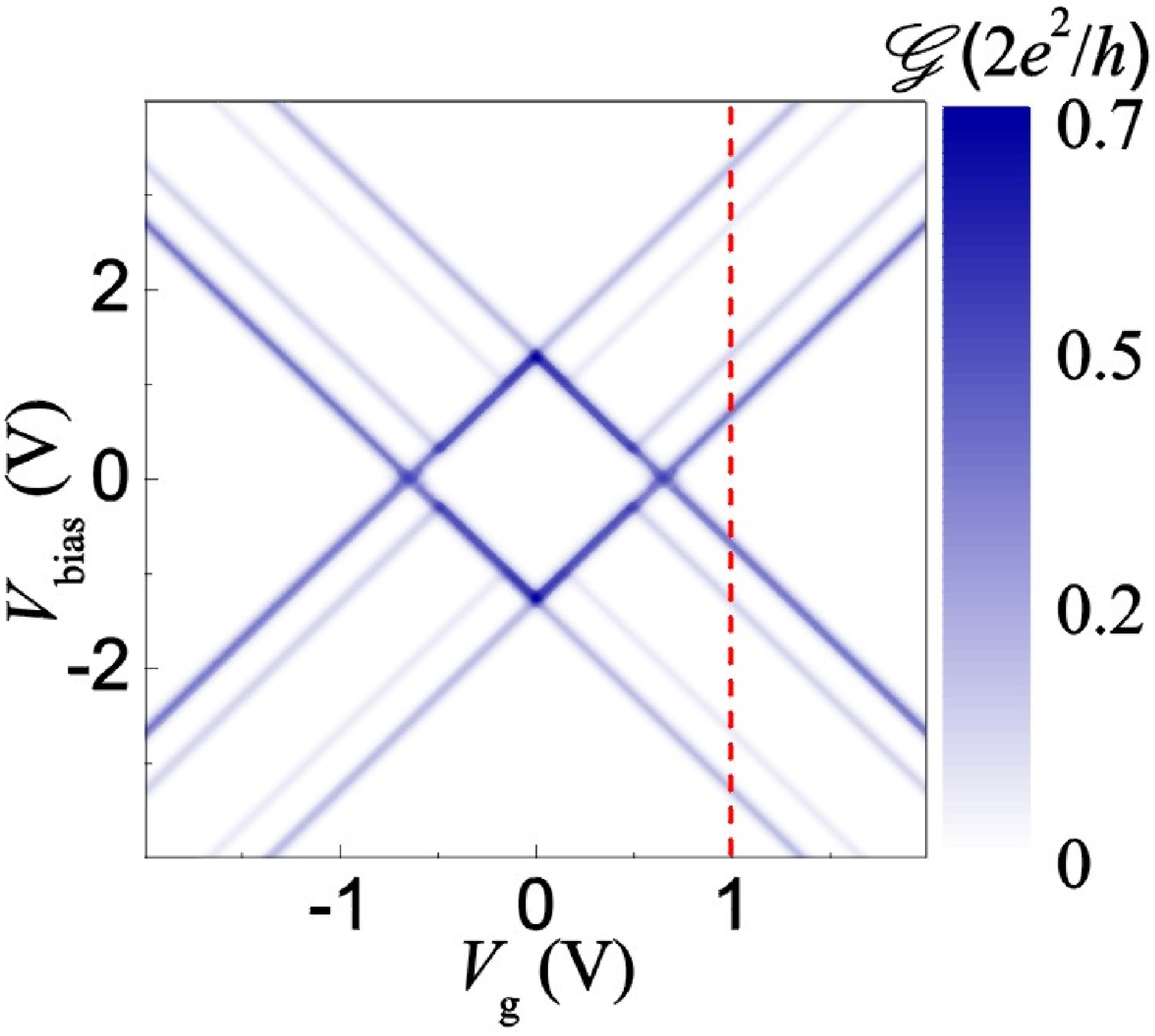, width=0.63\columnwidth} \caption{(Color) The
stability diagram (the contour plot of the differential conductance) calculated by
our \emph{Ansatz} for the two level model with parameters as in Fig.~\ref{fig-1x}.
The latter is indicated with a dash line at \mbox{$V_{\textrm{g}}=1.0~$V}.
}
\label{prb-3x}
\end{center}
\end{figure}

In the Fig.\,\ref{fig-1x} the typical curves of the differential conductance as a
function of the bias voltage at fixed gate voltage obtained by the two methods are
shown together: there is basically no difference in the results obtained by these
two methods. In the Fig.\,\ref{prb-3x} the contour plot of the differential
conductance obtained by our \emph{Ansatz} is shown. We do not present here the
contour plot obtained by the master equation method because it looks exactly the
same.

It is quite clear from the presented figures that our \emph{Ansatz} and the master
equation method give essentially the same results in the limit of weak coupling to
the leads. The systematic investigation of the deviations between the two methods at
stronger tunneling will be presented in a separate publication.

It is important that our \emph{Ansatz} can be applied straightforwardly to
multilevel systems in the case when the exact eigenstates of an isolated system are
unknown and the usage of the master equation method is not easy. In this paper we
consider the simplest example of such a system, namely a double site case.

\subsubsection{Double quantum dot (two sites)}

\begin{figure}[b]
\begin{center}
\epsfxsize=0.6\hsize \epsfbox{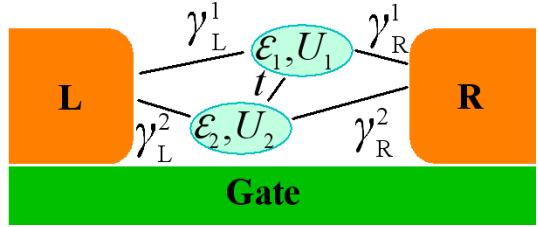} \caption{(Color) The general
configuration of a double site junction. The levels $\epsilon_{1,2}$ with charging
energies $U_{1,2}$ are connected via $t$ and coupled to the electrodes via the
linewidth injection rates $\gamma^{i}_{\alpha}$.} \label{gnrl-DQD}
\end{center}
\end{figure}

We now return to the investigation of the \DSJ\ system (Fig.~\ref{gnrl-DQD}) with Coulomb
interaction on each site. The Hamiltonian is expressed as follows,
\beann
H=H_{\textrm{D}}+H_{t}+\sum_{\alpha}(H_{\alpha}+H_{\alpha
\textrm{D}}),
\eeann
where
\beann
& \displaystyle
H^{\phantom{\dagger}}_{\textrm{D}}=\sum_{i,\sigma}\epsilon^{\phantom{\dagger}}_{i,\sigma}d^{\dagger}_{i,\sigma}d^{\phantom{\dagger}}_{i,\sigma}
+\frac{U_{i}}{2}n^{\phantom{\dagger}}_{i,\sigma}n^{\phantom{\dagger}}_{i,\bar{\sigma}}, \\
& \displaystyle
H^{\phantom{\dagger}}_{t}=\sum_{i\neq
j,\sigma}\frac{t}{2}(d^{\dagger}_{i,\sigma}d^{\phantom{\dagger}}_{j,\sigma}
+d^{\dagger}_{j,\sigma}d^{\phantom{\dagger}}_{i,\sigma}), \\
& \displaystyle
H^{\phantom{\dagger}}_{\alpha,\sigma}=\sum_{k,\sigma}\epsilon^{(\alpha)}_{k,\sigma}
c^{\dagger}_{\alpha,k,\sigma}c^{\phantom{\dagger}}_{\alpha,k,\sigma}, \\
& \displaystyle
H^{\phantom{\dagger}}_{\alpha \textrm{D},
\sigma}=\sum_{k,\sigma}\left(V^{\phantom{\dagger}}_{\alpha,k,\sigma}
c^{\dagger}_{\alpha,k,\sigma}d^{\phantom{\dagger}}_{i,\sigma}
+V^{*\phantom{\dagger}}_{\alpha,k,\sigma}d^{\dagger}_{i,\sigma}
c^{\phantom{\dagger}}_{\alpha,k,\sigma}\right),
\eeann
with $i,j=1,2$ indicate the site, $t$ is the constant for electron hopping between
different sites.

With the help of the EOM, and by means of the truncation approximation on the
double-particle GFs, we obtain the closed form for the retarded GFs as follows
\begin{subequations}
\begin{align}
\label{EoM-DS-1}
(\omega-\epsilon^{\phantom{\dagger}}_{i,\sigma}-\Sigma^{\textrm{r}}_{i,\sigma})
G^{(U,t)\textrm{r}}_{i,\sigma;j,\tau} 
=\delta^{\phantom{\dagger}}_{i,j}\delta^{\phantom{\dagger}}_{\sigma,\tau}
+U_{i}G^{(2)(U,t)\textrm{r}}_{i,\sigma;j,\tau}
+t~G^{(U,t)\textrm{r}}_{i,\sigma;j,\tau},\\ \label{EoM-DS-2}
(\omega-\epsilon^{\phantom{\dagger}}_{i,\sigma}-U_{i}-\Sigma^{\textrm{r}}_{i,\sigma})
G^{(2)(U,t)\textrm{r}}_{i,\sigma;j,\tau} 
=\langle
n^{\phantom{\dagger}}_{i,\bar{\sigma}}\rangle\delta^{\phantom{\dagger}}_{i,j}\delta^{\phantom{\dagger}}_{\sigma,\tau}+t~
n^{\phantom{\dagger}}_{i,\bar{\sigma}}G^{(U,t)\textrm{r}}_{i,\sigma;j,\tau},
\end{align}
\end{subequations}
where the DSJ retarded GFs are defined as
\begin{eqnarray}
& G^{(U,t)\textrm{r}}_{i,j;\sigma,\tau}=\langle\langle d^{\phantom{\dagger}}_{i,\sigma}|
d^{\dagger}_{j,\tau}\rangle\rangle^{\textrm{r}}, \\
& G^{(2)(U,t)\textrm{r}}_{i,j;\sigma,\tau}=\langle\langle
n^{\phantom{\dagger}}_{i,\bar{\sigma}}d^{\phantom{\dagger}}
_{i,\sigma}|d^{\dagger}_{j,\tau}\rangle\rangle^{\textrm{r}}.
\end{eqnarray}
Here $\bar{i}$ means `NOT $i$', and $\Sigma^{\textrm{r}}_{i, \sigma}$ are the electron
self-energy from leads.

From Eqs.~(\ref{EoM-DS-1}), (\ref{EoM-DS-2}) and performing the same \textit{Ansatz}
as in the case of \SSJ, we can obtain the \DSJ\ lesser GFs with Coulomb-interaction
effects as follows
\bea \label{eq:Gkeldysh-DS-new}
G^{(U,t)<}(\omega)=(1+G^{(U,t)\textrm{r}}\Sigma^{\textrm{r}}_{t}) G^{(U)<}
(1+\Sigma^{\textrm{a}}_{t}G^{(U,t)\textrm{a}})
+G^{(U,t)\textrm{r}}\Sigma^{<}_{t}G^{(U,t)\textrm{a}},
\eea
with
\beann \Sigma^{\textrm{r}}_{t}=\Sigma^{\textrm{a}}_{t}=
\begin{pmatrix}
0 & t & 0 & 0\\
t & 0 & 0 & 0\\
0 & 0 & 0 & t\\
0 & 0 & t & 0
\end{pmatrix},
\eeann and $\Sigma^{<}_{t}=0$. $G^{(U)<}$ is the DSJ lesser GF
with the same form as Eq.~(\ref{eq:Gkeldysh-SS-new}), but taking
\begin{equation}
U=\begin{pmatrix}
U_{1} & 0 & 0 & 0\\
0 & U_{2} & 0 & 0\\
0 & 0 & U_{1} & 0\\
0 & 0 & 0 & U_{2}
\end{pmatrix}, \
\Gamma_{\alpha}=
\begin{pmatrix}
\gamma^{1}_{\alpha} & 0 & 0 & 0\\
0 & \gamma^{2}_{\alpha} & 0 & 0\\
0 & 0 & \gamma^{1}_{\alpha} & 0\\
0 & 0 & 0 & \gamma^{2}_{\alpha}
\end{pmatrix},
\end{equation}
where $\gamma^{i}_{\alpha}$ indicates the line width function of lead $\alpha$ to
site $i$, and $U_{i}$ is the charging energy at site $i$.
$G^{\textrm{r}/\textrm{a}}$ and $G^{(2)\textrm{r}/\textrm{a}}$ are the GF matrix
from Eqs.~(\ref{EoM-DS-1}) and (\ref{EoM-DS-2}). Here, in order to distinguish
different GFs, we introduce the subscript `$(U,t)$' for the one with both Coulomb
interaction $U$ and inter-site hopping $t$, while `$(U)$' for the one only with
Coulomb interaction.

For our models, the lesser GFs in Eq.~(\ref{eq:Gkeldysh-SS-new}), (\ref{GF2-SS}) and
(\ref{eq:Gkeldysh-DS-new}), which are obtained with help of our \emph{Ansatz}, can
also be obtained by the EOM NEGF formula (\ref{Lfd}) within the same truncation
approximation.

The current can be generally written as~\cite{Meir92prl}
\beann
 J=\frac{\textrm{i}e}{2\hbar}\int{\frac{d\epsilon}{2\pi}}
\textrm{Tr}\{(\Gamma_{\textrm{L}}-\Gamma_{\textrm{R}})G^{(U,t)<}
+[f_{\textrm{L}}(\omega)\Gamma_{\textrm{L}}
-f_{\textrm{R}}(\omega)\Gamma_{\textrm{R}}](G^{(U,t)\textrm{r}}-
G^{(U,t)\textrm{a}})\},
\eeann
where the lesser GF is given by
Eq.~(\ref{eq:Gkeldysh-DS-new}). The differential conductance is defined as \beann
\mathcal{G}=\frac{\partial J}{\partial V_{\textrm{bias}}}, \eeann where the bias
voltage is defined as $V_{\textrm{bias}}=(\mu_{\textrm{R}}-\mu_{\textrm{L}})/e$.

\paragraph*{(i) Serial configuration}

\begin{figure}[t]
\vskip 0.1cm
\centerline{\psfig{file=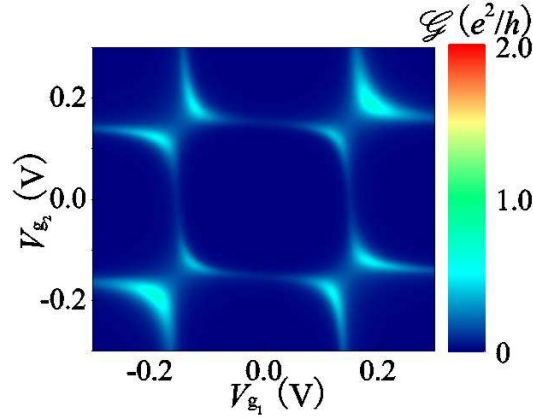,width=0.6 \columnwidth}}
\caption{\label{stab-diagram-DS2}(Color) The stability diagram of a serial
\DSJ~with $\epsilon_{1,\sigma}=\epsilon_{2,\sigma}=-0.15~$eV, $U_{1}=U_{2}=0.3~$eV,
$t=0.05~$eV, $\gamma^{1}_{\textrm{L}}=\gamma^{2}_{\textrm{R}}=0.02~$eV,
$\gamma^{2}_{\textrm{L}}=\gamma^{1}_{\textrm{R}}=0$ ,$ V_{\textrm{bias}}=0.005V$.
The maximums of conductance are observed when the levels of the first site
($\epsilon_{1,\sigma}$ or $\epsilon_{1,\sigma}+U$) are overlapped with the levels of
the second site ($\epsilon_{2,\sigma}$ or $\epsilon_{2,\sigma}+U$), and with the
Fermi energy in the leads. The splitting of the four maximums is due to the hopping
between the dots.}
\end{figure}

By taking $\gamma^{2}_{L}=\gamma^{1}_{R}=0$, we obtain a serial DSJ, which could
describe the kind of molecular quantum junctions like the ones studied in
Ref.~\cite{Elbing05pnas}. First, at small bias voltages, the conductance with the
two gate voltages $V_{\textrm{g}_{1}}$ and $V_{\textrm{g}_{2}}$ was calculated, and
the relative stability diagram was obtained as shown in Fig.~\ref{stab-diagram-DS2}.
Because of the double degeneracy (spin-up and spin-down) considered for each site
and electrons hopping between the dots, there are eight resonance-tunnelling
regions. This result is consistent with the master-equation
approach.~\cite{vanderWiel02rmp}

\begin{figure}
\centerline{\psfig{file=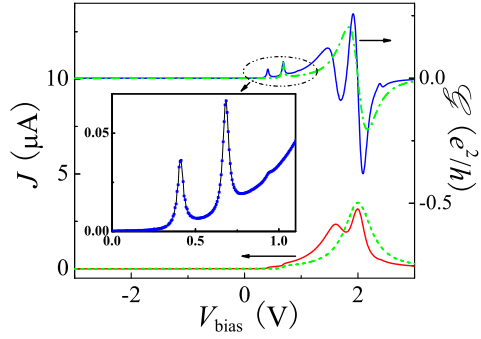,width=0.6\columnwidth}}
\caption{\label{stab-diagram-DS3}(Color) Current and conductance
vs.~bias-voltage of a DSJ far from equilibrium with parameters
$\epsilon_{1,\sigma}=0.5~$eV, $\epsilon_{2,\sigma}=-0.5~$eV, $U_{1}=U_{2}=U=0.2~$eV,
$t=0.07~$eV, $\gamma^{1}_{\textrm{L}}=\gamma^{2}_{\textrm{R}}=0.03~$eV,
$V_{\textrm{g}_{2}}=-V_{\textrm{g}_{1}}=V_{\textrm{bias}}/4$ and
$V_{\textrm{R}}=-V_{\textrm{L}}=V_{\textrm{bias}}/2$. The red curve represents the
current, while the blue the conductance. The inset is the blow-up for the
conductance peak split. The dash and dot-dash curves are for current and conductance
with $U=0$, respectively.}
\end{figure}

Further, we studied the nonequilibrium current for large bias-voltages
(Fig.~\ref{stab-diagram-DS3}). Because $\epsilon_{1,\sigma}$ and $\epsilon_{2,\sigma}$
are taken as asymmetric, for the case without Coulomb interaction, the $I$-$V$ curve is
asymmetric for $\pm V_{\textrm{bias}}$, and there are one step and one maximum for the
current. The step contributes to one peak for the conductance. When we introduce the
Coulomb interaction to the system, the one conductance peak is split into several: two
peaks, one pseudo-peak and one dip, while the current maximum comes to be double split
(see Fig.~\ref{stab-diagram-DS3}). The origin of this is in the effective
splitting of the degenerate level when one of the spin states is occupied and the other
is empty. When both spin states are occupied, the degeneracy is restored.

\begin{figure}[b]
\centerline{\psfig{file=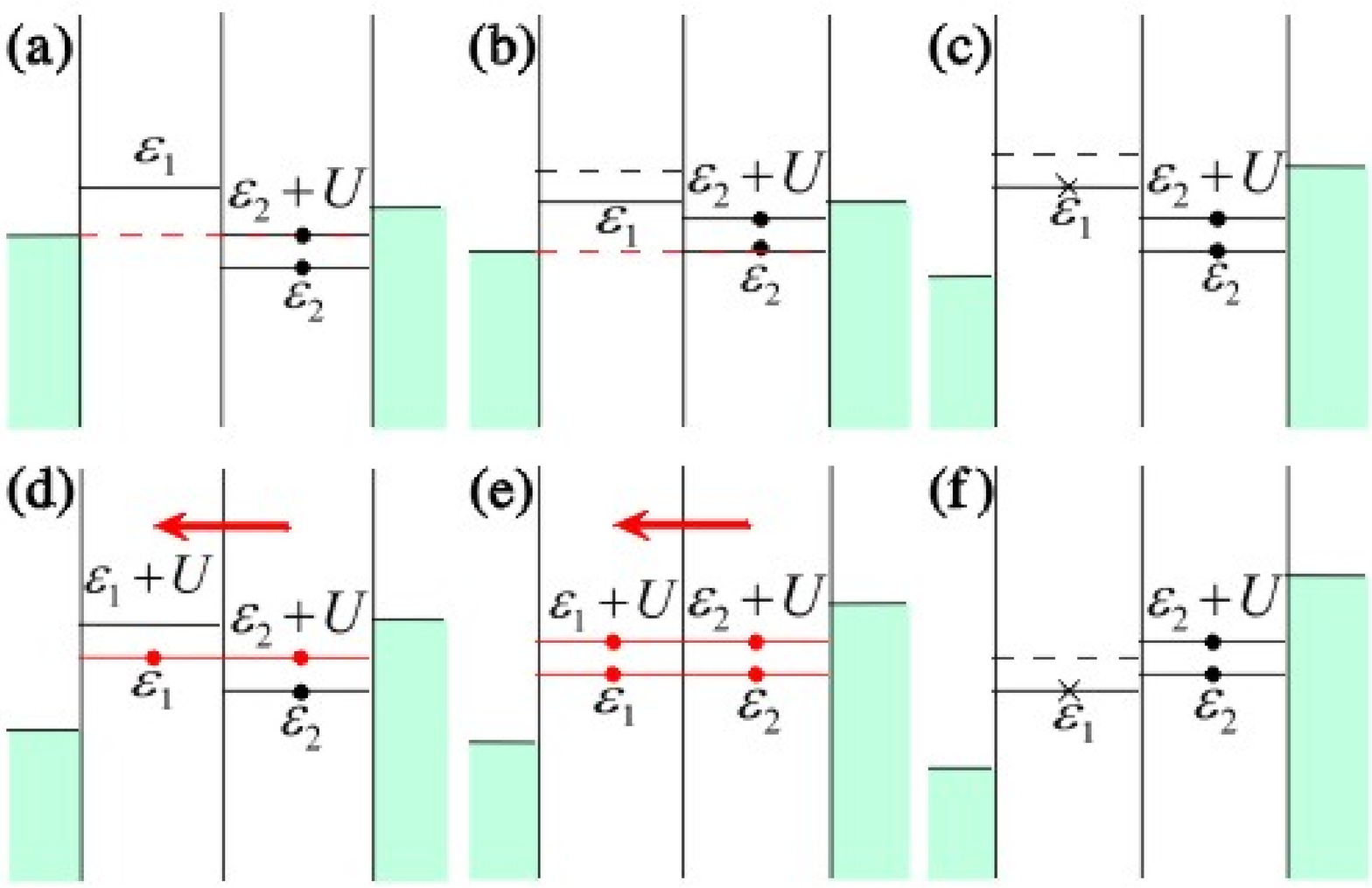,width=0.6\columnwidth}}
\caption{\label{fig-Dynamic}(Color) The processes involved in the transport
characteristics in figure \ref{stab-diagram-DS3}.
$\epsilon_{1}\equiv\epsilon_{1,\sigma}$, $\epsilon_{2}\equiv\epsilon_{2,\sigma}$,
The red line indicates electron resonant-tunnelling. a) The first conductance peak.
b) The second conductance peak. c) The pseudo-peak of conductance. d) The first
current maximum, and the red line indicates resonant tunnelling of electrons. e) The
second current maximum for electron resonant tunnelling. f) The dip of conductance.}
\end{figure}

This process can be illustrated by the help of Fig.~\ref{fig-Dynamic}. At zero
bias-voltage, $\epsilon_{2,\sigma}$ is occupied and $\epsilon_{1,\sigma}$ is empty.
Then we start to increase the bias voltage. a) The level $\epsilon_{2,\sigma}+U$ is
first opened for transport. It will contribute the first peak for conductance. b)
Further, the levels $\epsilon_{2,\sigma}$ and $\epsilon_{1,\sigma}$ come into the
transport window between the left and the right Fermi levels, resulting in the
second peak. c) When the level $\epsilon_{1,\sigma}+U$ comes into play, only a
pseudo-peak appears. This is because there is only a little possibility for
electrons to occupy the level $\epsilon_{1,\sigma}$ under positive bias voltage. d)
Levels $\epsilon_{2,\sigma}+U$ and $\epsilon_{1,\sigma}$ meet, which results in
electron resonant-tunnelling and leads to the first maximum of the current. Then a
new level $\epsilon_{1,\sigma}+U$ appears over the occupied $\epsilon_{1,\sigma}$
due to the Coulomb interaction. e) The meeting of $\epsilon_{2,\sigma}$ and
$\epsilon_{1,\sigma}$ results in electron resonant tunnelling. It means that
$\epsilon_{1,\sigma}$ will be occupied, which leads to the appearance of a new level
$\epsilon_{1,\sigma}+U$. Then $\epsilon_{2,\sigma}+U$ meets $\epsilon_{1,\sigma}+U$
and another resonant tunnelling channel is opened for electrons. The two channels
result in the second current maximum. f) finally, the level $\epsilon_{1,\sigma}+U$
disappears if the level $\epsilon_{1,\sigma}$ is empty. This means that a dip
appears in the conductance.

It should be noted that the characteristics of serial DSJ in
Fig.~\ref{stab-diagram-DS3} have showed some reasonable similarities to experiments
of a single-molecule diode.~\cite{Elbing05pnas}

\paragraph*{(ii) Parallel configuration}

\begin{figure}[b]
\centerline{\psfig{file=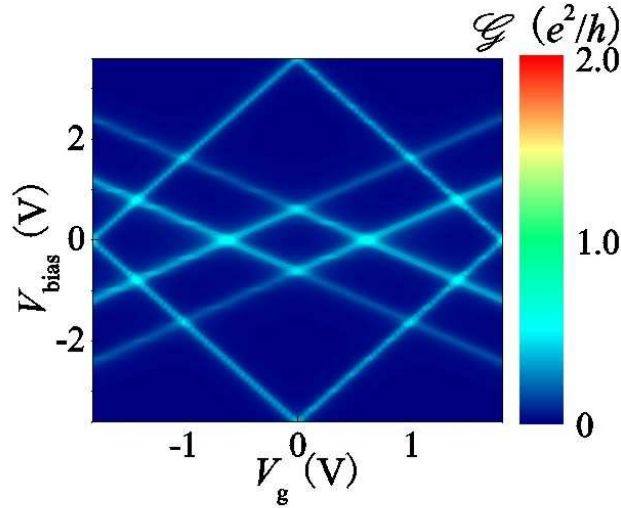, width=0.7\columnwidth}}
\caption{\label{nesting}(Color) Nested stability diagram of a parallel DSJ
with parameters $\epsilon_{1,\sigma}=-1.8~$eV, $\epsilon_{2,\sigma}=-0.3~$eV,
$U_{1}=3.6~$eV, $U_{2}=0.6~$eV $t=0.001~$eV,
$\gamma^{1}_{\textrm{L}}=\gamma^{1}_{\textrm{R}}=0.04~$eV,
$\gamma^{2}_{\textrm{L}}=\gamma^{2}_{\textrm{R}}=0.05~$eV,
$V_{\textrm{g}_{2}}=V_{\textrm{g}_{1}}/2=V_{\textrm{g}}/2$ and
$V_{\textrm{R}}=-V_{\textrm{L}}=V_{\textrm{bias}}/2$. See discussion in the text.}
\end{figure}

If on the other hand, the two sites are
symmetrically connected to the electrodes,
possibly with a small inter-dot hopping, but
with charging energies $U_{1}$ and
$U_{2}$ fixed to different scales for transport.
The resulting stability diagram contains both interference effects for parallel
pathways and an overlap of $U_{1}$ and $U_{2}$ stability diagrams,
which we refer to a nesting characteristic. (see Fig.~\ref{nesting}).

The physics of the \textit{weak} lines in the figure can be understood by the help of charging
effects. For simplicity, here we would ignore the site index $i$. In the region of large
positive gate voltage at zero bias voltage, $\epsilon_{\uparrow}$ and
$\epsilon_{\downarrow}$ are all empty, which means that the two levels are degenerate.
Therefore adding a bias voltage, first, there will be two channels ($\epsilon_{\uparrow}$
and $\epsilon_{\downarrow}$) opened for current (thick lines). After then, one level
$\epsilon_{\sigma}$ (spin-up or spin-down) is occupied, while the other obtains a shift
for Coulomb interaction: $\epsilon_{\bar{\sigma}}\rightarrow \epsilon_{\bar{\sigma}}+U$.
Therefore, when the bias voltage is further increased to make the Fermi-window boundary
meeting level $\epsilon_{\bar{\sigma}}+U$, only one channel is opened for the current,
which results in the \textit{weak} lines in Fig.~\ref{nesting}, which is the characteristic
of CB. The similar case appears in the region of large negative gate voltages.

Finally, we here introduced a powerful \textit{Ansatz} for the lesser Green
function, which is consistent with both the Dyson-equation approach and the
equation-of-motion approach. By using this \textit{Ansatz} together with the
standard equation-of-motion technique for the retarded and advanced Green functions,
we obtained the NEGF for both the single and the double site junctions in the
Coulomb blockade regime \emph{at finite voltages} and calculated the transport
observables. The method can be applied to describe self-consistently transport
through single molecules with strong Coulomb interaction and arbitrary coupling to
the leads.

To test our method, we here analyzed the CB stability diagrams for a SSJ and a DSJ.
Our results are all consistent with the results of experiments and the
master-equation approach. We showed, that the improved lesser Green function gives
better results for weak molecule-to-contact couplings, where a comparison with the
master equation approach is possible.

For the serial configuration of a DSJ, such as a donor/acceptor rectifier, the
$I$-$V$ curves maintain a diode-like behavior, as it can be already inferred by
coherent transport calculations.~\cite{Pump07ss} Besides, we predict that as a
result of charging effects, one conductance peak will be split into three peaks and
one dip, and one current maximum into two. For a DSJ parallel configuration, due to
different charging energies on the two dot sites, the stability diagrams show
peculiar nesting characteristics.

\subsection{Nonequilibrium vibrons}

Though the electron-vibron model described in the Section II has a long history, the
many questions it implies are not answered up to now. While the isolated
electron-vibron model can be solved exactly by the so-called polaron or Lang-Firsov
transformation \cite{Lang63jetp,Hewson74jjap,Mahan90book}, the coupling to the leads
produces a true many-body problem. The inelastic resonant tunneling of {\em single}
electrons through the localized state coupled to phonons was first considered in
Refs.~\cite{Glazman88jetp,Wingreen88prl,Wingreen89prb,Jonson89prb}. There, the exact
solution in the single-particle approximation was derived, ignoring completely the
Fermi sea in the leads. At strong electron-vibron couplings and weak couplings to
the leads, satellites of the main resonant peak are formed in the spectral function
(Fig.\,\ref{V-SPA-T1}). The number of the relevant side-bands is determined by the
well known Huang-Rhys factor \cite{Huang50procrsoc} $g=(\lambda/\omega_0)^2$. The
question which remains is whether these side-bands can be observed in the
differential conductance, when the coupling to all electrons in the leads should be
taken into account simultaneously. New theoretical treatments were presented
recently in
Refs.\,\cite{Kral97,Lundin02prb,Zhu03prb,Braig03prb,Flensberg03prb,Aji03condmat,Mitra04prb,
Frederiksen04prl,Frederiksen04master,Hartung04master,Galperin04nanolett,Galperin04jcp,
Paulsson05prb,Galperin05nanolett,
Cizek04prb,Cizek05czechjp,Ryndyk05prb,Gutierrez05nanolett,Gutierrez05prb,Koch05prl,Arseyev05jetplett,
Arseyev06jetplett,Paulsson06nanolett,Paulsson06jpconf,Zazunov06prb1,Zazunov07prb,
Koch06prb,Koch06prb2,Galperin06prb,Gutierrez06prb,Jauho06jpcs, Ryndyk06prb,
Frederiksen07prb,Frederiksen07prb2,Galperin07jpcm,Song07prb,Ryndyk07prb,
Raza08prb,Raza08condmat,Gagliardi08njp,DAmico08njp,Ryndyk08preprint}.

In parallel, the theory of inelastic scanning tunneling spectroscopy was developed
\cite{Persson87prl,Gata93prb,Tikhodeev01surfsci,Mii02surfsci,Mii03prb,Tikhodeev04prb,Raza07condmat}.
For a  recent review of the electron-vibron problem and its relation to charge
transport at the molecular scale see  Ref.~\cite{Galperin07jpcm}. Note the related
problem of quantum shuttle
\cite{Gorelik98prl,Fedorets02epl,Novotny03prl,Fedorets04prl}.

Many interesting results by the investigation  of quantum transport  in the strong
electron-vibron coupling limit has been achieved with the help of the master equation
approach \cite{Braig03prb,Mitra04prb,Koch05prl,Koch06prb,Koch06prb2}. This method,
however, is valid only in the limit of very weak molecule-to-lead coupling and
neglects all spectral effects, which are the most important at finite coupling to
the leads.

\subsubsection{Nonequilibrium Dyson-Keldysh method}

\paragraph*{(i) The model electron-vibron Hamiltonian}

We use the minimal transport model described in the previous sections. For
convenience, we present the Hamiltonian here once more. The full Hamiltonian is the
sum of the molecular Hamiltonian $\hat H_M$, the Hamiltonians of the leads $\hat
H_{R(L)}$, the tunneling Hamiltonian $\hat H_T$ describing the molecule-to-lead
coupling, the vibron Hamiltonian $\hat H_V$ including electron-vibron interaction
and coupling of vibrations to the environment (describing dissipation of vibrons)
\begin{equation}
 \hat H=\hat H_M+\hat H_V+\hat H_L+\hat H_R+\hat H_T.
\end{equation}

A molecule is described by a set of localized states $|\alpha\rangle$ with energies
$\epsilon_\alpha$ and inter-orbital overlap integrals $t_{\alpha\beta}$ by the following
model Hamiltonian:
\begin{equation}
 \hat H^{(0)}_M=\sum_\alpha\left(\epsilon_\alpha+e\varphi_\alpha(t)\right)
 d^{\dag}_\alpha d^{\phantom{\dag}}_\alpha +\sum_{\alpha\neq\beta}t_{\alpha\beta}
 d^{\dag}_\alpha d^{\phantom{\dag}}_\beta.
\end{equation}

Vibrations and the electron-vibron coupling are described by the Hamiltonian
\cite{Frederiksen04prl,Frederiksen04master,Hartung04master,Ryndyk06prb} ($\hbar=1$)
\begin{equation}
 \hat H_{V}=\sum_q\omega_qa_q^\dag a^{\phantom{\dag}}_q
 +\sum_{\alpha\beta}\sum_q\lambda^q_{\alpha\beta}(a^{\phantom{\dag}}_q+a_q^\dag)
 d^{\dag}_\alpha d^{\phantom{\dag}}_\beta.
\end{equation}
Here vibrations are considered as localized phonons and $q$ is an index labeling
them, not the wave-vector. The first term describes free vibrons with the energy
$\omega_q$. The second term represents the electron-vibron interaction. We include
both diagonal coupling, which describes a change of the electrostatic energy with
the distance between atoms, and the off-diagonal coupling, which describes the
dependence of the matrix elements $t_{\alpha\beta}$ over the distance between atoms.

The Hamiltonians of the right (R) and left (L) leads read
\begin{equation}
 \hat H_{i=L(R)}=\sum_{k\sigma}(\epsilon_{ik\sigma}+e\varphi_i)
 c^{\dag}_{ik\sigma}c^{\phantom{\dag}}_{ik\sigma},
\end{equation}
$\varphi_i(t)$ are the electrical potentials of the leads. Finally, the tunneling
Hamiltonian
\begin{equation}
 \hat H_T=\sum_{i=L,R}\sum_{k\sigma,\alpha}\left(V_{ik\sigma,\alpha}
 c^{\dag}_{ik\sigma}d_\alpha+{\rm h.c.}\right)
\end{equation}
describes the hopping between the leads and the molecule. A direct hopping between
two leads is neglected.

\paragraph*{(ii) Keldysh-Dyson equations and self-energies}

We use the nonequilibrium Green function (NGF) method, as introduced in Section III.
The current in the left ($i=L$) or right ($i=R$) contact to the molecule is
described by the  expression
\begin{equation}\label{J}
 J_{i=L,R}=\frac{ie}{\hbar}\int\frac{d\epsilon}{2\pi}{\rm Tr}\left\{
 {\bf\Gamma}_i(\epsilon-e\varphi_i)\left({\bf G}^<(\epsilon)
 +f^0_i(\epsilon-e\varphi_i)
 \left[{\bf G}^R(\epsilon)-{\bf G}^A(\epsilon)\right]\right)\right\},
\end{equation}
where $f^0_i(\epsilon)$ is the equilibrium Fermi distribution function with chemical
potential $\mu_i$, and the level-width function is
%
$${\bf\Gamma}_{i=L(R)}(\epsilon)=
\Gamma_{i\alpha\beta}(\epsilon) =2\pi\sum_{k\sigma}
V_{ik\sigma,\beta}V^*_{ik\sigma,\alpha}\delta(\epsilon-\epsilon_{ik\sigma}).$$

The lesser (retarded, advanced) Green function matrix of a nonequilibrium molecule
${\bf G}^{<(R,A)}\equiv G_{\alpha\beta}^{<(R,A)}$ can be found from the
Dyson-Keldysh equations in the integral form
\begin{eqnarray}
& {\bf G}^R(\epsilon)={\bf G}^R_0(\epsilon)
  +{\bf G}^R_0(\epsilon){\bf\Sigma}^R(\epsilon)
  {\bf G}^R(\epsilon), \\[0.3cm]
& {\bf G}^<(\epsilon)=
  {\bf G}^R(\epsilon){\bf\Sigma}^<(\epsilon){\bf G}^A(\epsilon),
\end{eqnarray}
or from the corresponding equations in the differential form (see e.g.
Refs.\,\cite{Ryndyk05prb,Ryndyk06prb} and references therein).

Here
\begin{equation}
{\bf\Sigma}^{R,<}=
{\bf\Sigma}^{R,<(T)}_{L}+{\bf\Sigma}^{R,<(T)}_{R}+{\bf\Sigma}^{R,<(V)}
\end{equation}
is the total self-energy of the molecule composed of the tunneling (coupling to the
left and right leads) self-energies
\begin{equation}
{\bf\Sigma}_{j=L,R}^{R,<(T)}\equiv
 {\Sigma}_{j\alpha\beta}^{R,<(T)}=\sum_{k\sigma}\left\{V^*_{jk\sigma,\alpha}
 {G}^{R,<}_{jk\sigma}V_{jk\sigma,\beta}\right\},
\end{equation}
and the vibronic self-energy
${{\bf\Sigma}}^{R,<(V)}\equiv{\Sigma}^{R,<(V)}_{\alpha\beta}$.

For the retarded tunneling  self-energy ${\bf\Sigma}_i^{R(T)}$ one obtains
\begin{equation}\label{SigmaRT}
{\bf\Sigma}^{R(T)}_{i}(\epsilon)={\bf\Lambda}_i(\epsilon-e\varphi_i)-\frac{\mathrm{i}}{2}
{\bf\Gamma}_{i}(\epsilon-e\varphi_i),
\end{equation}
where ${\bf\Lambda}_i$ is the real part of the self-energy, which usually can be
included in the single-particle Hamiltonian $\hat H^{(0)}_M$, and ${\bf\Gamma}_i$
describes level broadening due to coupling to the leads. For the corresponding
lesser function one finds
\begin{equation}
{\bf\Sigma}^{<(T)}_i(\epsilon)=\mathrm{i}{\bf\Gamma}_i(\epsilon-e\varphi_i)
f^0_i(\epsilon-e\varphi_i).
\end{equation}

In the standard self-consistent Born approximation, using the Keldysh technique, one
obtains for the vibronic self-energies
\cite{Mitra04prb,Frederiksen04prl,Frederiksen04master,Hartung04master,
Galperin04nanolett,Galperin04jcp,Galperin07jpcm,Ryndyk06prb}
\begin{eqnarray}\label{SigmaRA}
& \displaystyle {\bf\Sigma}^{R(V)}(\epsilon)=\frac{i}{2}\sum_q\int\frac{d\omega}{2\pi}
 \left({\bf M}^q{\bf G}^{R}_{\epsilon-\omega}{\bf M}^qD^K_{q\omega}+
 \right. \nonumber \\[0.2cm]
& \displaystyle \left. +{\bf M}^q{\bf G}^K_{\epsilon-\omega}{\bf M}^qD^{R}_{q\omega}
 -2D^{R}_{q\omega=0}
 {\bf M}^q{\rm Tr}\left[{\bf G}^<_{\omega}{\bf M}^q\right]\right), \\[0.3cm]
 \label{SigmaK}
& \displaystyle {\bf\Sigma}^{<(V)}(\epsilon)=i\sum_q\int\frac{d\omega}{2\pi}{\bf M}^q
 {\bf G}^<_{\epsilon-\omega}{\bf M}^qD^<_{q\omega},
\end{eqnarray}
where ${\bf G}^K= 2{\bf G}^< + {\bf G}^R-{\bf G}^A$ is the Keldysh Green function, and
${\bf M}^q\equiv M^q_{\alpha\beta}$.

If vibrons are noninteracting, in equilibrium, and non-dissipative, then the vibronic
Green functions write:
\begin{equation}
  D^R_0(q,\omega)=
  \frac{1}{\omega-\omega_q+\mathrm{i}0^+}-\frac{1}{\omega+\omega_q+\mathrm{i}0^+},
\end{equation}
\begin{align}
  D^<_0(q,\omega)= & -2\pi
  \mathrm{i}\left[(f_B^0(\omega_q)+1)\delta(\omega+\omega_q) 
  +f_B^0(\omega_q)\delta(\omega-\omega_q)\right],
\end{align}
where the equilibrium Bose distribution function is
\begin{equation}
f_B^0(\omega)=\frac{1}{\exp\left(\omega/T\right)-1}.
\end{equation}

In the Migdal model the retarded vibron function is calculated from the
Dyson-Keldysh equation
\begin{equation}
 D^R(q,\omega)=\frac{2\omega_q}{\omega^2-\omega^2_q-2\omega_q{\Pi}^R(q,\omega)},
\end{equation}
where $\Pi(q,\omega)$ is the polarization operator (boson self-energy). The equation
for the lesser function (quantum kinetic equation in the integral form) is
\begin{equation}\label{kinPi}
 (\Pi^{R}_{q\omega}-\Pi^A_{q\omega})D^<_{q\omega}-(D^{R}_{q\omega}-D^A_{q\omega})
 \Pi^{<}_{q\omega}=0,
\end{equation}
this equation in the stationary case considered here is algebraic in the frequency domain.

The polarization operator is the sum of two parts, environmental and electronic:
$\Pi^{R,<}_{q\omega}=\Pi^{R,<(\rm env)}_{q\omega}+\Pi^{R,<(\rm el)}_{q\omega}$.

The environmental equilibrium part of the polarization operator can be approximated
by the simple expressions
\begin{eqnarray}
& \displaystyle \Pi^{R(\rm env)}(q,\omega)=-\frac{i}{2}\gamma_q{\rm sign}(\omega),
\\[0.3cm] & \displaystyle \Pi^{<(\rm env)}(q,\omega)=-i\gamma_qf_B^0(\omega){\rm
sign}(\omega),
\end{eqnarray}
where $\gamma_g$ is the vibronic dissipation rate, and $f_B^0(\omega)$ is the
equilibrium Bose-Einstein distribution function.

The electronic contribution to the polarization operator within the SCBA is
\begin{align}\label{PiRA} \Pi^{R(\rm el)}(q,\omega)=
-i\int\frac{d\epsilon}{2\pi}{\rm Tr} & \left({\bf M}^q{\bf
G}^{<}_{\epsilon}{\bf M}^q{\bf G}^{A}_{\epsilon-\omega} 
+{\bf M}^q{\bf
G}^R_{\epsilon}{\bf M}^q{\bf
G}^{<}_{\epsilon-\omega}\right),\\[0.3cm]
\Pi^{<(\rm el)}(q,\omega)= -i\int\frac{d\epsilon}{2\pi}{\rm Tr} &
\left({\bf M}^q {\bf G}^<_{\epsilon}{\bf M}^q{\bf
G}^>_{\epsilon-\omega}\right). \end{align}

We obtained the full set of equations, which can be used for numerical calculations.

\subsubsection{Single-level model: spectroscopy of vibrons}

The isolated single-level electron-vibron model is described by the Hamiltonian
\begin{equation}
  \hat H_{M+V}=(\epsilon_0+e\varphi_0)d^{\dag}d+\omega_0a^{\dag}a+\lambda\left(a^{\dag}+a\right)d^{\dag}d,
\end{equation}
where the first and the second terms describe the free electron state and the free
vibron, and the third term is electron-vibron minimal coupling interaction.

The electrical potential of the molecule $\varphi_0$ plays an important role in
transport at finite voltages. It describes the shift of the molecular level by the
bias voltage, which is divided between the left lead (tip), the right lead
(substrate), and the molecule as $\varphi_0=\varphi_R+\eta(\varphi_L-\varphi_R)$
\cite{Datta97prl}. We assume the simplest linear dependence of the molecular
potential ($\eta=const$), but its nonlinear dependence \cite{Rakshit05} can be
easily included in our model.

Here we assume, that the vibrons are in equilibrium and are not excited by the
current, so that the self-consistent Born approximation is a good starting point.
The vibron Green function are assumed to be equilibrium with the broadening defined
by the external thermal bath, see for details
Refs.\,\cite{Frederiksen04master,Galperin04nanolett,Galperin04jcp,Galperin07jpcm,Ryndyk06prb}.

\begin{figure}[b]
\begin{center}
\epsfxsize=0.5\hsize
\epsfbox{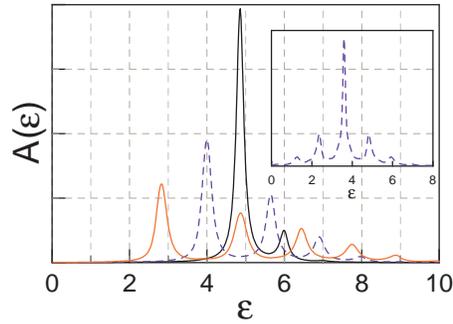}
\caption{(Color online) Spectral function at different electron-vibron couplings: $\lambda/\omega_0=0.4$
(black), $\lambda/\omega_0=1.2$ (blue, dashed), and $\lambda/\omega_0=2$ (red); at $\epsilon_0/\omega_0=5$,
$\Gamma_L/\omega_0=\Gamma_R/\omega_0=0.1$.  In the insert the spectral function at
$\lambda/\omega_0=1.2$ is shown at finite voltage, when the level is partially filled.
Energies are in units of $\hbar\omega_0$.}
\label{fig2b}
\end{center}
\end{figure}

For the single-level model all equations are significantly simplified. Combining $J_L$
and $J_R$ the expression for the current can be written for energy independent
$\Gamma_{L(R)}$ (wide-band limit) as
\begin{equation}\label{J_qc}
J=\frac{e}{h}\frac{\Gamma_L\Gamma_R}{\Gamma_R+\Gamma_L} \int d\epsilon
A(\epsilon)\left[f^0(\epsilon-e\varphi_L)- f^0(\epsilon-e\varphi_R)\right].
\end{equation}
It looks as simple as the Landauer-B\"{u}ttiker formula, but it is not trivial, because
the spectral density \mbox{$A(\epsilon)=-2{\rm Im}G^R(\epsilon)$} now depends on the
distribution function of the electrons in the fluctuating molecule and hence the applied
voltage, $\varphi_L=-\varphi_R=V/2$ \cite{Ryndyk05prb}. Indeed, $G^R(\epsilon)$ can be
found from (\ref{GR})
\begin{equation}\label{GR2}
G^R(\epsilon)=\frac{1}{\epsilon-\tilde\epsilon_0-\Sigma^{R(V)}(\epsilon)+
i(\Gamma_L+\Gamma_R)/2},
\end{equation}
where $\Sigma^{R(V)}(\epsilon)$ is a functional of the electron distribution function
inside a molecule. Actually, the lesser function $G^<(\epsilon)$ is used in the quantum
kinetic formalism as a distribution function. In the single-level case the usual
distribution function can be introduced through the relation
\begin{equation}
G^<(\epsilon)=iA(\epsilon)f(\epsilon).
\end{equation}

\begin{figure}[t]
\begin{center}
\epsfxsize=0.6\hsize \epsfbox{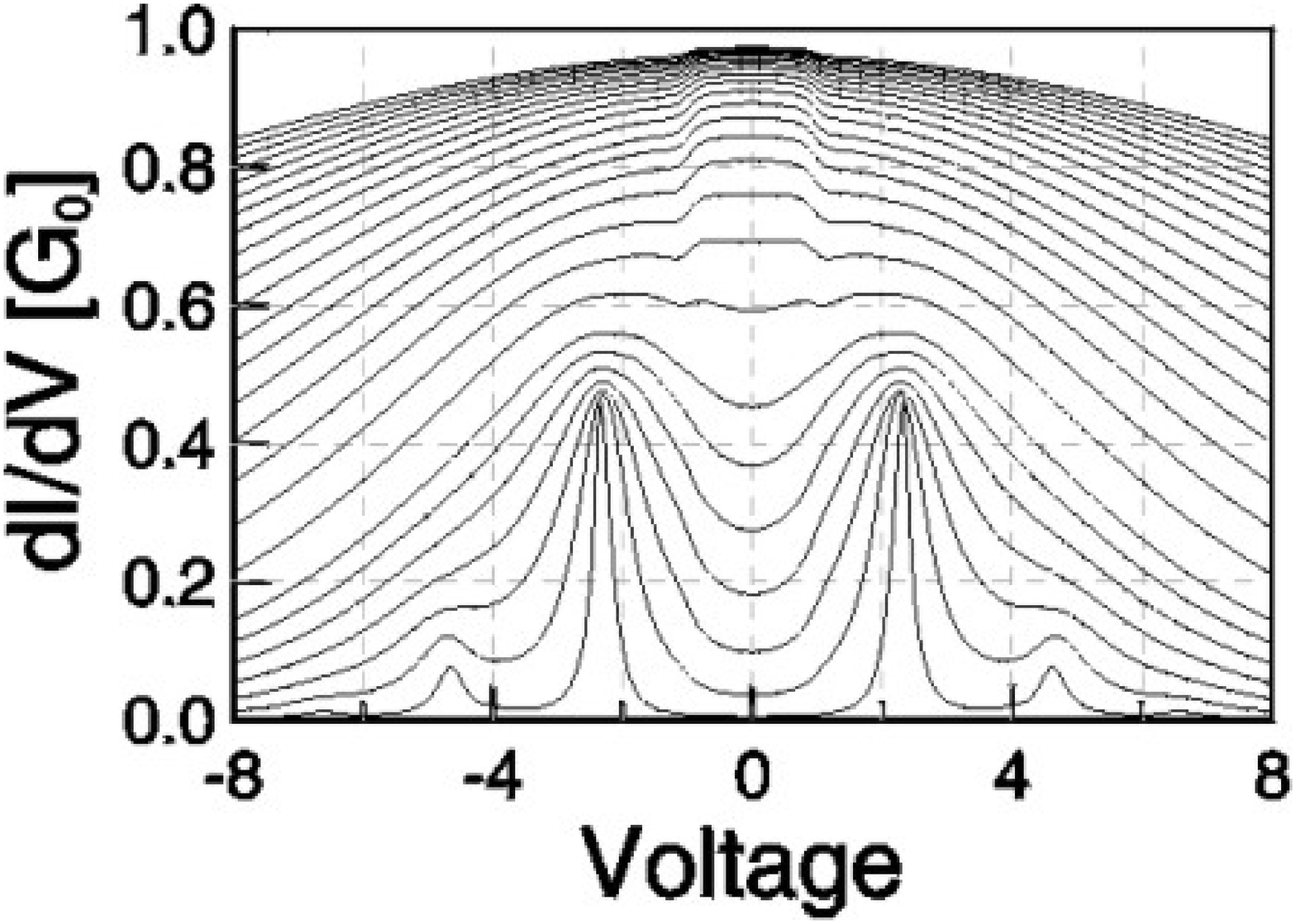}
\caption{Differential conductance of a {\em
symmetric} junction ($\eta=0.5$, $\Gamma_R=\Gamma_L$) at different molecule-to-lead
coupling, from \mbox{$\Gamma_L/\omega_0=0.1$} (lower curve) to $\Gamma_L/\omega_0=10$
(upper curve), \mbox{$\lambda/\omega_0=1$}, $\epsilon_0/\omega_0=2$. Voltage is in the
units of $\hbar\omega_0/e$.}
\label{fig3b}
\end{center}
\end{figure}

Note the essential difference between symmetric \mbox{($\Gamma_L=\Gamma_R$)} and
asymmetric junctions. It is clear from the noninteracting solution of the transport
problem. Neglecting for a moment the vibron self-energies, we obtain the noninteracting
distribution function
\begin{equation}
f(\epsilon)=\frac{\Gamma_Lf_L^0(\epsilon-e\varphi_L)+\Gamma_Rf_R^0(\epsilon-e\varphi_R)}
{\Gamma_L+\Gamma_R}.
\end{equation}
For strongly asymmetric junctions (e.g. $\Gamma_L\ll\Gamma_R$) the distribution function
remains close to the equilibrium function in the right lead $f_R^0(\epsilon-e\varphi_R)$,
thus essentially simplifying the solution. While for symmetric junctions the distribution
function has the double-step form and is very different from the equilibrium one.

A typical example of the spectral function at zero voltage is shown in Fig.~\ref{fig2b}.
At finite voltage it should be calculated self-consistently. In the insert the spectral
function of the symmetric junction at finite voltage is shown, it is changed essentially
because the distribution function is changed.

\begin{figure}[t]
\begin{center}
\epsfxsize=0.6\hsize
\epsfbox{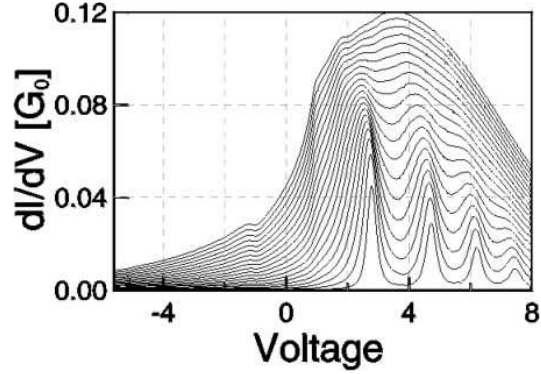}
\caption{Differential conductance of an {\em
asymmetric} junction ($\eta=0$, $\Gamma_R=20\Gamma_L$) at different molecule-to-lead
coupling, from \mbox{$\Gamma_R/\omega_0=0.2$} (lower curve) to $\Gamma_R/\omega_0=4$
(upper curve), $\lambda/\omega_0=2$, $\epsilon_0/\omega_0=5$. The voltage is in the units of
$\hbar\omega_0/e$}
\label{fig4b}
\end{center}
\end{figure}

Let us discuss a general picture of the vibronic transport in symmetric and asymmetric
single-molecule junctions, provided in experiments with the molecular bridges and
STM-to-molecule junctions, respectively. The differential conductance, calculated at
different molecule-to-lead coupling, is shown in Fig.~\ref{fig3b} (symmetric) and
Fig.~\ref{fig4b} (asymmetric). At weak coupling, the vibronic side-band peaks are
observed, reproducing the corresponding peaks in the spectral function. At strong
couplings the broadening of the electronic state hides the side-bands, and new features
become visible. In the symmetric junction, a suppression of the conductance at
$V\simeq\pm\hbar\omega_0$ takes place as a result of inelastic scattering of the
coherently transformed from the left lead to the right lead electrons. In the asymmetric
junction (Fig.~\ref{fig4b}), the usual IETS increasing of the conductance is observed at a
negative voltage $V\simeq-\hbar\omega_0$, this feature is weak and can be observed only
in the incoherent tail of the resonant conductance. We conclude, that the vibronic
contribution to the conductance can be distinguished clearly in both coherent and
tunneling limits.

\begin{figure}[t]
\begin{center}
\epsfxsize=0.6\hsize
\epsfbox{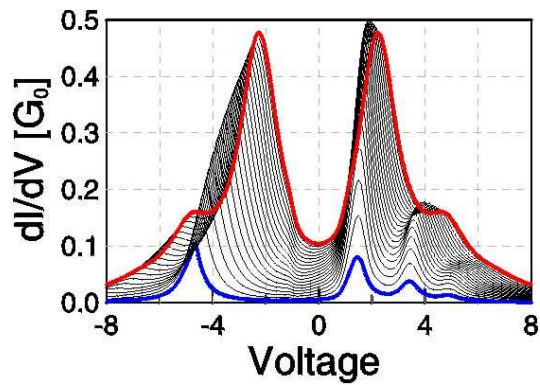}
\caption{(Color) Differential conductance
at different molecule-to-STM coupling (see the text), from {\em asymmetric} junction with
$\Gamma_L/\omega_0=0.025$, $\Gamma_R/\omega_0=0.5$ and $\eta=0.2$ (lower curve, blue
thick line) to {\em symmetric} junction with $\Gamma_L/\omega_0=\Gamma_R/\omega_0=0.5$
and $\eta=0.5$ (upper curve, red thick line), \mbox{$\lambda/\omega_0=1$},
$\epsilon_0/\omega_0=2$. Voltage is in the units of $\hbar\omega_0/e$}
\label{fig5}
\end{center}
\end{figure}

Now let us discuss the particular situation of STS experiments
\cite{Qiu04prl,Wu04prl,Repp05prl,Repp05prl2}. Here we concentrate mainly on the
dependence on the tip-to-molecule distance \cite{Wu04prl}. When the tip (left lead in our
notations) is far from the molecule, the junction is strongly asymmetric:
$\Gamma_L\ll\Gamma_R$ and $\eta\rightarrow 0$, and the conductance is similar to that
shown in Fig.\,\ref{fig4b}. When the tip is close to the molecule, the junction is
approximately symmetric: $\Gamma_L\approx\Gamma_R$ and $\eta\approx 0.5$, and the
conductance curve is of the type shown in Fig.\,\ref{fig3b}. We calculated the
transformation of the conductance from the asymmetric to symmetric case
(Fig.~\ref{fig5}). It is one new feature appeared in asymmetric case due to the fact that
we started from a finite parameter $\eta=0.2$ (in the Fig.~\ref{fig4b} $\eta=0$), namely a
single peak at negative voltages, which is shifted to smaller voltage in the symmetric
junction. The form and behavior of this peak is in agreement with experimental results
\cite{Wu04prl}.

In conclusion, at weak molecule-to-lead (tip, substrate) coupling the usual vibronic
side-band peaks in the differential conductance are observed; at stronger coupling to the
leads (broadening) these peaks are transformed into step-like features. A
vibronic-induced decreasing of the conductance with voltage is observed in
high-conductance junctions. The usual IETS feature (increasing of the conductance) can be
observed only in the case of low off-resonant conductance. By changing independently the
bias voltage and the tip position, it is possible to determine the energy of molecular
orbitals and the spectrum of molecular vibrations. In the multi-level systems with strong
electron-electron interaction further effects, such as Coulomb blockade and Kondo effect,
could dominate over the physics which we address here; these effects have to be included
in a subsequent step.

\subsubsection{Multi-level model: nonequilibrium vibrons}

Basically there are two main nonequilibrium effects: the electronic spectrum
modification and excitation of vibrons (quantum vibrations). In the weak
electron-vibron coupling case the spectrum modification is usually small (which is
dependent, however, on the vibron dissipation rate, temperature, etc.) and the main
possible nonequilibrium effect is the excitation of vibrons at finite voltages. We
have developed an analytical theory for this case~\cite{Ryndyk06prb}. This theory is
based on the self-consistent Born approximation (SCBA), which allows to take easily
into account and calculate nonequilibrium distribution functions of electrons and
vibrons.

If the mechanical degrees of freedom are coupled strongly to the environment
(dissipative vibron), then the dissipation of molecular vibrations is determined by
the environment. However, if the coupling of vibrations to the leads is weak, we
should consider the case when the vibrations are excited by the current flowing
through a molecule, and the dissipation of vibrations is also determined essentially
by the coupling to the electrons. Here , we show that the effects of vibron emission
and vibronic instability are important especially in the case of electron-vibron
resonance.

We simplify the equations and obtain some analytical results in the {\em vibronic
quasiparticle approximation}, which assumes weak electron-vibron coupling limit and
weak external dissipation of vibrons:
\begin{equation}
 \gamma^*_q=\gamma_q-2{\rm Im}\Pi^{R}(\omega_q)\ll\omega_q.
\end{equation}
So that the spectral function of vibrons can be approximated by the Dirac $\delta$,
and the lesser function reads
\begin{equation} \label{qpv<}
 D^<(q,\omega)=-2\pi
 i\left[(N_q+1)\delta(\omega+\omega_q)+
 N_q\delta(\omega-\omega_q)\right],
\end{equation}
where $N_q$ is (nonequilibrium) number of vibrations in the $q$-th mode. So, in this
approximation the spectrum mo\-di\-fi\-ca\-tion of vibrons is not taken into
account, but the possible excitation of vibrations is described by the
nonequilibrium $N_q$. The dissipation of vibrons is neglected in the spectral
function, but is taken into account later in the kinetic equation for $N_q$. A
similar approach to the single-level problem was considered recently in
\cite{Tikhodeev01surfsci,Mii02surfsci,Mii03prb,Tikhodeev04prb,
Mitra04prb,Galperin04nanolett,Galperin04jcp}. The more general case with broadened
equilibrium vibron spectral function seems to be not very interesting, because in
this case vibrons are not excited. Nevertheless, in the numerical calculation it can
be easy taken into consideration.

From the general quantum kinetic equation for vibrons, we obtain in this
limit
\begin{equation}
 N_q=\frac{\gamma_qN^0_q-{\rm Im}\Pi^{<}(\omega_q)}
 {\gamma_q-2{\rm Im}\Pi^{R}(\omega_q)}.
\end{equation}

This expression describes the number of vibrons $N_q$ in a nonequilibrium state,
$N^0_q=f_B^0(\omega_q)$ is the equilibrium number of vibrons. In the linear
approximation the polarization operator is independent of $N_q$ and $-2{\rm
Im}\Pi^{R}(\omega_q)$ describes additional dissipation. Note that in equilibrium
$N_q\equiv N^0_q$ because ${\rm Im}\Pi^{<}(\omega_q)=2{\rm Im}\Pi^{R}(\omega_q)
f_B^0(\omega_q)$. See also detailed discussion of vibron emission and absorption
rates in Refs.\,\cite{Tikhodeev01surfsci,Mii02surfsci,Mii03prb,Tikhodeev04prb}.

For weak electron-vibron coupling the number of vibrons is close to equilibrium and is
changed because of {\em vibron emission} by nonequilibrium electrons, $N_q$ is roughly
proportional to the number of such electrons, and the distribution function of
nonequilibrium electrons is not change essentially by the interaction with vibrons
(perturbation theory can be used). The situation changes, however, if nonequilibrium
dissipation $-2{\rm Im}\Pi^{R}(\omega_q)$ is {\em negative}. In this case the number
of vibrons can be essentially larger than in the equilibrium case ({\em vibronic
instability}), and the change of electron distribution function should be taken into
account self-consistently.

In the stationary state the {\em nonlinear} dissipation rate
\begin{equation}\label{negdiss}
 \gamma^*_q=\gamma_q-2{\rm Im}\Pi^{R}(\omega_q)
\end{equation}
is positive, but the nonequilibrium contribution to dissipation $-2{\rm
Im}\Pi^{R}(\omega_q)$ remains negative.

Additionally to the vibronic quasiparticle approximation, the {\em electronic
quasiparticle approximation} can be used when the coupling to the leads is weak. In
this case the lesser function can be parameterized through the number of electrons
$F_\eta$ {\em in the eigenstates of the noninteracting molecular Hamiltonian
$H^{(0)}_{M}$}
\begin{equation}\label{Gres}
 G^<_{\alpha\beta}=i\sum_{\gamma\eta}A_{\alpha\gamma}S_{\gamma\eta}F_\eta S^{-1}_{\eta\beta},
\end{equation}
we introduce the unitary matrix $\bf S$, which transfer the Hamiltonian ${\bf H}\equiv
H^{(0)}_{M\alpha\beta}$ into the diagonal form ${\bf\tilde H}={\bf S}^{-1}{\bf H}{\bf
S}$, so that the spectral function of this diagonal Hamiltonian is
\begin{equation}
 \tilde A_{\delta\eta}(\epsilon)=2\pi\delta(\epsilon-\tilde\epsilon_\delta)
 \delta_{\delta\eta},
\end{equation}
where $\tilde\epsilon_\delta$ are the eigenenergies.

\begin{figure}
\centerline{\includegraphics[width=.5\linewidth]{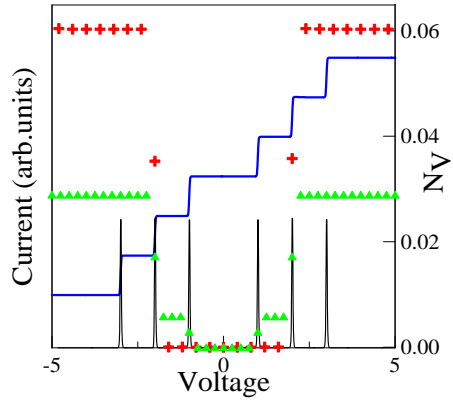}}
\caption{(Color) Vibronic emission in
the symmetric multilevel model: voltage-current curve, differential conductance, and
the number of excited vibrons in the off-resonant (triangles) and resonant (crosses)
cases (details see in the text).}
\label{fig2}
\end{figure}

Note that in the calculation of the self-energies and polarization operators we can
not use $\delta$-approximation for the spectral function (this is too rough and
results in the absence of interaction out of the exact electron-vibron resonance). So
that in the calculation we use actually (\ref{Gres}) with broadened equilibrium
spectral function. This approximation can be systematically improved by including
nonequilibrium corrections to the spectral function, which are important near the
resonance. It is important to comment that for stronger electron-vibron coupling {\em
vibronic side-bands} are observed in the spectral function and voltage-current curves
at energies $\tilde\epsilon_\delta\pm n\omega_q$, we do not consider these effects in
the rest of our paper and concentrate on resonance effects.

After correspondingly calculations we obtain finally
\begin{equation}\label{Nq}
 N_q=\frac{\gamma_qN^0_q-\sum_{\eta\delta}\kappa_{\eta\delta}(\omega_q)F_\eta(F_\delta-1)}
 {\gamma_q-\sum_{\eta\delta}\kappa_{\eta\delta}(\omega_q)(F_\eta-F_\delta)},
\end{equation}
where coefficients $\kappa_{\eta\delta}$ are determined by the spectral function and
electron-vibron coupling in the diagonal representation
\begin{equation}\label{Kappa}
 \kappa_{\eta\delta}(\omega_q)=\int\frac{d\epsilon}{2\pi}\tilde M^q_{\eta\delta}
 \tilde A_{\delta\delta}(\epsilon-\omega_q)\tilde M^q_{\delta\eta}\tilde
 A_{\eta\eta}(\epsilon),
\end{equation}
\begin{equation}
 F_\eta=\frac{\tilde\Gamma_{L\eta\eta}f^0_{L\eta}\!+\!\tilde\Gamma_{R\eta\eta}f^0_{R\eta}\!+\!
 \sum_{q\eta}\left[\zeta^{-q}_{\eta\delta}F_\delta N_q\!+\!
 \zeta^{+q}_{\eta\delta}F_\delta(1\!+\!N_q)\right]}
 {\tilde\Gamma_{L\eta\eta}\!+\!\tilde\Gamma_{R\eta\eta}\!+\!
 \sum_{q\eta}\left[\zeta^{-q}_{\eta\delta}(1\!-\!F_\delta\!+\!N_q)\!+\!
 \zeta^{+q}_{\eta\delta}(F_\delta\!+\!N_q)\right]},
\end{equation}
\begin{equation}
 \zeta^{\pm q}_{\eta\delta}=\tilde M^q_{\eta\delta}\tilde
 A_{\delta\delta}(\tilde\epsilon_\eta\pm\omega_q)
 \tilde M^q_{\delta\eta},
\end{equation}
here $\tilde\Gamma_{i\eta\eta}$ and $f^0_{i\eta}$ are the level width matrix in the
diagonal representation and Fermi function at energy $\tilde\epsilon_\eta-e\varphi_i$.

These kinetic equations are similar to the usual golden rule equations, but are more
general.

Now let us consider several examples of vibron emission and vibronic instability.

\begin{figure}
\centerline{\includegraphics[width=.5\linewidth]{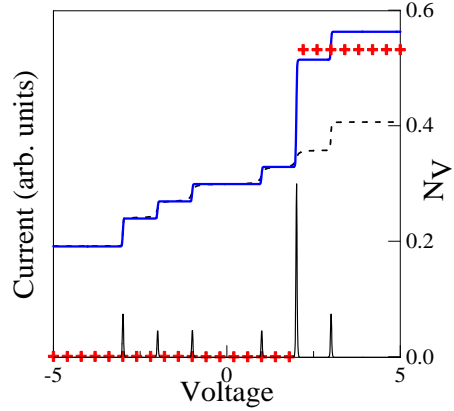}}
\caption{(Color) Vibronic instability
in an asymmetric multilevel model: voltage-current curve, differential conductance,
and the number of excited vibrons (crosses). Dashed line show the voltage-current
curve without vibrons (details see in the text).}
\label{fig3}
\end{figure}

First we consider the most simple case, when the instability is not possible and only
vibron emission takes place. This corresponds to a negative imaginary part of the
electronic polarization operator: ${\rm Im}\Pi^{R)}(\omega_q)<0$. From the Eq.
(\ref{Kappa}) one can see that for any two levels with the energies
$\tilde\epsilon_\eta>\tilde\epsilon_\delta$ the coefficient $\kappa_{\eta\delta}$ is
larger than $\kappa_{\delta\eta}$, because the spectral function $\tilde
A_{\delta\delta}(\epsilon)$ has a maximum at $\epsilon=\tilde\epsilon_\delta$. The
contribution of $\kappa_{\eta\delta}(\omega_q)(F_\eta-F_\delta)$ is negative if
$F_\eta<F_\delta$. This takes place in equilibrium, and in nonequilibrium for
transport through {\em symmetric} molecules, when higher energy levels are populated
after lower levels. The example of such a system is shown in Fig.\,\ref{fig2}. Here we
consider a simple three-level system ($\tilde\epsilon_1=1$, $\tilde\epsilon_2=2$,
$\tilde\epsilon_3=3$) coupled symmetrically to the leads
($\Gamma_{L\eta}=\Gamma_{R\eta}=0.01$). The current-voltage curve is the same with and
without vibrations in the case of symmetrical coupling to the leads and in the weak
electron-vibron coupling limit (if we neglect change of the spectral function). The
figure shows how vibrons are excited, the number of vibrons $N_V$ in the mode with
frequency $\omega_0$ is presented in two cases. In the off-resonant case (green
triangles) $N_V$ is very small comparing with the resonant case
($\omega_0=\tilde\epsilon_2-\tilde\epsilon_1$, red crosses, the vertical scale is
changed for the off-resonant points). In fact, if the number of vibrons is very large,
the spectral function and voltage-current curve are changed. We shall consider this in
a separate publication.

\begin{figure}[t]
\centerline{\includegraphics[width=.5\linewidth]{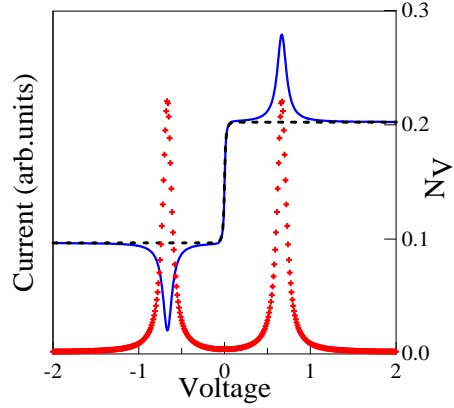}}
\caption{(Color) Floating level
resonance: voltage-current curve and the number of excited vibrons (crosses). Dashed
line show the voltage-current curve without vibrons (details see in the text).}
\label{fig4}
\end{figure}


Now let us consider the situation when the imaginary part of the electronic
polarization operator can be positive: ${\rm Im}\Pi^{R}(\omega_q)>0$. Above we
considered the normal case when the population of higher energy levels is smaller than
lower levels. The opposite case $F_2>F_1$ is known as inversion in laser physics. Such
a state is unstable if the total dissipation $\gamma^*_q$ (\ref{negdiss}) is negative,
which is possible only in the nonstationary case. As a result of the instability, a
large number of vibrons is excited, and in the stationary state $\gamma^*_q$ is
positive. This effect can be observed for transport through {\em asymmetric}
molecules, when higher energy levels are populated {\em before} lower levels. The
example of a such system is shown in Fig.\,\ref{fig3}. It is the same three-level
system as before, but the first and second levels are coupled not symmetrically to the
leads ($\Gamma_{L1}=0.001$, $\Gamma_{R1}=0.1$, $\Gamma_{L2}=0.1$,
$\Gamma_{R2}=0.001$). The vibron couple resonantly these levels
($\omega_q=\tilde\epsilon_2-\tilde\epsilon_1$). The result is qualitatively different
from the symmetrical case. The voltage-current curve is now asymmetric, a large {\em
step} corresponds to the resonant level with inverted population.

Note the importance of the off-diagonal electron-vibron coupling for the resonant
effects. If the matrix $\bf\tilde M$ in the eigen-state representation is diagonal,
there is no resonant coupling between different electronic states.

Finally, let us consider the important case, when initially symmetric molecule becomes
asymmetric when the external voltage is applied. The reason for such asymmetry is
simply that in the external electric field left and right atoms feel different
electrical potentials and the position of the levels
$\epsilon_\alpha=\epsilon^{(0)}_\alpha+e\varphi_\alpha$ is changed (float) with the
external voltage. The example of a such system is shown in Fig.\,\ref{fig4}. Here we
consider a two-level system, one level is coupled electrostatically to the left lead
$\tilde\epsilon_1\propto \varphi_L$, the other level to the right lead
$\tilde\epsilon_2\propto \varphi_R$, the tunneling coupling to the leads also is not
symmetrical ($\Gamma_{L1}=0.1$, $\Gamma_{R1}=0.001$, $\Gamma_{L2}=0.001$,
$\Gamma_{R2}=0.1$). The frequency of the vibration, coupling these two states, is
$\omega_0=1$. When we sweep the voltage, a {\em peak} in the voltage-current curve is
observed when the energy difference $\tilde\epsilon_1-\tilde\epsilon_2\propto eV$ is
going through the resonance $\tilde\epsilon_1-\tilde\epsilon_2\approx\omega_0$.

\subsection{Coupling to a vibrational contiuum: dissipation and renormalization}

\subsubsection{The model Hamiltonian}

In the previous section we have dealt with a simple, but nevertheless physically
rich, model describing the interaction of an electronic level with some specific
vibrational mode confined to the quantum dot. We have seen how to apply in this case
the  Keldysh non-equilibrium techniques described in
Section~III within the self-consistent Born and Migdal approximations. The latter are
however appropriate for the weak coupling limit to the vibrational degrees of
freedom. In the opposite case of strong coupling, different techniques must be
applied. For equilibrium problems, unitary transformations combined with variational
approaches can be used, in non-equilibrium only recently some attempts were made to
deal with the problem.~\cite{Galperin06prb}

In this section we will consider the  case of a multi-level electronic system in
interaction with a bosonic bath~\cite{Gutierrez05nanolett,Gutierrez05prb}. We will
use unitary transformation techniques to deal with the problem, but will only focus
on the low-bias transport, so that strong non-equilibrium effects can be
disregarded. Our interest is to explore how the {\it qualitative} low-energy
properties of the electronic system are modified by the interaction with the bosonic
bath.  We will see that the existence of a continuum of vibrational excitations (up
to some cut-off frequency) dramatically changes the analytic properties of the
electronic Green function and may lead in some limiting cases to a qualitative
modification of the low-energy electronic spectrum. As a result, the $I$-$V$
characteristics at low bias may display ``metallic'' behavior (finite current) even
if the isolated electronic system does exhibit a band gap. The model to be discussed
below has been motivated by the very exciting electrical transport measurements on
short poly(dG)-poly(dC) DNA molecular wires carried out at the group of N. Tao some
time ago~\cite{Xu04nanolett}. Peculiar in these experiments was the large measured
currents -up to 150 nA at 0.8 V- at low voltages, which stood in strong contrast to
the usually accepted view that DNA should behave as an insulator at low applied
bias. Further, a power-law length scaling of the linear conductance with increasing
wire length was demonstrated, indicating that long-range charge transport was
possible. Since the experiments were carried out in an aqueous solution, the
possibility of a solvent-induced modification of the low-energy transport properties
of the wire lied at hand, although additional factors like internal vibrations could
also play a role.

\begin{figure}[tb]
\centerline{\includegraphics[width=.7\linewidth]{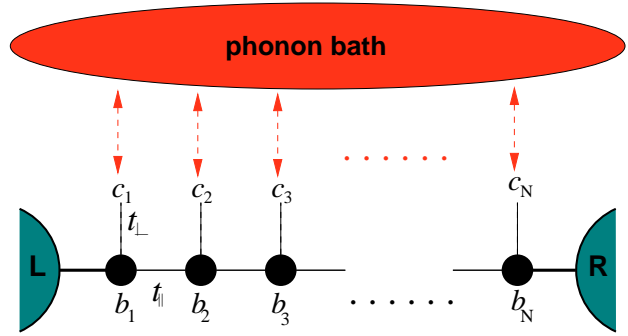}}
\caption{\label{fig:fig1}%
(Color)  Schematic drawing of a DNA molecular wire in contact with a dissipative environment.
  The central chain (extended states) with $N$ sites is connected to semiinfinite left (L)
and right (R) electronic reservoirs.
The bath only interacts  with the side chain sites ({\sf c}), which
we call  for simplicity backbone  sites, but which collectively stay for non-conducting,
localized electronic states.
The Hamiltonian associated with this model is given by
Eqs.~(\ref{eq:eq1}), (\ref{eq:eq2}), and (\ref{eq:eq3}) in the main text.
}
\end{figure}

The proposed model is based on an earlier work~\cite{Cuniberti02prb} and assumes,
within a minimal tight-binding picture, that the DNA electronic states can be
qualitatively classified into extended (conducting) and localized (non-conducting)
states. The former may correspond e.g. to the $\pi$-orbital stack of the base pairs,
the latter to energetically deeper lying (w.r.t. the frontier orbitals) base-pair
states or sugar-phosphate  backbone states. A further assumption is that any
modification of the conducting states through the environment only takes place
through a coupling to the non-conducting set. The tight-binding electronic
Hamiltonian for $N$ sites can then be written as (see also Fig.~\ref{fig:fig1}):
\begin{eqnarray}
{\cal H}_\textrm{el} &=& \epsilon_b\sum_{j} b^{\dagger}_{j}b_{j} - t_{||} \sum_{j}
\lsb b^{\dagger}_{j}b_{j+1} + \Hc \rsb + \epsilon \sum_{j} c^{\dagger}_{j}c_{j}
\nonumber \\ &-& t_{\perp}\sum_{j} \lsb b^{\dagger}_{j}c_{j} + \Hc \rsb = {\cal
H}_\textrm{C} + {\cal H}_{\textrm{b}} + {\cal H}_{\textrm{C-b}}. \label{eq:eq1}
\end{eqnarray}

Hereby ${\cal H}_{\textrm{C}}$ and ${\cal H}_\textrm{b}$ are the Hamiltonians of the
extended and localized states (called in what follows ``backbone'' states for
simplicity), respectively, and ${\cal H}_{\textrm{C-b}}$ is the coupling between
them. $t_{||}$ and $t_{\perp}$ are hopping integrals along the central chain
(extended states) and between the localized states and the central chain,
respectively. If not stated otherwise, the on-site energies will be later set equal
to zero to simplify the calculations. Notice that this model displays a gap in the
electronic spectrum roughly proportional to the transversal coupling $t_{\perp}$.
This can be easely seen by looking at the limit $N\to\infty$ which leads to a
periodic system. In this case, the Hamiltonian can be analytically diagonalized and
two energy dispersion curves are obtained, which are given by $E_{\pm}(k)= t_{||}
\cos(k) \pm \sqrt{t^2_{\perp}+ t^2_{||} \cos^2(k)}$. The direct gap between the two
bands is simply $\delta=2\sqrt{t^2_{\perp}+ t^2_{||}}$.  Since this model further
shows electron-hole symmetry, two electronic manifolds (bands in the limit of
$N\to\infty$) containing $N$ states  each, are symmetrically situated around the
Fermi level, which is taken as the zero of energy.

The gap is obviously temperature independent and furthermore it is expected that
transport at energies $E<\delta$ will be strongly suppressed due to the absence of
electronic states to suport charge propagation. As a result, the linear conductance
should display a strong exponential dependence as a function of the chain length
$N$.  In view of this behavior, an immediate issue that arises is how stable this
electronic structure, \ie~two electronic manifolds separated by a gap, is against
the coupling to an environment. This is an issue which reaches farther than the
problem of charge transport in DNA wires, since it addresses the interaction of an
open  quantum mechanical system with a countable number of electronic energy levels
to a  continuum of states (``universe''). A generic example of such a situation is
the measurement process in quantum mechanics. It is well-known that the interaction
with complex environments is a source of dissipation and decoherence in quantum
mechanical systems.~\cite{Weiss99book} Concerning more specifically the case of DNA
(and proteins), there is broad experimental evidence that  the molecule dynamics
follows the solvent dynamics over a broad temperature range. Especially,
conformational changes, low-energy vibrational excitations and the corresponding
temperature dependences turn out to be very sensitive to the solvents dynamics.
\cite{Caliskan04jcp} We will thus consider the vibrational degrees of freedom of
counterions and hydration shells of the solvent as a dynamical bath able to break
the electronic phase coherence and additionally to act as a dissipative environment.
We do not consider specific features of the environment but represent it in a
generic way by a bosonic bath of $M$ harmonic oscillators. Then, the previous
Hamiltonian can be extended to:
\begin{equation}
{\cal H}_\textrm{W}={\cal H}_\textrm{el} +\sum_{\alpha} \Omega_{\alpha}
B^{\dagger}_{\alpha} B_{\alpha} +\sum_{\alpha,j} \lambda_{\alpha}
c^{\dagger}_{j}c_{j} (B_{\alpha}+B^{\dagger}_{\alpha}) = {\cal H}_\textrm{el} +{\cal
H}_{\textrm{B}}+{\cal H}_{\textrm{c-B}}, \label{eq:eq2}
\end{equation}
where ${\cal H}_{\textrm{B}}$ and ${\cal H}_{\textrm{c-B}}$ are the phonon bath
Hamiltonian and the (localized) state-bath interaction, respectively. $B_{\alpha}$
is a  bath phonon operator and $\lambda_{\alpha}$ denotes the electron-phonon
coupling. Note that we assume a local coupling of the bath modes to the electronic
density at the side chain. Later on, the thermodynamic limit ($M\to \infty)$ in the
bath degrees of freedom will be carried out and the corresponding bath spectral
density introduced, so that at this stage we do not need to further specify the set
of bath frequencies {$\Omega_{\alpha}$} and coupling constants {$\lambda_{\alpha}$}.
Obviously, the bath can be assumed to be in thermal equilibrium and be described by
a canonical partition function.

To complete the formulation of the model, we have to include the interaction of the
system with electronic reservoirs in order to describe charge transport along the
same lines as before. We assume, as usual, a tunnel-type Hamiltonian with the form:
\begin{eqnarray}
{\cal H}&=&{\cal H}_\textrm{W}+\sum_{{\bf k}\in \textrm{L,R}, \sigma} \epsilon_{{\bf
k}\sigma} d^{\dagger}_{{\bf k}\sigma}d_{{\bf k}\sigma}
+\sum_{{\bf k}\in \textrm{L}, \sigma} ( V_{{\bf k},1} \, d^{\dagger}_{{\bf
k}\sigma} \, b_{1} + \Hc) \nonumber \\ &&+\sum_{{\bf k}\in R, \sigma} (
V_{{\bf k},N} \, d^{\dagger}_{{\bf k}\sigma} \, b_{N} + \Hc)  =
{\cal H}_\textrm{W} + {\cal H}_{\textrm{L/R}}+{\cal H}_{\textrm{L-C}}+{\cal
H}_{\textrm{R-C}}. \label{eq:eq3}
\end{eqnarray}

The Hamiltonian of Eq.~(\ref{eq:eq3}) is the starting point of our investigation.
For a weak charge-bath coupling, a perturbative approach similar to the second order
Born approximation, as described in the previous section can be applied. We expect,
however, qualitative new effects rather in the opposite limit of strong coupling to
the bath. To deal with this problem, a unitary transformation, the Lang-Firsov (LF)
transformation, can be performed on the Hamiltonian  of Eq.~(\ref{eq:eq3}), which
allows to eliminate the linear charge-vibron interaction ${\cal H}_{\textrm{c-B}}$.
In the limiting case of an isolated system with a single electron (or hole) this
transformation becomes exact and allows for a full decoupling of electronic and
vibronic propagators, see e.g. Ref.~\cite{Mahan90book}. In the present case,
this transformation is not exact and further approximations have to be introduced in
order to make the problem tractable.

The generator of the LF transformation is given by $$S=\sum_{\alpha,j} (
{\lambda_{\alpha}} / {\Omega_{\alpha}}) c^{\dagger}_{j}
c_{j}(B_{\alpha}-B^{\dagger}_{\alpha})$$ and $S^{\dagger}=-S$. In the transformed
Hamiltonian $\bar{\cal{H}} = \ee^{S} {\cal{H}} \ee^{-S}$ the linear coupling to the
bath is eliminated. One should notice that in $\bar{\cal{H}}$ only the ``backbone''
part of the Hamiltonian is modified since the conducting state operators $b_\ell $
as well as the lead operators $d_{\bf{k}\sigma}$ are invariant with respect to the
above transformation. The new Hamiltonian reads:
\begin{eqnarray}
\bar{{\cal H}}&=&{\cal H}_\textrm{C} + {\cal H}_{\textrm{L/R}} + {\cal H}_\textrm{B}
+ {\cal H}_{\textrm{L/R-C}} +(\epsilon-\Delta)\sum_{j} c^{\dagger}_{j}c_{j} - {
t_{\perp}\sum_{j} \lsb b^{\dagger}_{j}c_{j}{\cal X} + \Hc \rsb }, \label{eq:eq4}
\nonumber \\ {\cal X}&=& \exp{\lsb \sum_{\alpha}
\frac{\lambda_{\alpha}}{\Omega_{\alpha}}(B_{\alpha}-B^{\dagger}_{\alpha})\rsb}, \;
\; \; \Delta = \sum_{\alpha} \frac{\lambda_{\alpha}^2}{\Omega_{\alpha}}.
\end{eqnarray}

As a result of the LF we get a shift of the onsite energies (polaron shift or
reorganization energy in electron transfer theory) and a renormalization of both the
tunneling  and of the transversal coupling Hamiltonian via the bosonic operators
${\cal X}$. There is also an additional  electron-electron interaction term which we
will not be concerned with in the remaining of this section and is thus omitted.
Since we are mainly interested in qualitative statements, we will assume the
wide-band approximation in the coupling to the electrodes which is equivalent to
substituting the electrode self-energies by a purely imaginary constant, i.e.
$\Sigma_{\rm L,R}\approx -\ii \Gamma_{\rm L,R}$. We are thus not interested in
specific features of the electrode electronic structure.

To further proceed, let us now introduce two kinds of retarded thermal Green
functions related to the central chain $G_{j \ell}(t)$ and to the ``backbones''
$P_{j \ell}(t)$, respectively (taking $\hbar=1$):
\begin{eqnarray}
\label{eq:eq5}
G_{j \ell}(t,t')&=&-i\Theta(t-t')\lab \lsb b_{j}(t),b^{\dagger}_{\ell}(t')\rsb _{+}
\rab, \\ P_{j \ell}(t,t')&=&-i \Theta(t-t') \lab \lsb c_{j}(t){\cal
X}(t),c^{\dagger}_{\ell}(t'){\cal X}^{\dagger}(t')\rsb _{+} \rab, \nonumber
\end{eqnarray}
where $\Theta$ is the Heaviside function. Notice that the $P$-Green function doe not
have a pure electronic character but also contains the bath operators ${\cal X}$.
For a full out-of-equilibrium calculation, the full Keldysh formalism including
lesser- and greater-GF would also be needed. However, as we will briefly show below,
the final expression for the electrical current at low applied voltages and for
small transversal coupling $t_{\perp}$ will only include the retarded propagators.

We now use the equation of motion technique (EOM) to obtain an expression for the GF
$G_{j \ell}(t)$. We first remark that in the time domain two EOM can be written,
depending on which time argument in the double-time GF the time derivative will act.
One thus obtains in general:
 \begin{eqnarray}
 \ii \partial_{t}  G(t,t')=\lab \lsb b (t),b^{\dagger}(t') \rsb _{+} \rab \delta(t-t') +   (( \lsb b(t),H\rsb |b^{\dagger}(t')
 )).\nonumber \\
G(t,t')[-\ii \partial_{t'} ]=\lab \lsb b (t),b^{\dagger}(t') \rsb _{+} \rab \delta(t-t') -   ((  b(t) | \lsb b^{\dagger}(t'),H \rsb
 )).\nonumber
 \end{eqnarray}

The EOM for the GF $G_{j \ell}(t)$ reads then in the energy space:
\begin{eqnarray}
&&\sum_{n}\lsb G^{-1}_{0}(E)\rsb _{\ell n} G_{nj}(E)= \delta_{\ell
j}-t_{\perp}((c_\ell {\cal X}|b^{\dagger}_{j}))
\label{eq:eqA1}
\end{eqnarray}
\vspace{-0.5cm}
\begin{eqnarray}
\Big[G^{-1}_{0}(E)\Big]_{\ell n} &=& (E-\epsilon_b)\delta_{n
\ell}+t_{||}(\delta_{n, \ell+1}+\delta_{n, \ell-1}) -
\Sigma_{\textrm{L}}\delta_{\ell 1}\delta_{n1}-\Sigma_{\textrm{R}}\delta_{\ell
N}\delta_{nN} \nonumber \\ \Sigma_{\textrm{L(R)}}&=&\sum_{{\bf k}\in L(R)}
\frac{|V_{{\bf k},1(N)}|^2}{E-\epsilon_{{\bf k}}+\ii 0^{+}} \approx -\ii \Gamma_{\rm L,R} \nonumber
\end{eqnarray}

In the next step, EOM for the ``right'' time argument $t'$ of the GF $Z_{\ell
j}^{{\cal X}}(t,t')((c_{\ell}(t) {\cal X}(t)|b^{\dagger}_{j}(t')))$ can be written.
This leads to:
\begin{eqnarray}
\sum_{m} Z_{\ell m}^{{\cal X}}(E) \lsb G^{-1}_{0}(E)\rsb _{mj}=-t_{\perp}
((c_\ell {\cal X}|c^{\dagger}_{j}{\cal X}^{\dagger}))
=-t_{\perp} P_{\ell j}(E) \label{eq:eqA2}
\end{eqnarray}

Inserting Eq.~(\ref{eq:eqA2}) into Eq.~(\ref{eq:eqA1}) we arrive at the matrix
equation:
\begin{eqnarray}
{\bf G}(E)={\bf G}_{0}(E)+ {\bf G}_{0}(E) {\bf\Sigma}_{\textrm{B}}(E) {\bf
G}_{0}(E), \nonumber
\end{eqnarray}
which can be transformed into a Dyson-like equation when introducing the
irreducible part ${\bf \Sigma}_{\textrm{B}}(E)={\bf
\Sigma}^{\textrm{irr}}_{\textrm{B}}(E)+{\bf
\Sigma}^{\textrm{irr}}_{\textrm{B}}(E){\bf G}_{0}(E) {\bf
\Sigma}^{\textrm{irr}}_{\textrm{B}}(E)+\dots$:
\begin{eqnarray}
{\bf G}(E)={\bf G}_{0}(E)+ {\bf G}_{0}(E) {\bf
\Sigma}^{\textrm{irr}}_{\textrm{B}}(E) {\bf G}(E),
\end{eqnarray}
or equivalently:
\begin{eqnarray}
{\bf G}^{-1}(E)&=&{\bf G}_0^{-1}(E)
- t^2_{\perp} {\bf P}(E) \label{eq:eq6} \\ {\bf G}_0^{-1}(E)&=& E{\bf 1}-{\cal
H}_\textrm{C}-\Sigma_{\textrm{L}}(E)- \Sigma_{\textrm{R}}(E). \nonumber
\label{eq:eqA4}
\end{eqnarray}

${\bf \Sigma}^{\textrm{irr}}_{\textrm{B}}(E)=t^2_{\perp} {\bf P}(E)$ is the crucial
contribution to the GF since it contains the influence of the bosonic bath.  Note
that ${\bf \Sigma}^{\textrm{irr}}_{\textrm{B}}(E)$ includes the transversal hopping
$t_{\perp}$ to all orders, the leading one being $t^2_{\perp}$.

In the next step, an expression for the electrical current flowing through the
system must be derived.  Using the results of Sec.~2, we can directly write the
following expression:
\begin{align}
 I= \frac{2 e}{h}\int & dE \,\textrm{Tr}(f_{\rm L}(E)-f_{\rm R}(E)) \; t(E) \nonumber \\
& +t^2_{\perp}\frac{2 e}{h}\int dE \, \left\lbrace  \textrm{Tr} [{\bf \Sigma}^{>}_{\rm
L} \, {\bf P}^{<} -{\bf \Sigma}^{<}_{\rm L} \, {\bf P}^{>}  ] - (L \leftrightarrow R
)\right\rbrace.
\end{align}

The first summand has the same form as Landauer's expression for the current with an
effective transmission function $t(E)= \textrm{Tr} [{\bf G}^{\dagger} {\bf
\Gamma}_{\rm R} {\bf G}  {\bf \Gamma}_{\rm R} ]$. However, the reader should keep in
mind that the GFs appearing in this expression do contain the full dressing by the
bosonic bath and hence, $t(E)$ does not describe elastic transport. The remaining
terms contain explicitly contributions from the bath. It can be shown after some
transformations that the leading term is proportional to $(t^2_{\perp})^2$ so that
within a perturbative approach in $t_{\perp}$ and at low bias it can be
approximately neglected. We therefore remian with the exression $I= \frac{2
e}{h}\int dE \,\textrm{Tr}(f_{\rm L}(E)-f_{\rm R}(E)) \; t(E)$ to obtain the
current.

To remain consistent with this approximation, the bath selfenergy should also be
treated to order $t^2_{\perp}$, more explicitly:
\begin{align}
&P_{\ell j}(t,t')=((c_{\ell}(t) {\cal X}(t) |c^{\dagger}_{j}(t'){\cal X}^{\dagger}(t')))\nonumber \\
& \approx -\ii \theta(t-t') \left\lbrace \lab c_{\ell}(t) c^{\dagger}_{j}(t') \rab \lab {\cal X}(t)
{\cal X}^{\dagger}(t') \rab + \lab c^{\dagger}_{j}(t') c_{\ell}(t)\rab
\lab {\cal X}^{\dagger}(t') {\cal X}(t)\rab \right\rbrace \nonumber \\
& \approx -\ii \delta_{\ell j} \theta(t-t')\left\lbrace  \lab c_{j}(t) c^{\dagger}_{j}(t')
\rab \lab {\cal X}(t) {\cal X}^{\dagger}(t') \rab + \lab c^{\dagger}_{j}(t') c_{j}(t)\rab
\lab {\cal X}^{\dagger}(t') {\cal X}(t)\rab  \right\rbrace \nonumber \\
& = -\ii \delta_{\ell j}
\theta(t-t') \ee^{-{\ii}(\epsilon-\Delta)\,t}  \left\lbrace  (1-f_{\sf c}) \ee^{-\Phi(t)} +
f_{\sf c} \ee^{-\Phi(-t)}   \right\rbrace.
\end{align}

In the previous expression we have replaced the full averages of the ``backbone''
operators by their zero order values (free propagators). $\ee^{-\Phi(t)}= \lab{\cal
X}(t) {\cal X}^{\dagger}(0)\rab_{\textrm{B}}$ is a dynamical bath correlation
function to be specified later on. The average $\lab \cdot \rab_{\textrm{B}}$ is
performed over the bath degrees of freedom. $f_{\sf c}$ is the Fermi function at the
backbone sites. In what follows we consider the case of  empty sites by setting
$f_{\sf c}=0$. The Fourier transform $P_{\ell j}(E)$ reads then:
\begin{equation}
P_{\ell j}(E)= -\ii \delta_{\ell j}\int_{0}^{\infty} \mathrm{d}t \,
\ee^{{\ii}(E+{\ii}0^{+})t}\, \ee^{-{\ii}(\epsilon-\Delta)\,t} \,
\; \left [(1-f_{\sf c}) \ee^{-\Phi(t)} + f_{\sf c} \ee^{-\Phi(-t)} \right ]
\label{eq:eq7}
\end{equation}

In order to get closed expressions for the bath thermal averages it is appropriate
to introduce a bath spectral density~\cite{Weiss99book} defined by :
\begin{eqnarray}
J(\omega)= \sum_{\alpha} \lambda^{2}_{\alpha} \delta(\omega-\Omega_{\alpha}) = J_0
( \frac{\omega}{\oc})^s \ee^{-\omega/\oc} \Theta(\omega),
\label{eq:eq8}
\end{eqnarray}
where $\oc$ is a cut-off frequency related to the bath memory time
$\tau_\textrm{c}\sim \oc^{-1}$. It is easy to show that the limit $\oc\to \infty$
corresponds to a Markovian bath, \ie~ $J(t)\sim J_0 \delta(t)$. Using this {\it
Ansatz}, $\Phi(t)$ can be written as:
\begin{eqnarray}
\Phi(t)&=& \int_{0}^{\infty} \mathrm{d}\omega \, \frac{J(\omega)}{\omega^2}\lsb
1-\ee^{-{\ii}\omega t}+2\frac{1-\cos{\omega t}}{\ee^{\beta\omega}-1}\rsb .
\label{eq:eq9}
\end{eqnarray}
Although the integral can be performed analytically~\cite{Weiss99book}, we consider
$\Phi(t)$ in some limiting cases where it is easier to work directly with
Eq.~(\ref{eq:eq9}).

\subsubsection{Limiting cases}

We use now the results of the foregoing section to discuss the electronic
transport properties of our model in some
limiting cases for which analytic expressions can be derived. We
will discuss the mean-field approximation and  the weak-coupling regime in
the electron-bath interaction as well as to elaborate on the strong-coupling limit.
 Farther, the cases of ohmic ($s=1$) and  superohmic ($s=3$) spectral densities are
 treated.
\begin{figure}[t]
\vskip 0.7cm
\centerline{\includegraphics[width=.7\linewidth]{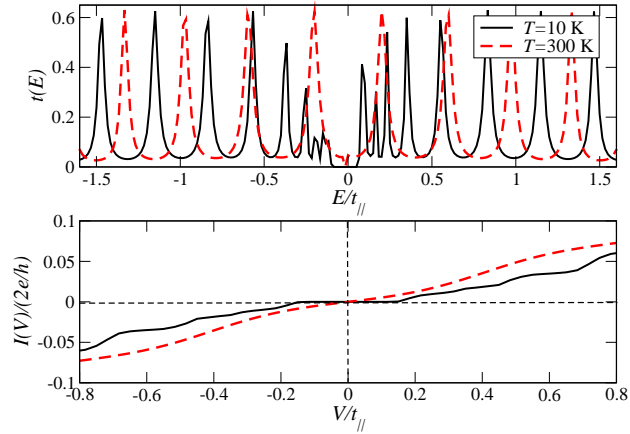}}
\caption{\label{fig:fig2}%
(Color) Electronic transmission and corresponding current in the mean-field
approximation for two different temperatures. Parameters: $N=20, J_0/\oc= 0.12,
t_{\perp}/t_{||}=0.5, \Gamma_{\textrm{L/R}}/t_{||}=0.5$.
}
\end{figure}

\paragraph*{(i) Mean-field approximation}
The mean-field approximation is the simplest one and neglects bath fluctuations contained in $P(E)$.
The MFA can be introduced by writing the phonon operator
${\cal X}$ as $\lab{\cal X}\rab_{\rm B} +\delta {\cal X}$ in ${\cal
H}_{\textrm{C-c}}$ in Eq.~(\ref{eq:eq4}), \ie~${\cal
H}^{\textrm{MF}}_{\textrm{C-b}}=-t_{\perp}\sum_{j} \lsb
b^{\dagger}_{j}c_{j}\lab{\cal X}\rab_{\textrm{B}} + \Hc \rsb + O(\delta {\cal
X})$. As a result a real, static and temperature dependent term
in Eq.~(\ref{eq:eq6}) is found:
\begin{eqnarray}
{\bf G}^{-1}(E)={\bf G}_0^{-1}(E)- t^2_{\perp} \frac{|\lab {\cal
X}\rab_{\textrm{B}}|^{2}}{E-\epsilon+\Delta+\ii 0^{+}}{\bf 1},
\label{eq:eq11}
\end{eqnarray}
where $\labs \lab {\cal X}\rab_\textrm{B} \rabs^2=\ee^{-2\kappa(T)}$ and
$\kappa(T)$ is given by:
\begin{eqnarray}
\kappa(T)= \int_{0}^{\infty} \frac{\mathrm{d}\omega}{\omega^2} J(\omega)
\coth{\frac{\omega}{2 \KB T}}.
\end{eqnarray}
The effect of the MF term is thus to scale the bare transversal hopping
$t_{\perp}$ by the exponential temperature dependent factor $\ee^{-\kappa(T)}$.

In the case of an ohmic bath, $s=1$, the integrand in $\kappa(T)$ scales as $1/\omega^{p},\,
p=1,2$ and
has thus a logarithmic divergence
at the lower integration limit. Thus,
the MF contribution would vanish. In other words, no gap would
exist on this approximation level.

In the superohmic case ($s=3$) all integrals are regular. One obtains
$\Delta=\int \mathrm{d}\omega \,
\omega^{-1} J(\omega)= \Gamma(s-1)J_0=2J_0$, with $\Gamma(s)$ being the Gamma
function and $\kappa(T)$ reads:
\begin{eqnarray}
\kappa(T) = \frac{2J_0}{\oc} \left [ \, 2 \lrb
\frac{\KB T}{\oc}\rrb ^{2} \, \zeta_{\textrm{H}} \lrb 2,\frac{\KB
T}{\oc}\rrb -1 \right ].  \label{eq:eq13}
\end{eqnarray}
$ \zeta_{\textrm{H}}(s,z)=\sum_{n=0}^{\infty} (n+z)^{-s}$ is the Hurwitz
$\zeta$-function, a generalization of the Riemann
$\zeta$-function.~\cite{Gradshteyn00book}

It follows from Eq.~(17) that $\kappa(T)$ behaves like a constant for low
temperatures ($\KB T/\oc < 1$), $\kappa(T)\sim J_0/\oc$, while it scales linear with
$T$ in the high-temperature limit ($\KB T/\oc > 1$), $\kappa(T)\sim J_0/\oc(1+2 \KB
T/\oc)$).

For $J_0\neq 0$ and at zero temperature the hopping integral is roughly reduced to
$t_{\perp}\ee^{-\frac{J_0}{\oc}}$ which is similar to  the renormalization of the
hopping in Holstein's polaron model~\cite{Holstein59apny}, though here it is
$t_{\perp}$ rather than $t_{||}$ the term that is rescaled. At high temperatures
$t_{\perp}$ is further reduced ($\kappa(T)\sim T$) so that the gap in the electronic
spectrum finally collapses and the system becomes ``metallic'', see
Fig.~\ref{fig:fig2}. An appreciable temperature dependence can only be observed in
the limit $J_0/\oc < 1$; otherwise the gap would collapse already at zero
temperature due to the exponential dependence on $J_0$. We further remark that the
MFA is only valid in this regime ($J_0/\oc < 1$), since for $J_0/\oc \gg 1$
multiphonon processes in the bath, which are not considered in the MFA, become
increasingly relevant and thus a neglection of bath fluctuations is not possible.

\paragraph*{(ii) Beyond MF: weak-coupling limit}

As a first step beyond the mean-field approach let's first consider the
weak-coupling limit in ${\bf P}(E)$. For $J_0/\oc < 1$ and not too high temperatures
($\KB T/\oc < 1$) the main contribution to the integral in Eq.~(\ref{eq:eq9})
comes from long times $t\gg \oc^{-1}$.
\begin{figure}[t]
\vskip 0.7cm
\centerline{\includegraphics[width=.7\linewidth]{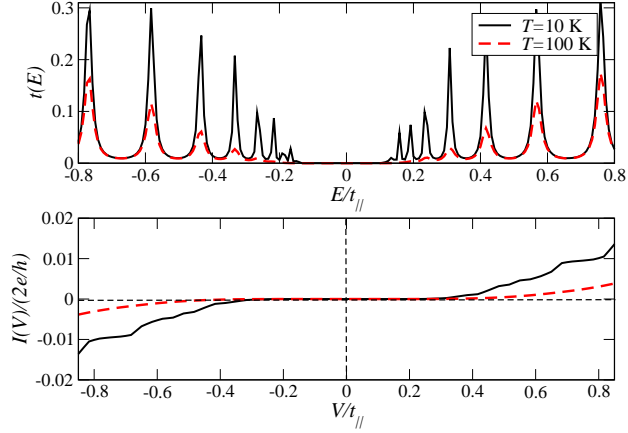}}
\caption{\label{fig:fig3}%
(Color) Electronic transmission and corresponding current in the weak-coupling limit with
ohmic dissipation ($s=1$) in the bath. Parameters: $N=20, J_0/\oc= 0.2,
t_{\perp}/t_{||}=0.6, \Gamma_{\textrm{L/R}}/t_{||}=0.5$ }
\end{figure}
With the change of variables $z=\omega t$, $\Phi(t)$ can be written as:
\begin{eqnarray}
\Phi(t)&=&J_0 \oc^{-s} t^{1-s} \int_{0}^{\infty} \mathrm{d}z \, z^{s-2} \ee^{-\frac{z}{\oc
t}} \nonumber \\ & &\times \left (1-\ee^{-\ii z}+
2\frac{1-\cos{z}}{\ee^{z\frac{\beta\oc}{\oc t}}-1}\right). \label{eq:eq14}
\end{eqnarray}
As far as $\oc t \gg \beta\oc $ this can be simplified to:
\begin{eqnarray}
\Phi(t)&\approx & J_0 \oc^{-s} t^{1-s} \int_{0}^{\infty}  \mathrm{d}x \, z^{s-2}
\ee^{-\frac{z}{\oc t}} \nonumber \\ & & \times \lrb 1-\ee^{-\ii z}+
2\frac{\beta\oc}{\oc t}\frac{1-\cos{z}}{z} \rrb.
\end{eqnarray}
Since in the long-time limit the low-frequency bath modes are giving the most
important contribution we may expect some qualitative differences in the ohmic
and superohmic regimes. For $s=1$ we obtain $\Phi(t)\sim \pi
\frac{J_0}{\oc}\frac{\KB T}{\oc}(\oc t)$ which leads to ($\Delta(s=1)=J_0$):
\begin{eqnarray}
{\bf G}^{-1}(E)&=&{\bf G}_0^{-1}(E)- t^2_{\perp} \frac{1}{E+J_0+\ii
\pi\frac{J_0}{\oc} \KB T}{\bf 1}, \label{eq:eq15}
\end{eqnarray}
\ie~there is only a pure imaginary contribution from the bath. For the simple case of
$N=1$ (a two-states model) one can easily see that the gap
approximately scales as $\sqrt{\KB T}$; thus it grows with increasing
temperature. This is shown in Fig.~\ref{fig:fig3}, where we also see that the
intensity of the transmission resonances strongly goes down with increasing
temperature. The gap enhancement is induced by the suppression of
the transmission peaks of the frontier orbitals, i.~e.~those closest to the Fermi energy.

For $s=3$ and $\KB T/\oc < 1$, $\Phi(t)$ takes a nearly temperature independent
value proportional to $J_0/\oc$. As a result the gap is slightly reduced
($t_{\perp}\to t_{\perp}\ee^{-J_0/\oc}$) but, because of the weak-coupling
condition, the effect is rather small. \\ From this discussion we can conclude that
in the weak-coupling limit ohmic dissipation in the bath induces an enhancement of
the electronic gap while superohmic dissipation does not appreciably affect it. In
the high-temperature limit $\KB T/\oc > 1$ a short-time expansion can be performed
which yields similar results to those of the strong-coupling limit (see next
section),~\cite{Ao96prb} so that we do not need to discuss them here. Note farther
that the gap obtained in the weak-coupling limit is an ``intrinsic'' property of the
electronic system; it is only quantitatively modified by the interaction with the
bath degrees of freedom. We thus trivially expect a strong exponential dependence of
$t(E=E_{{\rm F}})$, typical of virtual tunneling through a gap. Indeed, we find
$t(E=E_{{\rm F}})\sim \exp{(-\beta \, L)}$ with $\beta\sim 2-3 \, \AA^{-1}$.

\begin{figure}[t]
\vskip 0.7cm
\centerline{
\includegraphics[width=.7\linewidth]{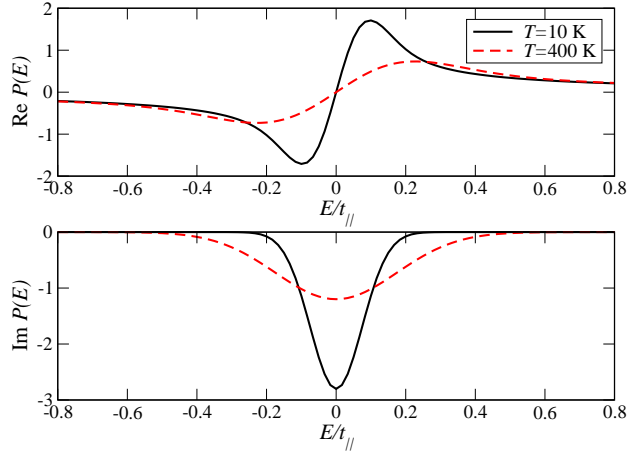}%
} \caption{\label{fig:fig4}(Color) Temperature dependence of the real and imaginary parts
of $P(E)$ for $N=20$, $J_0/\oc=10, t_{\perp}/t_{||}=0.4,
\Gamma_{\textrm{L/R}}/t_{||}=0.5$. With increasing temperature the slope of the real
part near $E=0$ decreases and the imaginary part broadens and loses intensity. A
similar qualitative dependence on $J_0$ was found (not shown). }
\end{figure}

\paragraph*{(iii) Beyond MF: strong coupling limit (SCL)}
In this section we elaborate  on the strong-coupling regime, as defined by the
condition $J_0/\oc > 1$. In
the SCL the main contribution to the time integral in Eq.~(\ref{eq:eq9}) arises
from short times. Hence a short-time expansion of $\Phi(t)$ may already give
reasonable results and it allows, additionally, to find an analytical
expression for ${\bf P}(E)$. At $t\ll \oc^{-1}$ we find,
\begin{eqnarray}
\label{eq:eq16} \Phi(t)&\approx&{\ii} \Delta \, t + (\oc t)^2 \, \kappa_{0}(T) \\
P_{\ell j}(E)&=&-{\ii}\delta_{\ell j}\, \, \int_{0}^{\infty} \mathrm{d}t \,
\ee^{{\ii}(E-\epsilon+{\ii}0^{+})t}\, \ee^{-(\oc t)^2 \kappa_{0}(T)} \nonumber \\
&=&-{\ii}\delta_{\ell j}\, \, \frac{\sqrt{\pi}}{2}\frac{1}{\oc
\sqrt{\kappa_{0}(T)}} \; \exp \lrb
{-\frac{(E-\epsilon+{\ii}0^{+})^2}{4\omega^2_\textrm{c} \kappa_{0}(T)}} \rrb
\nonumber \\ && \times \lrb 1+{\textrm{erf}}\lsb
\frac{{\ii}(E-\epsilon+{\ii}0^{+})}{2\oc \sqrt{\kappa_{0}(T)}}\rsb \rrb, \nonumber \\
\kappa_{0}(T)&=& \frac{1}{2\omega^2_\textrm{c}}\int_{0}^{\infty} \mathrm{d}\omega J(\omega)
\coth{\frac{\omega}{2 \KB T}}.  \nonumber
\end{eqnarray}

%
Before presenting the results for the electronic transmission, it is
useful to first consider the dependence of the real and imaginary parts of
${\bf P}(E)$ on temperature and on the reduced coupling constant $J_0/\oc$.
Both functions are shown in Fig.~\ref{fig:fig4}. We see that around the Fermi level at
$E=0$ the
real part is approximately linear, ${\rm Re} \, P(E)\sim E$ while the imaginary part shows a
Lorentzian-like behavior. The imaginary part loses intensity and becomes
broadened with increasing temperature or $J_0$, while the slope in the real
part decreases when $\KB T$ or $J_0$ are increased.
\begin{figure}[t]
\vskip 0.71cm
\centerline{
\epsfclipon
\includegraphics[width=.7\linewidth]{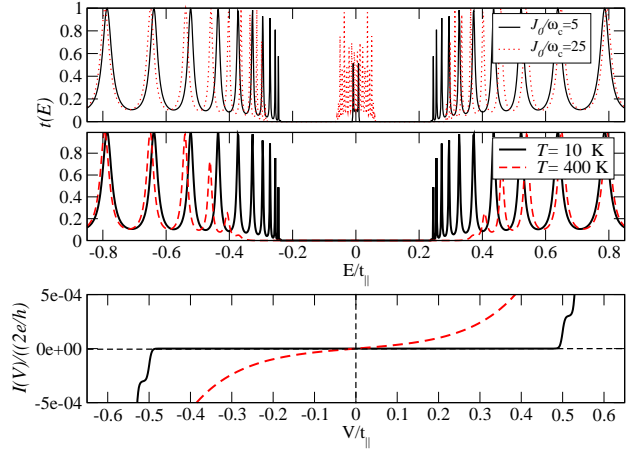}%
\epsfclipoff
}
\caption{\label{fig:fig5}%
(Color) Upper panel: $t(E)$ with ${\rm{Im}}\, P(E)=0$; the intensity of the resonances on
the central narrow band is strongly dependent on $J_0/\omega_c$ and $\KB T$ (not
shown). Temperature dependence of $t(E)$ with full inclusion of $P(E)$ (middle
panel) and corresponding current (lower panel) for $N=20$, $J_0/\oc=5,
t_{\perp}/t_{||}=0.5, \Gamma_{\textrm{\textrm{L/R}}}/t_{||}=0.2$. The pseudo-gap
increases with temperature. }
\end{figure}
If we neglect for the moment the imaginary part (the dissipative influence of the
bath), we can understand the consequences of the real part being nonzero around the
Fermi energy, \ie~in the gap region of the model of Ref.~\cite{Cuniberti02prb}.
 The solutions of the non-linear equation
${\rm{det}}|(E-t_{\perp}^{2} {\rm Re}\, P(E)){\bf 1}-{\cal H}_\textrm{C}|=0$ give the new
poles of the Green function of the system  in presence of the phonon bath.
For comparison, the equation determining the eigenstates {\it without} the bath
is simply ${\rm{det}}|(E-t_{\perp}^{2}/E){\bf 1}-{\cal H}_\textrm{C}|=0$. It
is just the $1/E$ dependence near $E=0$ that induces the appearance of two
electronic bands of states separated by a gap. In our present
study, however, ${\rm Re} \, P(E\to 0)$ has no singular behavior and additional
poles of the Green function  may be expected to appear in the low-energy sector.
This is indeed the
case, as shown in Fig.~\ref{fig:fig5}\ (upper panel). We find a third band of
states around the Fermi energy, which we may call a polaronic band because it results from the
strong interaction between an electron and the bath modes. The  intensity of this band as well as its band
width strongly depend
on temperature and on $J_0$. When $\KB T$ (or $J_0$) become large enough, these
states spread out and eventually merge with the two other side bands. This
would result in a transmission spectrum similar of a gapless system.

This picture is nevertheless not complete since the imaginary component of
$P(E)$ has been neglected. Its inclusion leads to a dramatic modification of
the spectrum, as shown in Fig.~\ref{fig:fig5}\ (middle panel). We now only see
two bands separated by a gap which basically resembles the semiconducting-type
behavior of the original model. The origin of this gap or rather
{\it pseudo-gap} (see below) is however quite different. It turns out that the
imaginary part of $P(E)$, being peaked around $E=0$,
strongly suppresses the transmission resonances
belonging to the third band. Additionally, the frontier orbitals on the side bands, \ie~ orbitals closest
to the gap region, are also strongly damped, this
effect becoming stronger with increasing temperature (${\rm{Im}}\,P(E)$
broadens). This latter effect has some similarities with the previously discussed
weak-coupling regime.
Note, however, that the new electronic manifold around the Fermi energy
does not appear in the weak-coupling
regime. We further stress that the density
of states around the Fermi level is not exactly zero (hence the term
pseudo-gap); the states on the polaronic  manifold, although strongly damped,
contribute nevertheless with a finite temperature dependent background to the
transmission. As a result, with increasing temperature, a crossover from
``semiconducting'' to ``metallic'' behavior in the low-voltage region of the $I$-$V$
characteristics takes place, see Fig.~\ref{fig:fig5}\ (lower panel). The slope
in the $I$-$V$ plot becomes larger when $t_{\perp}$ is reduced, since the side
bands approach each other and the effect of ${\rm{Im}}\,P(E)$ is reinforced.

\begin{figure}[t]
\centerline{\includegraphics[width=.7\linewidth]{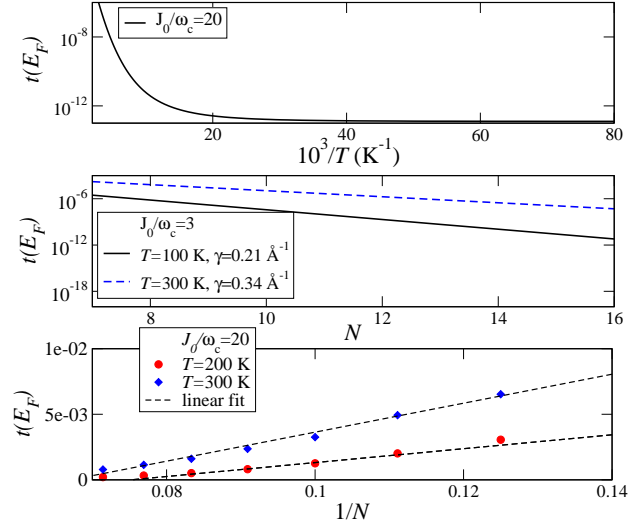}}
\caption{\label{fig:fig6}%
(Color) Upper panel: Arrhenius plot for $t(E_F)$.
Parameters: $ N=20,\, t_{||}=0.6 \, eV, \, t_{\perp}/t_{||}=0.2,
\,\Gamma_{\textrm{\textrm{L/R}}}/t_{||}=0.3$.
Middle and lower panels: Length dependence of $t(E_F)$ at different
temperatures for two different strengths of the electron-bath
coupling $J_0/\omega_{\rm c}$. The electronic coupling parameters are the same
as in the upper panel.
 }
 \end{figure}

In Fig.~\ref{fig:fig6} (top panel) an Arrhenius plot of the transmission at the
Fermi energy is shown, which suggests that activated transport is the physical
mechanism for propagation at low energies. Increasing the coupling to the phonon
bath makes the suppression of the polaronic  band around $E=0$ less effective
(${\rm{Im}}\,P(E\sim 0)$ decreases) so that the density of states around this energy
becomes larger. Hence the absolute value of the transmission will also increase.  On
the other side, increasing $t_{\perp}$ leads to a reduction of the transmission at
the Fermi level, since the energetic separation of the side  bands increases with
$t_{\perp}$.

A controversial issue in transport through molecular wires is the actual length
dependence of the electron transfer rates or correspondingly, of the linear
conductance. This is specially critical in the case of DNA nanowires
\cite{Treadway02chemphys,Meggers98jacs,Kelley99science}. Different functional
dependences have been found in charge transfer experiments ranging from strong
exponential behavior related to superexchange mediated electron transfer
\cite{Meggers98jacs} to algebraic dependences typical of thermal activated  hopping
\cite{Treadway02chemphys,Kelley99science}. As far as transport experiments are
concerned, the previously mentioned experiments at the group of N.
Tao~\cite{Xu04nanolett} reported an algebraic length dependence of the conductance
for poly(GC) oligomers in solution. We have investigated the length dependence of
$t(E_F)$ and found for the strong dissipative regime $J_0/\omega_{\rm c}> 1$, an
exponential law for energies close to $E_{\rm F}$, $t(E_{\rm F}) \sim \exp(-\gamma
N)$.  At the first sight, this might be not surprising since a gap in the spectrum
does exist. Indeed, in the absence of the bath, \ie~with an intrinsic gap, we get
decay lengths $\gamma_{{\rm coh}}$ of the order of 2 $\, {\rm \AA}^{-1}$. However,
as soon as the interaction with the bath is included, we find values of $\gamma$
much smaller than expected for virtual tunneling, ranging from $0.15 \, {\rm
\AA}^{-1}$ to $0.4 \, {\rm \AA}^{-1}$. Additionally, $\gamma$ is strongly dependent
on the strength of the electron-bath coupling $J_0/\oc$ as well as on temperature;
$\gamma$ is reduced when $J_0/\oc$ or $k_{\rm B} T$ increases, since in both cases
the density of states within the pseudo-gap increases. Remarkably, a further
increase of the electron-bath coupling eventually leads to an algebraic length
dependence, see lower panel of Fig.~~\ref{fig:fig6}.

The studies presented in this section indicate that the presence of a complex
environment, which induces decoherence and dissipation, can dramatically modify the
electronic response of a nanowire coupled to electrodes.  Electron transport on the
low-energy sector of the  transmission spectrum is supported by the formation of
(virtual) polaronic states. Though strongly damped, these  states manifest
nonetheless with  a finite density of states inside the bandgap and mediate
thermally activated transport.

\section{Conclusions and Perspectives}

In this chapter we have reviewed the method of nonequilibrium Green functions and
few selected  applications to problems related with charge transport at the
molecular scale. Hereby we have only focused on minimal model Hamiltonian
formulations which build a very appropriate starting point to illustrate the power
and range of validity of such techniques. We have showed how this approach can be
used to deal with a variety of physical systems, covering both noninteracting and
interacting cases. Thus, so different issues as coherent transport, Coulomb blockade
phenomena, charge-vibron interaction,  coupling to dissipative environments, and the
Kondo effect (not addressed in this review) can be in principle treated on the same
footing. Specially, the existence of well-developed diagrammatic techniques allows
for a systematic treatment of interactions in nanoscale quantum systems.  For the
sake of space, we did not deal with  applications of NGF techniques to
spin-dependent transport, a field that has been increasingly attracting the
attention of the physical community in the past years due to its potential
applications in quantum information theory and quantum
computation~\cite{Nielsen00book,Bouwmeester00book}. For the same reason, the
implementation of NGF into first-principle based approaches was not discussed
neither. This is nevertheless a crucial methodological issue, since system-specific
and realistic information about molecule-metal contact details, charge transfer
effects, modifications of the molecular electronic structure and configuration upon
contacting, the electrostatic potential distribution in a device, etc can only be
obtained via a full {\it ab initio} description of transport. For charge transport
through noninteracting systems this has been accomplished some years ago by
combining NGF with DFT methods. The inclusion of interactions, however, represents a
much stronger challenge and has been mainly carried out, within the self-consistent
Born-approximation, for the case of tunneling charges coupling to vibrational
excitations in the molecular region. Also there were some efforts to include
dynamical correlations into DFT-based approach. Much harder and till the present not
achieved at all is the inclusion of electronic correlation effects, responsible for
many-particle effects like Coulomb blockade or the Kondo effect, in a
non-equilibrium transport situation. DFT-based techniques, being essentially
mean-field theories, cannot deal in a straightforward way with such problems and
have to be improved, e.g.~within the LDA+U approaches. For the case of equilibrium
transport, a generalization of the Landauer formula including correlations has been
recently formulated as well as first attempts to go beyond the linear response
regime; for strong out-of-equilibrium situations this will be, in our view, one of
the most demanding issues that the theoretical ``transport'' community will be
facing in the coming years.

\section{Acknowledgments}

The authors acknowledge the collaboration with Miriam del Valle, Marieta Gheorghe,
Michael Hartung, Sudeep Mandal, and Soumya Mohapatra, with whom part of the work
reviewed here was done. We appreciate very useful and illuminating discussions with
Andrea Donarini,  Pino D'Amico, Dietrich F\"orster, Milena Grifoni, Joachim Keller,
Abraham Nitzan, Norbert Nemec, Danny Porath, Florian Pump, Klaus Richter, Eugen
Starikow, and Juyeon Yi. We thank Antti-Pekka Jauho, Alessandro Pecchia, Tom\'a\v{s}
Novotn\'y, Hassan Raza, and Kristian Thygesen for their comments.

Funding by the EU through grant FET-IST-2001-38951 (international collaboration
"DNA-based nanoelectronic devices (DNAnanoDEVICES)"), by the DFG through grant
CU~44/3-2 (international collaboration "Single molecule based memories"), by the
Volkswagen Stiftung under Grant No. I/78 340, by the DFG Graduated School
``Nonlinearity and Nonequilibrium in Condensed Matter'' GRK-638, by the DFG Priority
Program ``Quantum Transport at the Molecular Scale'' SPP1243, and by Collaborative
Research Center SFB 689 is acknowledged. We also thank the Vielberth Foundation, the
Minerva Foundation, the German Exchange Program (DAAD), the Humboldt foundation, and
the German-Israeli foundation (GIF) for financial support.

\tableofcontents

\end{document}